\documentclass[11pt]{article}
\usepackage[margin=1in]{geometry}
\usepackage{times}
\usepackage[
backend=biber,
style=alphabetic,
sorting=nyt,
maxnames=7,
maxalphanames=7,
backref=true,
url=false
]{biblatex}
\usepackage{tikz}
\usepackage{scalerel}
\usepackage{pict2e}
\usepackage{setspace}
\usepackage{fancyhdr}
\usepackage{tkz-euclide}
\usetikzlibrary{calc}
\usetikzlibrary{patterns,arrows.meta}
\usetikzlibrary{shadows}
\usetikzlibrary{external}
 
\usepackage{pgfplots}
\pgfplotsset{compat=newest}
\usepgfplotslibrary{statistics}
\usepgfplotslibrary{fillbetween}

\usepackage{xcolor}

\usepackage{nicefrac}

\usepackage[colorlinks=true,citecolor=blue,linkcolor=blue]{hyperref}

\hypersetup{
    colorlinks,
    linkcolor={red!50!black},
    citecolor={blue!50!black},
    urlcolor={blue!80!black}
}

\usepackage[]{amsmath,amssymb,amsfonts,latexsym,amsthm,enumerate,fullpage,xcolor,bbm,mathtools}
\usepackage{wrapfig}
\usepackage{float}
\usepackage[ruled]{algorithm2e}

\usepackage{enumitem}
\usepackage[nameinlink, noabbrev, capitalize]{cleveref}
\usepackage{comment}
\usepackage{nicefrac}
\crefname{prop}{Proposition}{Propositions}
\crefname{ineq}{inequality}{inequalities}
\creflabelformat{ineq}{#2(#1)#3}
\usepackage{todonotes}
\usepackage{multicol}
\newtheorem{globtheorem}{Theorem}

\newtheorem{globcorollary}{Corollary}

\newtheorem{counter}{Counter}[section]
\newtheorem{theorem}[counter]{Theorem}

\newtheorem{lemma}[counter]{Lemma}

\newtheorem{claim}[counter]{Claim}

\newtheorem{question}[counter]{Question}

\newtheorem{corollary}[counter]{Corollary}

\newtheorem{definition}[counter]{Definition}

\newtheorem{remark}[counter]{Remark}

\usepackage{subcaption}
\usepackage{graphicx}
\usepackage{color, colortbl}
\definecolor{LightCyan}{rgb}{0.88,1,1}
\definecolor{Gray}{gray}{0.9}
\usetikzlibrary{positioning}
\usetikzlibrary{decorations.pathmorphing}
\usetikzlibrary{automata,positioning}
\usetikzlibrary{arrows}
\usetikzlibrary{arrows.meta}
\usetikzlibrary{calc}

\usepackage{enumitem}
\newlength\caselen
\newlist{casesenum}{enumerate}{2}
\setlist[casesenum,1]{label=\textbf{Case~\arabic*.}, 
  itemindent=*,leftmargin=0pt}
\setlist[casesenum,2]{label=\textbf{Case~\roman*.}, 
  itemindent=*,leftmargin=\parindent}

\newcounter{constraint}
\crefname{constraint}{Constraint}{Constraints}
\Crefname{constraint}{Constraint}{Constraints}

\newcounter{step}
\crefname{step}{Step}{Steps}
\Crefname{step}{Step}{Steps}

\newenvironment{step}{
  \begin{list}{Step~\thestep.}{
    \usecounter{step}
    \setlength{\leftmargin}{2em}
    \setlength{\labelsep}{0.5em}
  }
}{\end{list}}

\newcommand{\N}{\mathbb{N}}

\newcommand{\poly}{\operatorname{poly}}

\newcommand{\U}{\mathbf{U}}
\newcommand{\V}{\mathbf{V}}
\newcommand{\X}{\mathbf{X}}

\newcommand{\A}{\mathbf{A}}
\newcommand{\B}{\mathbf{B}}

\newcommand{\bR}{\mathbf{R}}

\newcommand{\bZ}{\mathbf{Z}}

\newcommand{\bx}{\mathbf{x}}
\newcommand{\by}{\mathbf{y}}
\newcommand{\bz}{\mathbf{z}}

\newcommand{\bpi}{\boldsymbol{\pi}}

\renewcommand{\multicitedelim}{\addsemicolon\space}

\newcommand{\zo}{\{0,1\}}

\newcommand{\abs}[1]{\left\lvert #1 \right\rvert}

\usepackage{xspace}

\newcommand{\eps}{\varepsilon}

\newcommand{\Nbr}{\text{Nbr}}

\newcommand{\Adv}{\mathsf{Adv}}
\newcommand{\Res}{\ensuremath{\mathsf{Res}}}

\newcommand{\rest}{\ensuremath{\mathsf{rest}}}

\newcommand{\cA}{\mathcal{A}}
\newcommand{\cF}{\mathcal{F}}

\newcommand{\cP}{\mathcal{P}}

\DeclareMathOperator{\I}{\mathbf{I}}

\DeclareMathOperator*{\argmax}{\arg\!\max}

\DeclareMathOperator*{\E}{\mathbb{E}}

\newcommand{\assembly}[1][\relax]{%
\ifx\relax#1\textrm{assembly}\xspace%
\else\ensuremath{{\left(#1\right)}}\textrm{-assembly}%
\fi
}

\newcommand{\NOBF}[1][\relax]{%
\ifx\relax#1\textrm{NOBF source}\xspace%
\else\ensuremath{{\left(#1\right)}}\textrm{-NOBF source}%
\fi
}
\newcommand{\NOBFs}[1][\relax]{%
\ifx\relax#1\textrm{NOBF sources}\xspace%
\else\ensuremath{{\left(#1\right)}}\textrm{-NOBF sources}%
\fi
}

\usepackage{subfiles}
\newcommand{\dobib}{
    \printbibliography
}

\newcommand{\red}[1]{{\color{red} {#1}}}

\newcommand{\eshan}[1]{{\color{red} \footnotesize(Eshan: #1)}}
\newcommand{\mohit}[1]{{\color{orange} \footnotesize(Mohit: #1)}}
\newcommand{\nomi}[1]{{\color[RGB]{0, 100, 0} \footnotesize(Nomi: #1)}}
\newcommand{\rocco}[1]{{\color{blue} \footnotesize{Rocco: #1}}}
\newcommand{\rnote}[1]{\footnote{{\bf \color{blue}Rocco}: {#1}}}

\newcommand{\nocomments}{
    \renewcommand{\eshan}[1]{}
    \renewcommand{\mohit}[1]{}
    \renewcommand{\nomi}[1]{}
    \renewcommand{\rocco}[1]{}
    \renewcommand{\rnote}[1]{}
}

\nocomments 

\begin{document}
\renewcommand{\dobib}{}
\renewcommand*{\multicitedelim}{\addcomma\space}
 

 \title{\fontsize{15}{17}\selectfont Improved Bounds for Coin Flipping, Leader Election, and Random Selection}

  \author{  Eshan Chattopadhyay\thanks{Supported by a Sloan Research Fellowship, NSF CAREER Award 2045576, and NSF Award CCF-2514586.}\\ Cornell University\\ \texttt{eshan@cs.cornell.edu}  \and Mohit Gurumukhani\footnotemark[1] \\ Cornell University\\ \texttt{mgurumuk@cs.cornell.edu} \and  Noam Ringach \thanks{Supported by NSF GRFP grant DGE – 2139899,  NSF CAREER Award 2045576, a Sloan Research Fellowship, and NSF Award CCF-2514586.}   \\ Cornell University\\ \texttt{nomir@cs.cornell.edu} \and Rocco A. Servedio\thanks{Supported by NSF Award CCF-2106429 and NSF Award CCF-2211238.} \\ Columbia University \\ \texttt{rocco@cs.columbia.edu}}

 \date{}

 \maketitle
 \pagenumbering{Roman}

\begin{abstract}
Random selection is a fundamental task in fault-tolerant distributed computing where processors select a random outcome from some domain. Two special cases of this, leader election (where the processors designate a leader amongst themselves) and collective coin flipping (where the processors agree on a common random bit), have been especially widely studied. We study these problems in the full-information model, where processors communicate via a single broadcast channel, have access to private randomness, and face a computationally unbounded adversary that controls some  of the processors. Despite decades of study, key gaps remain in our understanding of the trade-offs between round complexity, communication per player in each round, and adversarial resilience.
We make progress by proving new lower bounds for coin flipping protocols and both new upper and lower bounds for leader election and random selection protocols.

We first show that any $k$-round coin flipping protocol, where each of $\ell$ players sends $1$ bit per round, can be biased by $O(\ell/\log^{(k)}(\ell))$ bad players. We obtain the same lower bound (with an additional $\log^{(k+1)}(\ell)$ factor in the numerator) for leader election as well.
This strengthens the previous best lower bounds [RSZ, SICOMP 2002], which ruled out coin flipping protocols resilient to $O(\ell / \log^{(2k-1)}(\ell))$ bad players and leader election protocols resilient to $O(\ell / \log^{(2k+1)}(\ell))$ bad players.
As a consequence, we establish that any protocol tolerating a linear fraction of corrupt players, while restricting player messages to 1 bit per round, must run for at least $\log^* \ell - O(1)$  rounds, improving on the prior best lower bound of $\frac{1}{2} \log^* \ell - \log^* \log^* \ell$.  We additionally show that the current best protocols that handle a linear number of corrupt players (from [RZ, JCSS 2001], [F, FOCS 1999]) are near optimal in terms of round complexity and communication per player in a round. 

We next initiate the study of one-round random selection protocols where each player sends 1 bit in the round. 
For all $m \ge (\log(\ell))^2$, we obtain an \emph{optimal} one-round protocol: We construct a protocol that is resilient to $O(\ell / m)$ bad players, outputting $m$ uniform random bits.
And, we show that any protocol that outputs $m$ uniform random bits can be corrupted using $O(\ell / m)$ bad players. As far as we are aware, this is the first provably optimal protocol for any task in the full information model.

As a consequence of our construction, we obtain a one-round leader election protocol resilient to $\ell / (\log(\ell))^2$ bad players, improving on the previous best protocol from [RZ, JCSS 2001] that is resilient to only $\ell / (\log(\ell))^3$ bad players and requires players to send many bits.
When $m = (\log(\ell))^2$, our resilience parameter matches that of the best one-round coin flipping protocol by Ajtai and Linial, which only outputs one bit. 
To obtain our lower bound, we introduce and study multi-output influence, a natural extension of the notion of influence of boolean functions to the multi-output setting.
\end{abstract}

\maketitle

  \newpage

\setcounter{tocdepth}{2}
\tableofcontents
 \newpage
\pagenumbering{arabic}
 



\section{Introduction} \label{sec: intro}

A fundamental task in fault tolerant distributed computing is \emph{random selection}, where processors collectively select a random outcome from a domain in the presence of adversarial faults. Two special cases of this, \emph{collective coin flipping}, where processors agree on a common random bit, and \emph{leader election}, where the task is to select a ``leader'' amongst them, are  extremely important. These tasks have been widely studied under various assumptions about the communication channels between the processors, the kind of faults allowed, the power of the adversary, and the kinds of randomness each processor has access to. We study these problems under the assumption that there is only a single common broadcast channel, that adversarial processors are computationally unbounded, and that each processor has access to private randomness. This extensively studied model was introduced by Ben-Or and Linial \cite{BL85coin}, and is known as the \emph{full information model} since all processors have access to the same information.

The standard way of modeling how processors coordinate their efforts is through the notion of a \emph{protocol}. All players (processors henceforth will be called players) agree on a fixed protocol $\pi$ beforehand and execute it, at the end of which they all agree on an outcome from the output domain. We assume that there is an adversary $\cA$ that selects a subset of players before the protocol begins and continues controlling them throughout the execution of the protocol. We refer to the coalition of controlled players as ``bad'' and the remaining players as ``good''.
A protocol consists of one or more rounds. In each round, all players should flip $r$ private random coins and broadcast those $r$ bits to everyone else. The identity of the sender of a bit is always known. This continues for $k$ rounds, after which the protocol determines the outcome. 
We assume that each round of the protocol is asynchronous, and we always consider the worst-case scenario in which all good players broadcast their bits at the beginning of the round. Then, based on their outputs, the adversary $\cA$ determines the $r$ bits output by each of the bad players. As a result, the outputs of the bad players are coordinated and depend on the outputs of the good players and the outputs of previous rounds. We note that the players are synchronized in between rounds; i.e., a round ends only when all outputs of all players are received. For formal definitions of these protocols, we refer the reader to \cref{subsubsec: defn leader election and collective coin flipping}.

Since bad players in this model are computationally unbounded, cryptography-based protocols, which are standard for the Byzantine generals problem and other related models where the adversary is (say) polynomially bounded, are not useful in our setting. Nevertheless, remarkable protocols do exist in this model that can guarantee that a good leader is always chosen with non-trivial probability, or that the outcome of the collective coin flip has small bias.

When constructing random selection protocols, the goal is to minimize the number of rounds, minimize the number of bits each player sends per round, and maximize the number of bad players the protocol can handle, while attaining the objective of making the selected outcome as ``random'' as possible. 
The ideal objective in random selection is that (even in the presence of bad players), the output distribution is statistically very close (or even just non-trivially close) to the uniform distribution over the output domain.
When the output domain is $\zo$, this exactly corresponds to the objective in collective coin flipping.
A slightly weaker objective in random selection is that the output distribution should not be too concentrated on a too-small set of outcomes from the domain.
This corresponds to the objective in leader election where the task is to elect a good player as a leader with non-trivial probability. \footnote{Concretely, if $b$ out of $[\ell]$ players are bad then the task is to ensure that the output distribution is not concentrated on any $S\subset [\ell]$ with $\abs{S} \le b$.}

\subsection{Related Work}

We first survey the best protocol constructions that were known prior to our work.

Given a leader election protocol, one can turn it into a collective coin flipping protocol at the expense of one extra round by making the elected leader flip a coin and output the result (see \cref{claim: leader to coin flip} for a proof). 
Starting with Ajtai and Linial's non-explicit function \cite{AL93resilient}, later made explicit by \cite{Chattopadhyay2019TwoSource,Meka17resilient, IMV23resilient, IV25resilient}, one can construct a one-round collective coin flipping protocol where each player sends $1$ random bit that is resilient to $O(\ell / (\log (\ell))^2)$ bad players. Leader election protocols that can guarantee a constant probability of electing a good leader in the presence of $O(\ell)$ bad players have also been widely studied: when the number of random bits per round is restricted to be $1$, a long line of works \cite{BL85coin, Saks89leader, AN93leader, CL95leader, ORV94leader, Zuckerman97sample, BN00leader, feige99lightestbin, Antonakopoulos06leader} constructed various explicit protocols, the best of which runs in $O(\log (\ell))$ rounds. Moreover, when the number of random bits per round is unrestricted, \cite{RZ01leader, feige99lightestbin} constructed explicit leader election protocols that run in $\log^*(\ell) + O(1)$ rounds.\footnote{Recall that $\log^*(\ell)$ is the minimum number of logs that need to be applied to $\ell$ until it attains value at most $1$.} 

The study of general random selection protocols with the weaker objective of not being too concentrated on a small set of outputs was initiated by \cite{GGL98selection}, and later works constructed protocols with stronger guarantees and reduced complexity \cite{SV08selection, GVZ06selection}. All these protocols last for many rounds, each player can send many bits per round, and the number of bad players considered is linear in $\ell$ and sometimes even more than $\ell / 2$. The work of \cite{KZ07NOSF} constructed a one round random selection protocol with all players sending $1$ bit that is resilient to $b$ bad players and outputs $\ell / (b^{1.58})$ random bits.

On the lower bound side, very few results are known and they have focused mainly on collective coin flipping since lower bounds here translate into lower bounds for leader election protocols with one fewer round (see \cref{cor: coin lb implies leader lb} for a formal claim). 
First, it is well known that no coin flipping or leader election protocol can handle $\ell / 2$ or more bad players \cite{Saks89leader}, regardless of the number of rounds and the number of random bits allowed per round. For one-round coin flipping protocols where players send 1 bit per round, \cite{KKL88influence} showed that every protocol can be biased towards some outcome by some set of $O(\ell / \log (\ell))$ bad players.
For multiple rounds, \cite{RSZ02leader} showed that any $k$-round coin flipping protocol where each player sends 1 bit per round can be biased towards some outcome by $O(\ell / \log^{(2k-1)}(\ell))$ bad players.\footnote{We use the notation $\log^{(i)} \ell$ to denote $i$-times-iterated logarithm, i.e.~$\overbrace{\log \cdots \log}^{i \text{~times}} \ell$.}
This also implies that in order to handle $\Theta(\ell)$ bad players when the number of bits per round is $1$, the number of rounds required is $\frac{1}{2}\log^*(\ell) - \log^*(\log^*(\ell))$. 
For random selection tasks, \cite{GGL98selection} showed that for any $\mu > 0$,  $b$ bad players can always ensure that the output lies in some set of density $\mu$ with probability at least $\mu^{1 - (b / \ell)}$. Kamp and Zuckerman showed that for any one round random selection protocol outputting $m$ bits and where all players send $1$ bit, $O\left(\frac{\ell}{m}\right)$ bad players can corrupt the protocol so that the output distribution is $0.1$-far from $\U_m$ \cite{KZ07NOSF}.

The work of \cite{FHHHZ19} gave lower bounds on collective coin flipping protocols under arbitrary product distributions, rather than the uniform distribution, on the Boolean cube. For further details, we refer the reader to the excellent survey of Dodis \cite{dodis2006fault} on protocols and lower bounds in the full information model.

We mention that one-round coin flipping protocols, random selection protocols that output many uniform random bits, and random selection protocols where the output is not too concentrated on a small set, naturally arise in extractor theory where these protocols are referred to as extractors / condensers for Non-Oblivious Symbol Fixing (NOSF) and Non-Oblivious Bit Fixing (NOBF) sources. These protocols have been a pivotal ingredient in a surge of works over the past decade that have resulted in constructions of near optimal two source extractors, two source condensers, affine extractors, sumset extractors, and more \cite{Chattopadhyay2019TwoSource, Li16TwoSource, BDT17TwoSource, BCDT19, CGL21affine, CL22sumset, Li23sumset}.

\subsection{Our Results}

We broadly obtain two kinds of results: (1) Improved lower bounds for coin flipping protocols and (2) Improved upper and lower bounds for one-round random selection protocols. For ease of exposition we will specialize and also describe our results in strand (2) for leader election protocols.

\subsubsection{Improved lower bounds for collective coin flipping}


Our lower bounds for collective coin flipping improve and subsume all previous $k$ round lower bounds (established by \cite{RSZ02leader}) for $k > 1$. 

Our first non-trivial improvement is for two-round protocols where players send $1$ bit per round. We show that $\frac{\ell}{\log(\log(\ell))}$ players suffice to corrupt any such protocol. Since the previous best bound was $\frac{\ell}{\log(\log(\log(\ell)))}$, due to \cite{RSZ02leader}, we thus exponentially improve the bound on the fraction of players that can corrupt such a protocol.
\footnote{We recall that for one-round protocols, there is a multiplicative gap of $\log(\ell)$ between the upper bound of $\frac{\ell}{(\log(\ell))^2}$ and the lower bound of $\frac{\ell}{\log(\ell)}$. Resolving this is a major question in Boolean Fourier analysis.}
Generally, we obtain:
\begin{globtheorem}[Informal version of \cref{thm: protocol lb}]\label{thm: intro protocol lb}
Let $\pi$ be a $k$-round coin flipping protocol over $\ell$ players where players send 1 bit per round.
Then, there exists an outcome $o\in \zo$ such that $b = O\left(\frac{\ell}{\log^{(k)}(\ell)}\right)$ bad players can bias $\pi$ to output $o$ with probability $\ge 0.99$.
\end{globtheorem}

For all $k > 1$, this directly improves upon the previous best lower bounds, due to \cite{RSZ02leader}, which showed that $O\left(\frac{\ell}{\log^{(2k-1)}(\ell)}\right)$ players can corrupt such protocols.

Arguably, the most important open problem in collective coin flipping and leader election is to figure out the optimal number of rounds for a protocol handling linear fraction of bad players. The above result lets us make progress on this question by giving us the following lower bound:
\begin{globcorollary}[Informal version of \cref{cor: round lb}]
\label{cor into: round lb}
Let $\pi$ be a $k$-round coin flipping protocol over $\ell$ players where players send 1 bit per round and where $k\le \log^*(\ell) - O(1)$. 
Then, there exists an outcome $o\in \zo$ such that $b = 0.01\ell$ players can bias $\pi$ to output $o$ with probability $\ge 0.99$.
\footnote{Since $k$-round leader election protocols imply ($k+1$)-round coin flipping protocols, this lower bound also applies to leader election protocols.}
\end{globcorollary}
This shows that any coin flipping protocol that can handle linear sized coalitions, where players send $1$ bit per round, requires at least $\log^*(\ell) - O(1)$ rounds. This improves upon the best previous lower bound of $\frac{1}{2} \log^*(\ell) - \log^*(\log^*(\ell))$ rounds due to \cite{RSZ02leader}. 

Our lower bound also essentially matches the number of rounds, $\log^*(\ell)$, used by the current best coin flipping protocols from \cite{RZ01leader, feige99lightestbin} that can handle linear sized coalition of bad players, given the additional freedom that players can send unlimited bits per round. This raises the natural question of whether we can further meaningfully show that the protocols of \cite{RZ01leader, feige99lightestbin} are optimal. We indeed show this by obtaining the following lower bound on protocols where players can send many bits per round:
\begin{globtheorem}[Informal version of \cref{thm: protocol lb longer messages}]
\label{thm: intro protocol lb longer messages}
Let $\pi$ be a $k$-round coin flipping protocol over $\ell$ players where each player can send $(\log^{(i)}(\ell))^{0.99}$ many bits in round $i$, and $k\le \log^*(\ell) - O(1)$.
Then, there exists an outcome $o\in \zo$ such that $b = 0.01\ell$ players can bias $\pi$ to output $o$ with probability $\ge 0.99$.
\footnote{Similarly, this lower bound also applies to leader election protocols.}
\end{globtheorem}

The coin flipping protocols from \cite{RZ01leader, feige99lightestbin} that can handle a linear sized coalition of bad players and take $\log^*(\ell) + O(1)$ rounds have the following property that in round $i$, they only require that all players send $O(\log^{(i)}(\ell))$ many bits. Hence, \Cref{thm: intro protocol lb longer messages} shows that the protocols from \cite{RZ01leader, feige99lightestbin} are essentially optimal, in the sense that if the number of rounds and the number of bits each player can send in round $i$ are even slightly lowered, then no protocol can handle a linear sized coalition of bad players. 

A weaker result of this flavor was obtained by \cite{RSZ02leader}; their result showed that if each player can send at most $(\log^{(2i-1)}(\ell))^{0.99}$ bits in round $i$, then any such protocol cannot handle a linear sized coalition of bad players. \cref{thm: intro protocol lb longer messages} in fact resolves an open problem raised in \cite{RSZ02leader} where they had exactly asked for the result that we obtain.

We remark that a key ingredient towards proving \cref{thm: intro protocol lb} is a new lemma regarding biasing most functions from a family of functions using a common set of bad players and a small specialized set of bad players specific to each function being biased.  For a formal statement, refer to \cref{lem: most functions from family can be corrupted with random and heavy set}.

We complement the lower bound result described above by constructing improved constant-round protocols in which each player can send one bit per round. We obtain:
\begin{globtheorem}[Informal version of \cref{thm: construct protocol}]
For any $k\ge 2$, there exists an explicit $k$-round coin flipping protocol over $\ell$ players where each player sends $1$ bit per round such that when the number of bad players is at most $O\left(\frac{\ell}{\log (\ell)\cdot(\log^{(k)} (\ell))^2}\right)$, the output bit is $0.01$-close to the uniform distribution.
\end{globtheorem}
When $k = 2$, this protocol can handle any $O\left(\frac{\ell}{\log (\ell)\cdot(\log(\log(\ell)))^2}\right)$-sized coalition of bad players, which is better than the best one-round protocol by \cite{AL93resilient} in this setting that is resilient to only an $O(\ell / (\log (\ell))^2)$-sized coalition of bad players. 

\subsubsection{Improved bounds for leader election}

We obtain improved upper and lower bounds for leader election protocols. Beginning with our lower bounds, our results subsume all previous lower bounds for multi-round leader election protocols.
Our first non-trivial result is for one-round protocols where the players send $1$ bit per round. We show that $\frac{\ell \log(\log(\ell))}{\log(\ell)}$ players suffice to corrupt any such protocol. This is a double exponential improvement upon the previous best lower bound of $\frac{\ell}{\log(\log(\log(\ell)))}$ due to \cite{RSZ02leader}.\footnote{Recall that given a leader election protocol, one can turn it into a collective coin flipping protocol at the expense of one extra round. So $k$-round collective coin flipping lower bounds translate to ($k-1$)-round leader election lower bounds. This was the exact strategy followed by \cite{RSZ02leader}.} 
We obtain the following general lower bound for $k$-round protocols:
\begin{globtheorem}[Informal version of \cref{thm: leader election lb}]\label{thm: intro leader protocol lb}
Let $\pi$ be a $k$-round leader election protocol over $\ell$ players where players send 1 bit per round.
Then, there exist $b = O\left(\frac{\ell\log^{(k+1)}(\ell)}{\log^{(k)}(\ell)}\right)$ bad players who can bias $\pi$ to elect a bad player as a leader with probability $0.99$.
\end{globtheorem}
This directly improves upon the previous best lower bounds, due to \cite{RSZ02leader}, which showed that $O\left(\frac{\ell}{\log^{(2k+1)}(\ell)}\right)$ players can corrupt such protocols, achieving a strictly better bound for all $k$.
For other lower bounds, we recall that our lower bounds from \cref{cor into: round lb} and \cref{thm: intro protocol lb longer messages} also yield leader election lower bounds. 

Turning to positive results, we construct the following one-round leader election protocol.
\begin{globtheorem}[Informal version of \cref{cor: leader election 1 round}
]\label{thm: intro leader protocol ub}
There exists an explicit one-round leader election protocol $\pi$ over $\ell$ players where players send 1 bit per round such that the following holds: For any set of $b$ bad players with $b\le O\left(\frac{\ell}{(\log(\ell))^2}\right)$, the protocol elects a good leader with probability $0.99$.
\end{globtheorem}
The $O\left(\frac{\ell}{(\log(\ell))^2}\right)$ number of bad players that this protocol can handle is exactly the number of bad players that the current best one-round coin flipping protocol by Ajtai and Linial can handle. Previously, the best one-round leader election protocol, constructed by \cite{RZ01leader}, required each player to send $O(\log(\ell))$ bits, and even then the protocol could only handle $O\left(\frac{\ell}{(\log(\ell))^3}\right)$ bad players.

\subsubsection{Optimal Bounds for one-round random selection}

Here we construct protocols and provide lower bounds for random selection in the natural setting of one-round protocols where each player sends $1$ bit per round. 

Our main result in this setting will be optimal (up to constants) protocols for outputting uniform random bits---the gold standard for random selection:
\begin{globtheorem}\label{thm inro: tight multioutput coin flipping}[Informal versions of \cref{thm: multioutput coin flipping protocol - general} and \cref{lem: improved coin flip condensing impossibility}]
For all $0 < \eps$ and $\ell, m$ such that $m \ge \log(\ell)^2$, the following holds:
\begin{enumerate}
\item
There exists an explicit one-round random selection protocol $\pi: \zo^{\ell}\to \zo^m$ where $b = O\left(\frac{\ell}{m}\right)$ such that for any coalition of $b$ bad players corrupting $\pi$, the output distribution on $m$ bits has statistical distance at most $\eps$ from $\U_m$.
\item
For any one-round random selection protocol $\pi: \zo^{\ell}\to \zo^m$, there exists a coalition of $O\left(\frac{\ell}{m}\right)$ bad players corrupting $\pi$ such that the output distribution of the corrupted protocol has statistical distance at least $1 - \eps$ from $\U_m$.
\end{enumerate}
\end{globtheorem}

As far as we are aware, this is the \emph{first} provably optimal protocol for any non-trivial task in the full information model.
This is quite surprising since in one-round leader election as well as in collective coin flipping, the multiplicative gap in the number of bad players between the lower bounds and the best known construction is at least $O(\log(\ell))$.

Our construction improves upon the previous best construction of Kamp and Zuckerman in this setting that outputted $m = O\left(\frac{\ell}{b^{\log(3)}}\right)$ uniform random bits \cite{KZ07NOSF}.
This also directly gives us a leader election protocol as in \cref{thm: intro leader protocol ub}.
For the setting of $m = O(\log(\ell)^2)$, the above protocol can handle $b = O\left(\frac{\ell}{(\log(\ell))^2}\right)$ bad players, which is exactly the number of bad players that the current best coin flipping protocol by Ajtai and Linial \cite{AL93resilient}. Hence, we improve upon their result by outputting many more bits while maintaining the ``resilience'' parameter.
Previous lower bounds in this setting were obtained by \cite{KZ07NOSF} that showed $O\left(\frac{\ell}{m}\right)$ bad players can corrupt the protocol so that the output distribution has statistical distance at least $0.1$ from the uniform distribution. From their techniques, it is not clear how to obtain a stronger result that rules out larger statistical distance or stronger random selection impossibility results. Ruling out such protocols is important since in the collective coin flipping literature, even protocols attaining statistical distance at most $0.9$ are considered good.

 
One key definition that we introduce to help prove our lower bounds is that of the \emph{multi-output influence} of a Boolean function.
This naturally generalizes the standard notion of influence of a Boolean function to larger output domains.
We prove a Poincar\'e style inequality for multi-output influence and use it to obtain our lower bounds.
See \cref{def overview: multioutput influence} for the formal definition, our bounds on it, and how we use it.
\dobib

\section{Proof Overview} \label{sec:proof-overview}

In this section, we give a brief overview of our main proofs.
In \cref{subsec:proof overview coin flipping protocols}, we sketch our proof of lower bounds for coin flipping protocols. In \cref{subsec:proof overview coin flipping protocols longer message}, we prove lower bounds against protocols where players can send many bits per round. 
In \cref{subsec overview: leader election lb}, we show how to obtain leader election lower bounds.
In \cref{subsec:proof overview biasing family of functions}, we prove a useful helper theorem that shows how any family of functions can be biased by a random set of variables and a small set of variables specific to each function. Then, in \cref{sec:warmup-protocols}, we construct new improved constant-round coin flipping protocols. 
We construct one-round random selection protocols in \cref{subsec: overview multioutput extract}.
Lastly, we obtain one-round random selection lower bounds in \cref{subsec: coin flip condensing impossibility}.

\subsection{Biasing Coin Flipping Protocols}
\label{subsec:proof overview coin flipping protocols}

We focus on the two-round setting which contains most of the core ideas behind our proof and allows us to simplify our analysis. 
For any two-round coin flipping protocol $\pi:(\zo^\ell)^2\to\zo$ we show:
\begin{theorem}[\cref{thm: protocol lb}, specialized]\label{thm:proof overview protocol lb}
    For any two-round coin flipping protocol $\pi$ over $\ell$ players where $\Pr[\pi=1]\geq0.01$, there exists $B\subseteq[\ell]$ with $\abs{B}=O\left(\frac{\ell}{\log\log \ell}\right)$ so that $\Pr[\pi|_B=1]\geq0.99$.
\end{theorem}

The main tool that we will use to prove this theorem is that for any family $\cF$ of functions from $\zo^\ell\to\zo$, there exists some common set of bad players $B_R\subset [\ell]$ such that for almost every function $f\in\cF$, there exist a small set of ``heavy'' bad players $B_H$, depending on $f,$ such that $B_R\cup B_H$ can bias $f$. Formally:
\begin{theorem}[Simplified version of \cref{lem: most functions from family can be corrupted with random and heavy set}]\label{thm:proof overview family can be corrupted by random and heavy}
    Let $\cF$ be any family of functions from $\zo^\ell\to\zo$ where $\Pr[f=1]\geq0.001$ for each $f\in\cF$. Then, there exists a common bad set $B_R\subseteq[\ell]$ with $\abs{B_R}=O\left(\frac{\ell}{\log \log \ell}\right)$ such that for a $0.999$ fraction of functions $f\in\cF$, there exists a heavy bad set $B_H=B_H(f)\subseteq[\ell]$ with $\abs{B_H} = (\log \ell)^{0.001}$ such that $B_R\cup B_H$ can $0.999$-bias $f$ towards $1$.
\end{theorem}

We will also require a simpler version of the above theorem that follows from the KKL theorem \cite{KKL88influence}:
\begin{theorem}[KKL]\label{thm:proof overview kkl}
    Let $f: \zo^{\ell} \to \zo$ be such that $\Pr[f = 1] \ge 0.001$. Then, there exists a set of bad players $B$ with $\abs{B}\le O\left(\frac{\ell}{\log \ell}\right)$ such that $\Pr[f|_B = 1] \ge 0.999$.
\end{theorem}

We will sketch a proof of \cref{thm:proof overview family can be corrupted by random and heavy} in \cref{subsec:proof overview biasing family of functions}. Let us see how \cref{thm:proof overview protocol lb} follows from this.

\begin{proof}[Proof Sketch for \cref{thm:proof overview protocol lb}]
In our proof we will separately find sets of bad players to corrupt in the first round and the second round, and at the end we will take the union of the two sets.
In particular, we let $B_I$ (initially empty) be the set of bad players that we will corrupt from the first round.

For $\alpha\in \zo^{\ell}$, let $\pi_{\alpha}: \zo^{\ell}\to \zo$ be the induced second round protocol when the players output $\alpha$ in the first round.
Since $\Pr[\pi = 1] \ge 0.01$, by a reverse Markov argument (see \cref{claim:reverse-markov}), we have that $\Pr[\pi_{\alpha} = 1] \ge 0.001$ for at least $0.009$ fraction of $\alpha$. Let $\cF$ be the family of functions consisting of these functions $\pi_{\alpha}$.
We apply \cref{thm:proof overview family can be corrupted by random and heavy} to find a common set of bad players $B_R$  such that for 99\% of $\pi_{\alpha}\in \cF$, there exists a set of bad players $g(\alpha)$ such that $B_R\cup g(\alpha)$ can $0.999$-bias $\pi_{\alpha}$ towards $1$. Moreover, $\abs{B_R}\le O\left(\frac{\ell}{\log \log \ell}\right)$ and $\abs{g(\alpha)} = h = (\log \ell)^{0.001}$.
For each $\alpha$ for which there does not exist such a set of bad players that along with $B_R$ can corrupt $\pi_{\alpha}$, we let $g(\alpha) = \bot$. Then, we know that $\Pr[g \ne \bot]\ge 0.008$.
We use \cref{thm:proof overview kkl} 
to find a set of bad players $B_{temp}^{(1)}$ with $\abs{B_{temp}^{(1)}} \le O\left(\frac{\ell}{\log \ell}\right)$ to bias the first round protocol so that $\Pr[g|_{B_{temp}^{(1)}} \ne \bot] \ge 0.999$. We add all the players from $B_{temp}^{(1)}$ to $B_I$.
We view $g$ as $(g_1, \dots, g_h)$ where each $g_i: \zo^{\ell}\to [\ell]\cup \{\bot\}$ where $g_i$ outputs the $i$-th largest element from $g(\alpha)$ and $g_i(\alpha) = \bot$ if and only if $g(\alpha) = \bot$.
We will repeatedly use \cref{thm:proof overview kkl} to find bad players that can bias each of $g_1, \dots, g_h$ so that over most inputs, each of their images lies in some set of size $c = \frac{\ell}{(\log \ell)^{1000}}$.
Towards this goal, we maintain sets $C_1, \dots, C_h$ that will satisfy the following invariant (initially each of these sets equals $[\ell]$):
\begin{quote}
        $(*)$~~~For every $\alpha$, if there exists $i\in [h]$ such that $g_i(\alpha)\not\in C_i$, then $g(\alpha) = \bot$. (Equivalently, if $g(\alpha) \neq \bot$, then $g_i(\alpha) \in C_i$ for all $i \in [h].$)
\end{quote}
We now find bad players to bias each of these $g_i$ and, while doing so, maintain that at the beginning of each iteration of the loop, $\Pr[g|_{B_I}\ne \bot] \ge 0.999$. Formally, we proceed as follows:
\begin{itemize}
    \item 
    While there exists $i\in [h]$ such that $\abs{C_i} > c \left(\textrm{Recall that } c = \frac{\ell}{(\log \ell)^{1000}}, h = (\log \ell)^{0.001}\right)$:
\begin{enumerate}
    \item 
    Let $\X_i$ be a random subset of $C_i$ with $\abs{\X_i} = \abs{C_i} / 2$. Since $\Pr[g_i|_{B_I}
    \ne \bot] \ge 0.999$, by our invariant property, $\Pr[g_i|_{B_I}\in C_i] \ge 0.999$.
    Hence, $\E[\Pr[g_i|_{B_I}\in \X_i]] \ge 0.49$. In particular, there exists $C'_i\subset C_i$ with $\abs{C'_i} = \abs{C_i} / 2$ such that $\Pr[g_i|_{B_I}\in C'_i] \ge 0.49$.
    We set $C_i := C'_i$. 

    \item 
    Now, we can only guarantee that $\Pr[g|_{B_I}\ne \bot] \ge 0.49$.
    To maintain our invariant, we apply \cref{thm:proof overview kkl} to the function $g|_{B_I}$ to find a set of bad players $B_{temp}^{(2)}$ with $\abs{B_{temp}^{(2)}} \le O\left(\frac{\ell}{\log \ell}\right)$ so that $\Pr[g|_{B_I\cup B_{temp}^{(2)}} \ne \bot] \ge 0.999$. We add all players from $B_{temp}^{(2)}$ to $B_I$.
\end{enumerate}
\end{itemize}

We let $B_H = \bigcup_{i=1}^{h} C_i$.
Our final set of bad players will be $B_R\cup B_H\cup B_I$.
We see that $\Pr[g|_{B_I} \ne \bot] \ge 0.999$. Whenever $g|_{B_I}(\alpha)\ne \bot$,  $\Pr[\pi_{\alpha}|_{B_R\cup B_H} = 1] \ge 0.999$. Hence, $\Pr[\pi_{B_R\cup B_H\cup B_I} = 1] \ge 0.999\cdot 0.999 \ge 0.99$ as desired.

We finally bound the number of bad players that we control. We know that 
$\abs{B_R}\le O\left(\frac{\ell}{\log \log \ell}\right).$
We also have that 
$
\abs{B_H}\le c\cdot h = \frac{\ell}{(\log \ell)^{1000}}\cdot (\log \ell)^{0.001} < \frac{\ell}{(\log \ell)^{999}}.
$
Since we decrease the size of each $C_i$ by a factor of $2$ each time we execute the loop, and we stop once every $i \in [h]$ has $|C_i| \leq c$, the total number of iterations of the loop is $h\cdot \log(\ell / c) = (\log \ell)^{0.001}\cdot (1000\log\log \ell) \le (\log \ell)^{0.002}$.
Since each time through the loop we add $O\left(\frac{\ell}{\log\ell}\right)$ players to $B_I$, we bound the size of $B_I$ as 
$
\abs{B_I}\le O\left(\frac{\ell}{\log \ell} + (\log \ell)^{0.002}\cdot \frac{\ell}{\log \ell}\right) \le \frac{\ell}{(\log \ell)^{0.99}}.
$
Hence, the total number of bad players we control is $\abs{B_R} + \abs{B_H} + \abs{B_I} \le O\left(\frac{\ell}{\log \log \ell}\right)$.
\end{proof}

The above ideas more or less work to extend our two-round lower bound to $k$ rounds; the main difference is that we inductively use our bounds to bias $(k-1)$-round protocols in place of the second use case of the KKL theorem above (see \cref{item: Second use of induction} in the ``Formal description of algorithm'' in \Cref{sec:biasing-protocols}).

\subsection{Biasing Coin Flipping Protocols with Longer Messages}
\label{subsec:proof overview coin flipping protocols longer message}

We now briefly sketch a proof for lower bounds for protocols where players can send more than one bit per round:

\begin{theorem}[Simplified version of \cref{thm: protocol lb longer messages}]
Let $\pi$ be a $k$-round coin flipping protocol over $\ell$ players where in round $i$, each player can send $r_i = (\log^{(i)}\ell)^{0.99}$ many bits. Then, there exists a set of bad players $B\subset [\ell]$ with $\abs{B}\le 0.01\ell$ so that $\Pr[\pi|_B = 1]\ge 0.99$.
\end{theorem}

\begin{proof}[Proof sketch]
    We essentially follow the same proof strategy as in the $k$-round version of \cref{thm:proof overview protocol lb}. To do this, we treat each round $i$ of the protocol as being over $r_i\cdot \ell$ bits and whenever our lower bound asks us to corrupt a bit, we corrupt the corresponding player that controls that bit. For instance, in the one-round version of this theorem, we treat the input as being over $\ell\cdot r_1$ bits. Then, we use our lower bounds to find a set of $\frac{\ell \cdot r_1}{\log(\ell \cdot r_1)}$ bits such that if bad players control them, then they can bias the protocol. So, we let $B$ be the set of all players that control each such bit. It must be the case that $\abs{B}\le \frac{\ell \cdot r_1}{\log(\ell \cdot r_1)}$. Since $\abs{r_1}\le (\log \ell)^{0.99}$, we end up controlling $o(\ell)$ players overall to corrupt $\pi$. 
    This generalizes to $k$-round protocols. 
\end{proof}

\subsection{Biasing Leader Election Protocols}\label{subsec overview: leader election lb}
Here we give a sketch of how to obtain our leader election lower bounds using coin flipping impossibility results. We will show this for one round and the idea for $k$ rounds will follow similarly.
\begin{theorem}[One-round version of \cref{thm: leader election lb}]\label{thm overview: leader election lb}
For all one-round leader election protocols $\pi$, there exist $b = O\left( \frac{\ell\log(\log(\ell))}{\log(\ell)}\right)$ bad players that can corrupt $\pi$ to elect a bad leader with probability $0.99$.
\end{theorem}

\begin{proof}
We will repeatedly apply the one round coin flipping impossibility result given by \cref{thm:proof overview kkl}.
Let $B$ be our collection of bad players, initialized to $\varnothing$, let $H_0 = [\ell]$, and let $t = \log(\log(\ell))$. 
For $i\in [t]$, we will maintain that the players in $B$ can corrupt $\pi$ to output from $H_i$ with probability $\ge 0.99$.
Using this guarantee, at step $i\in [t]$ we find $H_i\subset H_{i-1}$ with $\abs{H_i} = \abs{H_{i-1}} / 2$ such that the players in $B$ can corrupt $\pi$ to output from $H_{i-1}$ with probability at least $0.49$.  
We then consider a one-round coin flipping protocol that outputs $1$ if $B$ succeeds in corrupting $\pi$ to output from $H_i$ and outputs $0$ otherwise.
Using \cref{thm:proof overview kkl}, we corrupt this coin flipping protocol using $O\left(\frac{\ell}{\log(\ell)}\right)$ bad players, add them to $B$, and meet the guarantee for round $i$ that we output from $H_i$ with probability $0.99$.
The final size of $B$ will be $O\left(\frac{\ell}{\log(\ell)}\cdot t\right) = O\left(\frac{\ell\log(\log(\ell))}{\log(\ell)}\right)$, and the final size of $H_t$ will be $\ell / 2^t = \frac{\ell}{\log(\ell)}$.
We add $H_t$ to $B$ to infer the claim.
\end{proof}

\subsection{Biasing a Family of Functions}
\label{subsec:proof overview biasing family of functions}

We now sketch a proof of \cref{thm:proof overview family can be corrupted by random and heavy}. Towards proving it, we will require the following lemma which is about  biasing individual boolean functions:
\begin{lemma}[Simplified version of \cref{lem: most random set with small heavy set corrupts f}]\label{lem:proof overview most random sets with small heavy set corrputs f}
Fix $1\le h\le \ell^{0.99}$ and let $f: \zo^\ell\to \zo$ be such that $\Pr[f = 1] \ge 0.01.$
Then, for $0.99$ fraction of $B_R\subset [\ell]$ with $\abs{B_R} = O\left(\frac{\ell}{\log(h)}\right)$, there exists $B_H = B_H(B_R)\subset [\ell]$ with $\abs{B_H}\le h$ such that $B_R\cup B_H$ can $0.99$-bias $f$ towards $1$.
\end{lemma}

We prove this lemma later. Now we show how, using \cref{lem:proof overview most random sets with small heavy set corrputs f}, we can prove \cref{thm:proof overview family can be corrupted by random and heavy}.

\begin{proof}[Proof sketch of \cref{thm:proof overview family can be corrupted by random and heavy}]
We set $h = (\log \ell)^{0.001}$.
Let $G = (U, V)$ be a bipartite graph where the left part $U$ equals $\cF$, and the elements of the right part $V$ are all size $\frac{\ell}{\log(h)}$ subsets of $[\ell]$. 
We add an edge between $f\in U$ and a set $R\in V$ if there exists $H\subset [\ell]$ with $\abs{H}\le h$ such that $R\cup H$ can $0.99$-bias $f$ towards $1$. By \cref{lem:proof overview most random sets with small heavy set corrputs f}, the degree of each $f\in U$ is at least $0.99\cdot \abs{V}$. Therefore, there exists $B_R\in V$ such that $B_R$ has degree at least $0.99\cdot \abs{U}$. Such a $B_R$ satisfies the conditions of the theorem.
\end{proof}

We now focus on proving \cref{lem:proof overview most random sets with small heavy set corrputs f}. Towards proving this, we will need the following result regarding the influence of boolean functions that was proven in \cite{RSZ02leader} by slightly building up on the result of \cite{KKL88influence}.
\begin{lemma}[Simplified version of \cref{lem: kkl - heavy or average influence}]\label{lem:proof overview kkl - heavy or average influence}
Fix $1\le h\le \ell^{0.99}$. Let $f: \zo^\ell\to \zo$ be such that $0.01\le \Pr[f = 1]\le 0.99$. If $\I_i(f) \le \frac{1}{h}$ for all $i\in [\ell]$, then $\sum_{i=1}^\ell \I_i(f) \ge \frac{\log(h)}{2000}$, where $\I_i(f)$ is the influence of the coordinate $i$ on $f$.
\end{lemma}

We use \Cref{lem:proof overview kkl - heavy or average influence} to establish \cref{lem:proof overview most random sets with small heavy set corrputs f} by analyzing a semi-random process first studied in \cite{RSZ02leader}. Our analysis closely follows that of \cite{RSZ02leader}, and only differs in the final conclusion.

\begin{proof}[Proof sketch of \cref{lem:proof overview most random sets with small heavy set corrputs f}]
Let $r = \frac{C\ell}{\log(h)}$ where $C$ is a very large universal constant.
Consider the following semi-random process:
\begin{itemize}
    \item 
    Initialize $B_R = B_H = \emptyset$. Repeat the following for $r$ steps or until $\Pr[f|_{B_R\cup B_H} = 1] \ge 0.99$:
\end{itemize}
\begin{enumerate}
    \item (Heavy Case) If there exists $i\in [\ell]\setminus (B_R\cup B_H)$ such that the influence of $i$ on $f|_{B_R\cup B_H}$ is at least $\frac{2}{h}$, then add $i$ to $B_H$.

    \item (Random Case) Otherwise, pick a random $i\in [\ell]\setminus (B_R\cup B_H)$ and add $i$ to $B_R$.
\end{enumerate}
We say that the above process \emph{succeeds} if at the end, $\Pr[f|_{B_R\cup B_H} = 1] \ge 0.99$. We claim the following:
\begin{claim}[Simplified version of \cref{claim: process succeeds}]
\label{claim:proof overview process succeeds}
The above process succeeds with probability at least $0.999$. 
\end{claim}

We will prove this claim later. For now we show how \Cref{lem:proof overview most random sets with small heavy set corrputs f} follows from it.
First, we observe that $\abs{B_H}\le h$ (since each player added to $B_H$ ``pushes'' $f$ towards $1$ by a factor of $1/h$, this can happen at most $h$ times). Next, consider a random set $\bR\subset [\ell]$ with $\abs{\bR} = \ell$ and a random permutation $\bpi$ of $[r]$ so that $(\bR, \bpi)$ fixes an ordering of the elements of $\bR$. Consider the modified semi-random process where an initial randomly choice of  $(\bR, \bpi)$ is made, and then in the random case of the process, the earliest element from $\bR$ that is not in $B_R\cup B_H$ is chosen. At the end of the process, $B_R\subset \bR$, and if the process succeeds, $B_H\cup \bR$ can indeed bias $f$, as desired. By \cref{claim:proof overview process succeeds}, the process succeeds with probability at least $0.999$, so for a $0.999$ fraction of choices of $(R, \pi)$, the modified process succeeds. Consequently, there exists a $\pi^*$ such that for a $0.999$ fraction of $(R, \pi^*)$, the modified random process succeeds. Thus, for a $0.999$ fraction of sets $R$, there exists a set $B_H = B_H(R)$ (given by the process) such that $R\cup B_H$ can $0.99$-bias $f$ as desired. 
\end{proof}

We finally prove our claim that the process indeed succeeds with high probability:
\begin{proof}[Proof sketch of \cref{claim:proof overview process succeeds}]
For $j\in [r]$, let $v_j$ represent the variable chosen at step $j$ of the semi-random process.
Let $\X_j$ equal the influence of $v_j$ on $f|_{\{v_1, \dots, v_{j-1}\}}$ if the process hasn't stopped before step $j$, and let $\X_j$ equal $1$ otherwise.
Let $j \in [r]$ be a step such that $\Pr[f|_{\{v_1, \dots, v_{j-1}\}} = 1]\le 0.99$. Then, by \cref{lem:proof overview kkl - heavy or average influence}, in step $j$ there either exists a variable with influence $\frac{2}{h}$ or else a randomly chosen variable will have expected influence $\ge \frac{\log(h)}{2000\ell}$.
Since $\frac{1}{h}\ge \frac{\log(h)}{2000\ell}$ (recall that we assumed $h \leq \ell^{0.99}$), we always have that $\E[\X_j | \X_1, \dots, \X_{j-1}] \ge \frac{\log(h)}{2000\ell}$.

For $j\in [r]$, let $\bZ_j = \sum_{k\le j} \X_k$.
Then, $\bZ_1, \dots, \bZ_r$ forms a submartingale with 
\[
    \E[\bZ_r | \bZ_1, \dots, \bZ_{r-1}] = r\cdot \frac{\log(h)}{2000\ell} = C / 2000 \ge 10^6.
\]
We apply Azuma's inequality (see \cref{lem: azuma rsz}) to infer that $\bZ_r \ge 2$ with probability $\ge 0.999$.
Since $\bZ_r$ represents the sum total of `contributions' of influences towards $1$, whenever $\bZ_r \ge 2$ it must be that $f$ has been `pushed' towards $1$ by a total amount of $1$; so, $\Pr[f|_{B_R\cup B_H} = 1] \ge 0.999$ and the process succeeds. Since this happens with probability $\ge 0.999$, the process succeeds with at least this probability.
\end{proof}

\subsection{Constructing Coin Flipping Protocols} \label{sec:warmup-protocols}
We now move on to sketching the main ideas involved in constructing better coin-flipping protocols in the setting that each player sends one bit per round. For simplicity, we focus on the setting of two-round protocols, which will give a sense of  how our general construction (for $k$ rounds) works.
\begin{theorem}[Two-round version of \cref{thm: construct protocol}]
\label{thm:proof overview construct protocol}
For any $\gamma > 0$ there exists an explicit two-round coin flipping protocol over $\ell$ players such that when the number of bad players is at most $\frac{\gamma \ell}{(\log \ell)(\log\log\ell)^2}$, the output coin flip is $\eps$-close to uniform where $\eps = O\left(\gamma + \left(\log\ell\right)^{-0.2}\right)$.
\end{theorem}

To help construct such protocols, we use the explicit resilient functions of \cite{IV25resilient}:

\begin{theorem}[\cite{IV25resilient}]
\label{thm:proof overview explicit resilient}
There exists an explicit one-round protocol $f: \zo^n\to \zo$ s.t. when $b$ out of $n$ players are bad, the resultant output bit is $O\left( \frac{b (\log n)^2}{n} + n^{-0.99}\right)$ close to uniform.
\end{theorem}

Using this, our two-round protocol proceeds as follows:

\begin{proof}[Proof sketch of \cref{thm:proof overview construct protocol}]
We partition the $\ell$ players arbitrarily into parts $P_1, \dots, P_{\ell / \log \ell}$ where $\abs{P_i} = \log \ell$.
Players in the same part will act as a single entity that can make $\log \ell$ many coin flips.
We say entity $P_i$ is \emph{good} if all players that are part of the entity are good. We say the entity is \emph{bad} otherwise.
Since at most $\frac{\gamma \ell}{(\log \ell)(\log\log\ell)^2}$ players are bad, the number of bad entities is at most $\frac{\gamma \ell}{(\log \ell)(\log\log\ell)^2}$.

We now deploy the lightest bin protocol of \cite{feige99lightestbin} on the entities in the first round. In particular, we introduce `bins' $B_1, \dots, B_{\ell / (\log \ell)^3}$. We ask each entity to `vote' for a bin by outputting a random number between $1$ and $\ell / (\log \ell)^3$ (this can be done since each entity has access to $\log \ell$ random bits). Let $i^*\in [\ell / (\log \ell)^3]$ be such that among all the bins, bin $B_{i^*}$ ``is the lightest'' (has the smallest number of entities voting for it). Let $S\subset [\ell / \log \ell]$ be the set of entities that voted for $B_{i^*}$. 

In the second round, we apply the explicit resilient function from \cref{thm:proof overview explicit resilient} to the  entities in $S$ (asking them to output $1$ bit) and let that be the output of the protocol.

We now analyze this protocol. First, since there are $\ell / (\log \ell)^3$ bins and $\ell / \log \ell$ entities, the lightest bin $B_{i^*}$ will have at most $(\log \ell)^2$ players voting for it, i.e., $\abs{S}\le (\log \ell)^2$. Next, since the number of good entities is at least $\frac{\ell}{\log \ell} - \frac{\gamma \ell}{(\log \ell)(\log\log\ell)^2}$, by a Chernoff bound (see \cref{claim: lower tail chernoff}) and a union bound, with high probability, every bin will have roughly $\frac{\frac{\ell}{\log \ell} - \frac{\gamma \ell}{(\log \ell)(\log\log\ell)^2}}{\frac{\ell}{(\log \ell)^3}} = (\log \ell)^2 - \frac{\gamma(\log \ell)^2}{(\log \log \ell)^2}$  good entities.
Equivalently, the number of bad entities in bin $B_{i^*}$ is at most $\frac{\gamma(\log \ell)^2}{(\log \log \ell)^2}$ (out of roughly $(\log \ell)^2$ entities in it).
Hence, when we apply \cref{thm:proof overview explicit resilient} with $n \approx (\log \ell)^2$ players, the output coin flip will be $O(\gamma)$ close to the uniform distribution (the extra $(\log \ell)^{-0.2}$ error term is because of the error in the Chernoff bound).
\end{proof}

\subsection{Constructing One-Round Random Selection Protocols}\label{subsec: overview multioutput extract}
In this subsection, we sketch our one-round coin flipping protocol outputting many random bits. 
\begin{theorem}[\cref{thm: multioutput coin flipping protocol - general}, specialized]\label{thm overview: multioutput coin flipping protocol - general}
For $0 < \eps$, there exists $C_0 = C_0(\eps)$ and an explicit one-round coin flipping protocol $\pi: \zo^{\ell}\to \zo^m$ where $m\ge \log(\ell)^2$ such that for any coalition of $b = \frac{\ell}{C_0 m}$ bad players corrupting $\pi$, the output distribution has statistical distance at most $\eps$ from $\U_m$.    
\end{theorem}

To help construct such a protocol, we need the notion of resilient functions.
\begin{definition}[Resilient functions]
For $b, \ell, m\in \N$ and $\delta\in [0, 1]$, we say $\Res: \zo^{\ell} \to \zo^m$ is \emph{$(b, \delta)$-resilient} if for all $B\subset [\ell]$ with $\abs{B} = b$, we have $\Pr_{x\sim \zo^{\ell\setminus B}} \left[\Res(x, \cdot) \textrm{ is unfixed }\right] \le \delta$.
\end{definition}

We further say that $\Res$ is \emph{$\eps$-biased} if $\abs{\Res(\U_{\ell}) - \U_m}\le \eps$. Then $\Res$ forms a one-round coin flipping protocol handling $b$ bad players with distance $\eps + \delta$ from $\U_m$.
We use the following two resilient functions:
\begin{lemma}[\cref{lem: multioutput modified ajtai-linial}, simplified]
\label{lem overview: multioutput modified ajtai-linial}
For all $0 < \eps < 1/4$ there exists $C = C(\eps)$ such that the following holds.
For all $b, \ell\in \N$ where $b = \frac{\ell}{C(\log(\ell))^2}$, there exists an explicit function $\Res: \zo^{\ell} \to \zo^m$ where $m = 2\log(\ell)$ such that $\Res$ is $(b, \eps)$-resilient and is $\eps$-biased.
\end{lemma}

We construct this resilient function in  \cref{subsubsec overview: multioutput modified ajtai-linial} by modifying the Ajtai-Linial function \cite{AL93resilient}. 
The second resilient function we need is from \cite{IV25resilient}, that we saw earlier (\cref{thm:proof overview explicit resilient}).
Using these ingredients, we construct our desired one-round coin flipping protocol.
\begin{proof}[Proof of \cref{thm overview: multioutput coin flipping protocol - general}]
We let $\Res: \zo^{\ell/2}\to \zo^{\log(\ell)}$ be $(b, \eps/4)$ resilient function from \cref{lem overview: multioutput modified ajtai-linial} that is $\eps/4$ biased.
On input $x\in \zo^{\ell}$, we decompose $x = (x_1, x_2) \in (\zo^{\ell/2})^2$.
We further divide $x_2$ into $y_1, \dots, y_{\ell / 2m}$ where each $y_i\in \zo^m$.
We use the random bits of $\Res(x_1)$ to sample a random integer $j\in [\ell / 2m]$.
We finally output $y_j$.


We now analyze this construction. 
Let $G_y = \{i: y_i \textrm{ contains no bad players }\}$. 
As the number of bad players is at most $\frac{\ell}{C_0 m}$, size of $G_y$ is at least $\frac{\ell}{m}\left(\frac{1}{2} - \frac{1}{C_0}\right)$. 
So, if we truly randomly sampled $j\in [\ell/2m]$ and output $y_j$, then our output distribution will have distance at most $\frac{1}{2C_0}$ from the uniform distribution. We will let $C_0$ be large enough so that this is at most $\eps / 4$.
However, since we pick $j$ from $\Res(x_1)$, with probability at least $1 - \eps/4$ the integer $j$ will be picked from a distribution that has statistical distance at most $\eps/2$ from the uniform distribution. Furthermore, this distribution will be independent from the outcomes of the coin flips of the players in $x_2$. Hence, we conclude that the overall output distribution of our algorithm will have statistical distance at most $\eps$ from $\U_m$ as desired.
\end{proof}

\subsubsection{Constructing the Multi-output Ajtai-Linial Function}\label{subsubsec overview: multioutput modified ajtai-linial}
We now construct a resilient function 
that 
outputs $2\log(\ell)$ uniform bits as specified in \cref{lem overview: multioutput modified ajtai-linial}. We will need a  weaker variation of the above resilient function that outputs $\log(\ell) / 4$ bits:
\begin{lemma}[\cref{lem: base multioutput ajtai linial}, simplified]\label{lem overiew: base multioutput ajtai linial}
For all $0 < \eps < 1/4$ there exists $C = C(\eps)$ such that the following holds.
For all $b, \ell\in \N$ where $b = \frac{\ell}{C(\log(\ell))^2}$, there exists an explicit function $\Res: \zo^{\ell} \to \zo^m$ where $m = \log(\ell) / 4$ such that $\Res$ is $(b, \eps)$-resilient and is $\eps$-biased.
\end{lemma}
Using this, we increase the number of output bits to $2\log(\ell)$ by chopping the  input into $8$ parts, applying $\Res$ from above and outputting that. By similar analysis as earlier, the claim follows.
To construct this $\Res$ with guarantees from \cref{lem overiew: base multioutput ajtai linial}, we will construct a resilient function $\Res'$ that is allowed to output $\bot$.
\begin{lemma}[\cref{lem: base multioutput ajtai linial with bot}, simplified]\label{lem overview: base multioutput ajtai linial with bot}
For all $0 < \eps < 1/4$, there exists a constant $C = C(\eps)$ such that the following holds.
For all $b, \ell\in \N$ where $b = \frac{\ell}{C(\log(\ell))^2}$, there exists an explicit function $\Res': \zo^{\ell} \to \zo^m \cup \bot$ where $m = \log(\ell) / 4$ such that $\Res'$ is $(b, \eps)$-resilient, $\Pr[\Res' \ne \bot] \ge \frac{1}{100}$ and $\abs{(\Res'(\U_n) | (\Res'(\U_n)\ne \bot)) - \U_m}\le 100\cdot 2^{-m}$.
\end{lemma}
Let's see how using $\Res'$ we can construct our desired resilient function $\Res$.
\begin{proof}[Proof sketch of \cref{lem overiew: base multioutput ajtai linial}]
We divide the input $x \in \zo^{\ell}$ into $t = 100\log(2 / \eps)$ equal parts $(x_1, \dots, x_t)$ of length $\ell / t$ each.
We let $\Res': \zo^{\ell / t}\to \zo^m$ be the resilient function from \cref{lem overview: base multioutput ajtai linial with bot} with parameter $\eps' = \eps / t$.
For $i\in [t]$, let $y_i = \Res'(x_i)$.
If for all $i\in [t]$, it holds that $y_i = \bot$, we output $0^m$.
Otherwise, let $j\in [t]$ be the smallest $i$ such that $y_i \ne \bot$.
We output $y_j$.

On uniform input, the probability that all $y_i = \bot$  is at most $\left(1 - \frac{1}{100}\right)^t\le \exp(-t / 100) \le \eps/2$. Conditioning on this not happening, our output will be $100\cdot 2^{-m} \le \eps/2$ close to uniform, bringing our bias to $\le \eps$.
We set our parameter $b$ to be small enough so that each $\Res'$ can handle $b$ players.
Then similar to previous arguments, we apply a union bound to conclude our resilience parameter is at most $\eps'\cdot t \le \eps$.
\end{proof}

Lastly, we show how to construct $\Res'$. 
\begin{proof}[Proof sketch of \cref{lem overview: base multioutput ajtai linial with bot}]
To do this, we will modify the Ajtai-Linial resilient function construction \cite{AL93resilient} as derandomized  by \cite{IV25resilient} (see \cref{thm:proof overview explicit resilient} for formal claim). We slightly modify their construction and see that their analysis can be applied to our modified function as well.

Let $M = 2^m = (\ell)^{1/4}$.
Let's quickly review the construction of $f: \zo^{\ell}\to \zo$ from \cite{IV25resilient}. They pseudorandomly construct $u = \poly(\ell)$ many tribes functions (see \cref{def: tribes}) $T_1, \dots, T_u: \zo^{\ell}\to \zo$. They define $g(x) := \bigwedge_{i=1}^u T_i(x)$.
We keep the same set of tribes functions but slightly modify what we do with them.
For $j\in [M]$, we let $y_j = \bigwedge_{i = (j-1)\cdot (u / M) + 1}^{j\cdot (u / M)} T_i$.
If there exists a unique $z\in [M]$ such that $y_z = 0$ and for all $j\ne z, y_j = 1$, then we let $\Res'(x) = z$.
Otherwise, we output $\bot$.

To analyze the resilience property, we define $h: \zo^{\ell}\to \zo^u$ as $h(x) = (T_1(x), \dots, T_u(x))$. The analysis of \cite{IV25resilient} implicitly shows that this multi-output map $h$ is $(b, \eps)$-resilient.
Since our function $\Res'(x)$ can be computed solely given $h(x)$, if the bad players could decide between $\ge 2$ outcomes of $\Res'$, then they would also be able to decide between $\ge 2$ outcomes of $h$. Hence, $\Res'$ is also $(b, \eps)$-resilient.

Analyzing the bias property requires the most effort for \cite{IV25resilient}. 
The argument morally says: assume each tribe $T_i$ had independent uniform bits and show that the output has small bis. Then, use Janson's inequality to conclude that reality approximates this independent case. We use the same analysis as \cite{IV25resilient} and argue that reality approximates the independent case very well. For the sake of exposition, we omit this here and only sketch an argument that our bias is small in the independent case.
Let $p = \Pr[T_i = 0]$ and let $q = (1 - p)^u$. In the independent case, $q$ equals the probability that $g$ outputs $1$. The parameters are chosen carefully so that $\abs{q - \frac{1}{2}} \le O\left(\frac{(\log(\ell))^2}{\ell}\right)$.
For $i\in [M]$, let $q_i = (1 - p)^{u - u / M}$, the probability that all but $u / M$ tribes output $1$.
Then, $q_i - q$ is the probability that we output $i$. We can show that $\abs{q_i - q}\le \frac{q\ln(1/q)}{M} + O(1 / M^2)$.
So, the probability that we don't output $\bot$ is at least $q\ln(1 / q) + O(1 / M)\ge 0.01$ since $q\approx 1/2$. From above we can also easily show that conditioned on $\Res'(x)\ne \bot$, the output distribution has bias $\le O(1 / M)$. See \cref{subsubsec: base multioutput ajtai linial with bot} for all the details.
\end{proof}

\subsection{Biasing One-Round Random Selection Protocols}\label{subsec: coin flip condensing impossibility}

We will show the following impossibility result for one-round randomized selection protocols (with worse dependence on $\Delta$ than what we actually obtain in \cref{lem: improved coin flip condensing impossibility}).

\begin{theorem}[\cref{lem: technical coin flip condensing impossibility} restated]\label{lem:overview-technical coin flip condensing impossibility}
For all $\ell, m\in \N$ and $0 < \Delta < m$, and all one round random selection protocols $f: \zo^{\ell} \to \zo^m$, there exists a set  of bad players $B\subset [\ell]$ with $\abs{B} \le O\left(\frac{\ell}{m}\cdot 2^{2\Delta}\right)$ and $S\subset \zo^m$ with $\abs{S} \le 2^{m - \Delta}$ such that the players in $B$ can bias $f$ to output from $S$ with probability $0.99$.
\end{theorem}

Using this, we obtain the lower bound stated in \cref{thm inro: tight multioutput coin flipping} by setting $\Delta$ to be a large enough constant.



Our general strategy in creating $S$ and $B$ in \cref{lem:overview-technical coin flip condensing impossibility} is to repeatedly find an influential bad player and add it to $B$. 
However, to choose which player to add to $B$, the standard notion of influence does not suffice for us since we are considering multi-output functions. To fill this gap, we introduce the notion of multi-output influence and prove a Poincar\'e-style inequality on this version of influence. 
\begin{definition}[Multi-output influence]\label{def overview: multioutput influence}
    For $f:\zo^n\to \zo^m$ and $i\in[n]$, define the \emph{multi-output influence of $i$ on $f$} and the \emph{total multi-output influence} as 
    $\I_i(f) =\Pr[f(x)\neq f(x^{\oplus i})]$ and $\I(f)=\sum_{i=1}^n\I_i(f)$, respectively.
\end{definition}
For any multi-output function, we can lower bound the total multi-output influence as follows.
\begin{theorem}[General multi-output Poincar\'e]\label{thm overview: general multioutput poincare}
Let $f: \zo^n \to \zo^m$ be an arbitrary function. Then, $\I(f) \ge H(f(\U_n))$ where $H(\cdot)$ denotes Shannon entropy (\cref{def:Shannon entropy}).
\end{theorem} 
We prove this in \cref{sec:overview-multioutput poincare}. Using this, we obtain our random selection lower bound as follows:

\begin{proof}[Proof sketch of \cref{lem:overview-technical coin flip condensing impossibility}]
To create $S$ and $B$, we consider a procedure that runs for $u$ epochs where each epoch $i$ lasts for many rounds.
In a single round, we add one bad player to $B$ that allows us to bias $f$ the most. When we do this, though, we do not specify the behavior of the bad player and delay doing that until after the end of all $u$ epochs.
After multiple rounds within a given epoch, we will finish the epoch when sufficient progress has been made, where our definition of ``sufficient progress'' increases from epoch to epoch. 
Concretely, before the beginning of the first epoch, we pick a ``heavy set'' $S_H\subset \zo^m$ such that $y\in S_H$ if and only if $\Pr[f = y] \ge 2^{-m/2}$.
Then, our goal after $i$ epochs will be that with high probability over $x\in \zo^{[\ell]\setminus B}$, one of the following holds when the good players send $x$ as their random bits: 1) the bad players can force the outcome to be in $S_H$, or 2) the bad players can choose between at least $i+1$ outcomes from $\zo^m$.

\paragraph{End condition} 
We set the number of epochs $u = O\left(2^{\Delta}\right)$ so that at the end of all epochs, with high probability, either the bad players from $B$ can force the outcome to be in $S_H$ or can choose between at least $O\left(2^{\Delta}\right)$ outcomes.
At the end of all epochs, we consider the bipartite graph with vertex sets $V_L = \zo{[\ell]\setminus B}, V_R = \zo^m$ which has an edge between $x$ and $y$ if $f(x, \cdot)$ is unfixed over outcome $y$, among other outcomes.
The goal at the end then becomes to find a small ``almost dominating set'' $S\subset V_R$, i.e., a small set $S$ such that $\Nbr(S) \ge 0.99 \cdot \abs{V_L}$.
We will then specify the behavior of the bad players to always choose an outcome from $S$ if possible.
We construct $S$ as follows:
Firstly, we add $S_H$ to $S$.
From the end condition after $u$ epochs, the remaining uncovered vertices in $V_L$ will have out degree at least $O\left(2^{\Delta}\right)$.
We then greedily add outcomes from $V_R$ to $S$ that cover the largest number of uncovered vertices in $V_L$.
By an averaging argument, there will always exist a $y\in \V_R$ that covers an $O\left(2^{\Delta} / \abs{V_R}\right) = O\left(2^{\Delta - m}\right)$ fraction of uncovered vertices at each step.
Using this, one can show that the size of the almost dominating set $S$ we obtain will be at most $2^{m-\Delta-1} + \abs{S_H}$.
We also easily see that $\abs{S_H}\le 2^{m / 2}\le 2^{m - \Delta - 1}$, and hence $\abs{S}\le 2^{m - \Delta}$ as desired.

\paragraph{Definitions for each epoch and round} 
We now explain how we proceed within an epoch $i$.
At the beginning of epoch $i$, we are guaranteed that with some probability $1 - \eps_i$, bad players in $B$ can decide amongst $\ge i$ outcomes or choose an outcome from $S_H$.
Our goal is that at the end of epoch $i$, with probability $1 - \eps_{i+1}$, the bad players can decide amongst $\ge i+1$ outcomes or can choose an outcome from $S_H$.
Here, we will want that $\eps_{i+1} \le \eps_i + \frac{0.01}{u}$ so that the total error across all $u$ epochs is at most $0.01$.

To help keep track of these quantities, at round $r$ in epoch $i$, let $\alpha_i^r(\bot)$ be the probability over $x\sim \zo^{[\ell]\setminus B}$ that $f(x, \cdot)$ is unfixed over $< i$ outcomes, none of which are from $S_H$. 
Also let $\alpha_i^r(\#)$ be the probability that $f(x, \cdot)$ is unfixed over exactly $i$ outcomes, none of which are from $S_H$.
Lastly, let $\alpha_i^r(*, \top)$ be the probability that $f(x, \cdot)$ is unfixed over some outcome from $S_H$ or is unfixed over $\ge i+1$ outcomes.

\paragraph{Terminating condition for each epoch} 
We are guaranteed from the previous epoch that $\alpha_i^1(\bot) = \eps_i \le i\cdot \frac{0.01}{u}$.
We want to show that after some number $v_i$ of rounds in epoch $i$, $\alpha_i^{v_i}(*, \top) \ge 1 - \eps_i - \frac{0.01}{u}$ so that epoch $i$ ends at that round and we meet our goal.
We claim that if after some round $v$ we have $\alpha_i^v(\#) \le \frac{0.01}{u}$, then we indeed meet our desired goal and we should end the current epoch after round $v$.
To prove this claim, it suffices to show that for all $r$, we have $\alpha_i^r(\bot) \ge \alpha_i^{r+1}(\bot)$.
Indeed, then we would have that $\alpha_i^v(\bot) \le \alpha_i^1(\bot) = \eps_i$.
Since $\alpha_i^v(\top, *) + \alpha_i^v(\#) + \alpha_i^v(\bot) = 1$, we can conclude that $\alpha_i^v(\top, *)\ge 1 - \eps_i - \frac{0.01}{u}$, i.e., $\eps_{i+1} \le \eps_i + \frac{0.01}{u}$, as desired.

We now sketch why $\alpha_i^r(\bot) \ge \alpha_i^{r+1}(\bot)$ holds. The reason is simple: let $B_i^r$ be the set of bad players after round $r$ in epoch $i$ and let $j$ be the player added at that round so that $B_i^{r+1} = B_i^r \cup \{j\}$.
For $x\in \zo^{[\ell]\setminus B_i^{r+1}}$, if $f(x, \cdot)$ is unfixed over $< i$ outcomes, then so are both $f(x\circ 0, \cdot)$ and $f(x\circ 1, \cdot)$ where $x\circ 0$ and $x\circ 1$ are assignments over $\zo^{[\ell]\setminus B_i^r}$ obtained by setting the variable $j$ to $0$ and $1$ respectively.

\paragraph{Finding influential player in every round} 
To find influential players that will let us achieve our goal, we define function $h_i^r(x): \zo^{[\ell]\setminus B} \to \binom{\zo^m}{i}\cup \{@\}$ as follows: On input $x$, if $f(x, \cdot)$ is unfixed over exactly $i$ outcomes, none of which are from $S_H$, then we let $h_i^r(x)$ be the set of those $i$ outcomes. We let $h_i^r(x) = @$ otherwise.
We apply our multi-output influence bound from \cref{thm overview: general multioutput poincare} to $h_i^r$ and find a currently good player $j \in [\ell] \setminus B$ such that $\I_j(h_i^r) \ge \frac{H(h_i^r(\U))}{\ell-|B| } \geq \frac{H(h_i^r(\U))}{\ell},$ and add that player to our bad set $B$.
Since we did not terminate at round $r$, we must have that $\Pr[h_i^r\ne @] \ge \frac{0.01}{u}$. Also, using the fact that none of these outcomes are from $S_H$, we can show that $H(h_i^r(\U)) \ge \frac{0.01}{u}\cdot \frac{m}{4}$.
Therefore, $\I_j(h_i^r) \ge 0.0025\cdot \frac{m}{u \ell}$.

\paragraph{Bounding size of $B$} 

Since we add one bad player per round, we have that $\abs{B} \le \sum_{i=1}^u v_i$ where recall $v_i$ is the number of rounds that epoch $i$ runs for.
We claim that $v_i \le O\left(\frac{u\ell}{m}\right)$ for each $i$; if we show this, then $\abs{B} \le O\left(\frac{u^2\ell}{m}\right)$. Since $u = O\left(2^{\Delta}\right)$, then we would indeed obtain our desired bound on $\abs{B}$.

We lastly sketch our argument for why $v_i \le O\left(\frac{u\ell}{m}\right)$.
Fix any $i \in [u]$.
We use a potential function argument with the potential being $\phi(r) = (1 - \alpha_i^r(\bot)) + \alpha_i^r(\top, *)$.
We see that for all $r$, $0\le \phi(r)\le 2$, and we will show that $\phi(r+1) - \phi(r) \ge O\left(\frac{m}{u\ell}\right)$.
This will let us conclude that after $k = O\left(\frac{u \ell}{m}\right)$ steps, $\phi(k) \ge 2$.
This will imply $\alpha_i^r(\#) = 0$, and hence, the epoch must terminate after $O\left(\frac{u \ell}{m}\right)$ many rounds as desired.

To show $\phi(r+1) - \phi(r) \ge O\left(\frac{m}{u\ell}\right)$, 
we rewrite $\phi(r+1) - \phi(r) = \left(\alpha_i^r(\bot) - \alpha_i^{r+1}(\bot)\right) + \left(\alpha_i^{r+1}(\top, *) - \alpha_i^r(\top, *)\right)$.
We then carefully observe that whenever player $j$ influences $h_i^r$, it either causes $\alpha_i^r(\bot)$ to decrease or it causes $\alpha_i^r(\top, *)$ to increase for the next round. Among other observations, here we crucially use the fact that when $j$ exerts influence between two outcomes $y_1, y_2\in \binom{\zo^m}{i}$, then it must be that $y_1\ne y_2$ and so $\abs{y_1\cup y_2} \ge i+1$. 
Using the influence lower bound $\I_j(h_i^r) \ge 0.0025\cdot \frac{m}{u \ell}$ from earlier, the claim indeed follows.
\end{proof}

\subsubsection{Multi-output Poincar\'e}\label{sec:overview-multioutput poincare}

\begin{proof}[Proof sketch of \cref{thm overview: general multioutput poincare}]
We will need the edge isoperimetric inequality
\cite{harper1966optimal, bernstein1967maximally, lindsey1964assignment, hart1976note} that states that for all $f: \zo^n\to \zo$, we have that $\I(f) \ge 2\E[f]\log(1 / \E[f])$.

For all $y\in \zo^m$, we define an indicator function $\chi_y: \zo^n \to \zo$ as $\chi_y(x) = 1$ if  $f(x) = y$ and 0 otherwise.
We apply the Edge Isoperimetric inequality to $\chi_y$ to obtain that $\I(\chi_y)\ge 2\Pr[\chi_y = 1]\log(1 / \Pr[\chi_y = 1])$.

We now claim that $\I(f) = \frac{1}{2}\sum_{y\in \zo^m}\I(\chi_y)$. We prove this claim later. Let's see how using this claim, we can finish our proof. We compute:
\begin{align*}
\I(f) 
& = \frac{1}{2}\sum_{y\in \zo^m}\I(\chi_y)
\ge \sum_{y\in \zo^m}\Pr[\chi_y = 1]\log(1 / \Pr[\chi_y = 1])\\
& = \sum_{y\in \zo^m}\Pr[f(\U_n) = y]\log(1 / \Pr[f(\U_n) = y])
 = H(f(\U_n))
\end{align*}
We finally show that $\I(f) = \frac{1}{2}\sum_{y\in \zo^m}\I(\chi_y)$.
In fact, something stronger is true: for any $i\in [n]$, 
$\I_i(f) = \frac{1}{2}\sum_{y\in \zo^m}\I_i(\chi_y)$.
This holds because  whenever variable $i$ is influential between two outcomes $y_1, y_2$, its influence is accounted for in exactly $\I_i(\chi_{y_1})$ and $\I_i(\chi_{y_2})$.
\end{proof}



\paragraph{Organization}  We introduce the necessary preliminaries in \cref{sec:prelim}. In \cref{sec:family-baised}, we prove our main technical result on jointly  biasing most functions in any family of functions. We prove our main lower bound result on the resilience of coin flipping protocols in \cref{sec:biasing-protocols}. In \cref{sec:explicit-protocols}, we present new multiround coin flipping protocols. In \cref{sec: multioutput extract}, we construct one-round random selection protocols that output uniform random bits. 
In \cref{sec: random selection impossibility}, we prove our impossibility results for one-round random selection.
We finally conclude with some discussion and open problems in \cref{sec:open}.
We provide a high level comparison of our coin flipping protocol lower bound (\cref{thm: intro protocol lb}) to that of \cite{RSZ02leader} in \cref{sec: appendix compare to rsz}.

\dobib

\section{Preliminaries}\label{sec:prelim}

\subsection{Notation and Terminology}
For $0<\eps<1/2$ and $o \in \zo$, we say that $S\subset [\ell]$ can \emph{$(1-\eps)$-bias a function $f: \zo^{\ell} \to \zo$ towards $o$} if there exists an adversary $\cA_o: \zo^{\ell-\abs{S}}\to \zo^{\abs{S}}$ such that the function $g_o(y) := f(\cA_o(y), y)$ has $\Pr_{\by}[g_o(\by) = o] \ge 1-\eps$.
We will similarly use the notion that a set $S \subseteq [\ell]$ can $(1-\eps)$-bias a protocol $\pi$ towards $o$.

For a function $f: \zo^{\ell} \to \zo, S\subset [\ell],$ and $\cA: \zo^{\ell-\abs{S}} \to \zo^{\abs{S}}$, we define $f|_S: \zo^{\ell-\abs{S}}\to \zo$ to be $f|_S(x) = f(\cA(x), x)$. Often when we do this, $\cA$ will be implicitly defined by some other claim, and hence the notation reflects this by only mentioning $f$ and $S$ but not $\cA$. We  similarly extend this notation to protocols so that for $S\subset [\ell]$, a protocol $\pi$, and some adversarial function $\cA$, the protocol $\pi_S$ is well defined.

For a function $f: \zo^{\ell}\to \zo$, we sometimes write ``$\Pr[f=o]$'' to denote $\Pr_{\by \sim \zo^\ell}[f(\by)=o]$.  Similarly, for $\pi$ a $k$-round protocol over $\ell$ players in which each player outputs one bit per round, we sometimes write ``$\Pr[\pi=o]$'' to denote $\Pr_{\bx \sim {(\zo^{\ell})^k}}[\pi(\bx) = o]$.


\subsection{Probability}

\subsubsection{Useful Definitions}

\begin{definition}[Shannon entropy]\label{def:Shannon entropy}
    For a random variable $\X$ over $\Omega$, we define its Shannon entropy as 
    \begin{align*}
        H(\X)&=-\sum_{x\in\Omega}\Pr[\X=x]\cdot\log(\Pr[\X=x]).
    \end{align*}
\end{definition}

\begin{definition} \label{def:submartingale}
A \emph{submartingale} is a sequence of real valued random variables
$\bZ_0,\bZ_1,\dots,$ for which $\E[\bZ_i | \bZ_{i-1}] \geq \bZ_{i-1}.$
\end{definition}

We will also need the following definition of influence of a variable with respect to a boolean function:

\begin{definition}[Influence]
For $f: \zo^{n} \to \zo$ and $i \in [n]$, we say that the \emph{influence} of coordinate $i$ on $f$, denoted $I_i(f)$, is
\[
\Pr_{\bx \sim \U_{i-1},
\by \sim \U_{n-i}}[|f(\bx,1,\by) - f(\bx,0,\by)|],
\quad \text{or equivalently,}
\quad
\Pr_{\bz \sim \U_n}[f(\bz) \neq f(\bz^{\oplus i})].
\]
\end{definition}

\subsubsection{Helpful Claims}

We will make use of the following reverse Markov style inequality:
\begin{claim}[Reverse Markov]
\label{claim:reverse-markov}
Let $\X$ be a random variable taking values in $[0, 1]$. Then, for $0\le p < \E[X]$, it holds that
\[
\Pr[\X > p] \ge \frac{\E[\X] - p}{1 - p}.
\]
\end{claim}

We will also use the following lower tail Chernoff bound:
\begin{claim}[Lower tail Chernoff bound]
\label{claim: lower tail chernoff}
    For $n\ge 1$ and $p\in [0, 1]$, let $\X_1, \dots, \X_n$ be independent random variables such that for each $i\in [n]$, $\Pr[\X_i = 1] = p$ and $\Pr[\X_i = 0] = 1-p$.
    Let $\X = \sum_{i=1}^n \X_i$ and let $\mu = \E[\X] = pn$.
    Then, for all $\delta\in (0, 1)$:
    \[
        \Pr[\X \le (1-\delta)\mu] \le \exp(-\delta^2 \mu / 2).
    \]
\end{claim}

\subsection{Leader Election, Collective Coin Flipping, and Random Selection Protocols}

\subsubsection{Formalizing Protocols in the full information model}
\label{subsubsec: defn leader election and collective coin flipping}

We formalize the definition of random selection protocols in the full information model. Collective coin flipping protocols and leader election protocols are special cases of such protocols where the output domain is $\zo$ and $[\ell]$ respectively. 

\begin{definition}[Random Selection Protocol in the full information model]
A \emph{$k$-round random selection protocol with output domain $Y$ over $\ell$ players where each player sends $r$ random bits per round} is a function 
\[
\pi: \left(\left(\zo^r\right)^{\ell}\right)^k \to Y
\]
that takes in the input of each of the players during each round and outputs an element from the set $Y$ which is the outcome of the protocol.

Here is how the protocol operates in the presence of a set $B\subset [\ell]$ of bad players:
In round $i$, each of the players from $[\ell]\setminus B$ independently outputs a uniformly random element from $\zo^r$. Let their collective outputs be $\alpha_i\in \left(\zo^r\right)^{[\ell]\setminus B}$. Then, depending on $\alpha_1, \dots, \alpha_i$, the players in $B$ together output an element of $\left(\zo^r\right)^{B}$. Hence, we model the strategy of the bad players as a sequence of functions $\sigma = (\sigma_1, \dots, \sigma_k)$, where 
\[
\sigma_i: \left(\left(\zo^r\right)^{[\ell]\setminus B}\right)^{i} \to \left(\zo^r\right)^{B},
\]
where $\sigma_i$ takes in the inputs of the good players from the first $i$ rounds and maps it to the output of the bad players for round $i$.
For a fixed strategy $\sigma$, the outcome of the protocol can be modeled as follows: uniform random strings $\alpha_1, \dots, \alpha_k\in \left(\zo^r\right)^{[\ell]\setminus B}$ are chosen, and the outcome of the protocol is
\[
\pi(\alpha_1: \sigma_1(\alpha_1), \alpha_2: \sigma_2(\alpha_1, \alpha_2), \dots, \alpha_k: \sigma_k(\alpha_1, \dots, \alpha_k)).
\]
\end{definition}

We now specialize this definition to define collective coin flipping protocols.

\begin{definition}[Collective coin flipping protocol]
A \emph{collective coin flipping protocol} $\pi$ is a protocol in the full information model with output domain $Y = \zo$.
Furthermore, we say $\pi$ is \emph{$(b, \gamma)$-resilient} if in the presence of any set $B$ of bad players with $\abs{B}\le b$, we have that $\max_{o\in \zo} \Pr[\pi|_B = o] \le 1-\gamma$.
\end{definition}

\begin{remark}
Typically in the pseudorandomness literature, the quality of a coin flip is measured by its distance to the uniform distribution. The definition of resilience that we use, which is standard in the leader election and collective coin flipping literature, has a weaker requirement that each outcome has probability at most $1-\gamma$.    
Our lower bound results rule out, for any small $\gamma$, that there exist coin flipping protocols that are $(b, \gamma)$-resilient, with tradeoffs between the number of ``bad'' players $b$ and the number of rounds $k$. On the other hand, our positive results (collective coin flipping protocols) satisfy the stronger measure of quality that is standard in the pseudorandomness literature:  their output is $\eps$-close to the uniform distribution over $\zo$, for small $\eps$, even in the presence of bad players who are colluding using any strategies (with tradeoffs between the number of rounds, the number of bad players, and the closeness to the uniform distribution).
\end{remark}

In this paper we will mostly consider the case in which $r = 1$ and this will be the default assumption unless stated otherwise.
Note that when $k=1$, the protocol $\pi$ just becomes a function over $\zo^{\ell}$; such one-round coin flipping protocols which cannot be biased by any small set of bad players are also known as \emph{resilient functions}.

\begin{remark}
In this paper, 
when showing that some subset of bad players can corrupt any coin flipping protocol, we will often construct strategies in which some subset of bad players output a random string in some rounds. By an averaging argument we can always turn such a strategy into one in which each bad player outputs a deterministic string in every round, while maintaining the bias of the protocol.
\end{remark}

We also specialize the definition of protocols to define leader election protocols:

\begin{definition}[Leader election protocol]
A \emph{leader election protocol} $\pi$ is a protocol in the full information model with output domain $Y = [\ell]$, the number of players the protocol is operating on.
Furthermore, we say $\pi$ is \emph{$(b, \eps)$-resilient} if in the presence of any set $B$ of bad players with $\abs{B}\le b$, we have that $\Pr[\pi|_B \in B]\le 1-\eps$.
\end{definition}


\subsubsection{A Useful Claim}

We record the following well-known claim which states that any leader election protocol implies a collective coin flipping protocol. For completeness, we supply a proof.

\begin{claim}\label{claim: leader to coin flip}
Fix $\gamma \in (0, 1/2]$. Let $\pi$ be a $k$-round leader election protocol where each player sends $r$ bits per round and where in the presence of any $b$ bad players and their colluding strategy, $\pi$ guarantees that a good leader is chosen with probability at least $\gamma$. Then, there exists a $(k+1)$-round collective coin flipping protocol $\pi'$ that is $(b, \gamma/2)$-resilient in which each player sends $r$ bits per round.
\end{claim}

\begin{proof}
Let $\pi'$ be the protocol that executes the protocol $\pi$ in the first $k$ rounds and in round $k+1$ asks the elected leader to flip a coin. Formally, say the messages sent in the first $k$ rounds in $\pi'$ are $\alpha$. Let $i$ be the index of the player that is chosen as the leader by $\pi$ on input $\alpha$. Then, in round $k+1$, $\pi'$ outputs the first bit of the message sent by player $i$.

We see that whenever $\pi$ selects a good leader, $\pi'$ outputs a truly random coin toss, giving us the desired resilience parameter for $\pi'$.
\end{proof}
An immediate consequence is that  proving  lower bounds for coin flipping protocols also gives lower bounds for leader election protocols.
\begin{corollary}
\label{cor: coin lb implies leader lb}
Fix $\gamma\in (0, 1/2]$. Suppose that for every $k$ round coin flipping protocol $\pi$ over $\ell$ players, there exist $b$ bad players that can $(1-\gamma)$-bias $\pi$ towards a particular outcome $o\in \zo$. Then, for every $(k-1)$-round leader election protocol $\pi$, there exist $b$ bad players such that a good player is elected as a leader with probability at most $2\gamma$.
\end{corollary}

\dobib

\section{Biasing a Family of Functions}\label{sec:family-baised}

In this section, we show that for any family of functions mapping $\zo^\ell \to \zo$, there exists a ``common set'' $B_R$ of coordinates in $[\ell]$ such that for almost every function $f$ in the family, $B_R$ along with a small ``heavy set'' of coordinates (which may depend on $f$) can together bias $f$.  Moreover, crucially, neither $B_R$ nor the ``heavy set'' is too large.  
This result plays an essential role in our main lower bound for $k$-round coin flipping protocols, just as its simplified version, \Cref{thm:proof overview family can be corrupted by random and heavy}, played a crucial role in the sketch of our two-round lower bound given in \Cref{sec:proof-overview}.

Formally, we show the following:

\begin{theorem}\label{lem: most functions from family can be corrupted with random and heavy set}
Let $0 < \gamma, \delta < 1/2 , h\in \N, \ell\in \N$ be such that $h\log (h/2) < 40\ell/\gamma$ and $8\le h\le \ell$. 
Let $\cF$ be any family of functions from $\zo^\ell$ to $\zo$, and for each $f\in \cF$ let $o_f\in \zo$ be such that $\Pr[f = o_f] \ge \gamma$.
Then, there exists $B_R\subset [\ell]$ with $\abs{B_R}\le \frac{100\ell\log(1/\delta)}{\gamma \log(h)}$ such that for $(1-\delta)$ fraction of functions $f\in \cF$, there exists $B_H = B_H(f)\subset [\ell]$ with $\abs{B_H}\le h$ such that $B_R\cup B_H$ can $(1-\gamma)$-bias $f$ towards $o_f$.
\end{theorem}

We will use the following main lemma to prove \cref{lem: most functions from family can be corrupted with random and heavy set}:

\begin{lemma}\label{lem: most random set with small heavy set corrupts f}
As in \Cref{lem: most functions from family can be corrupted with random and heavy set} let $0 < \gamma, \delta < 1/2 , h\in \N, \ell\in \N$ be such that $h\log (h/2) < 40\ell/\gamma$ and $8\le h\le \ell$.
Let $f: \zo^\ell\to \zo$ and $o\in \zo$ be such that $\Pr[f = o] \ge \gamma.$
Then, for $(1-\delta)$ fraction of $B_R\subset [\ell]$ with $\abs{B_R} = \frac{100\ell\log(1/\delta)}{\gamma \log(h)}$, there exists $B_H = B_H(B_R)\subset [\ell]$ with $\abs{B_H}\le h$ such that $B_R\cup B_H$ can $(1-\gamma)$-bias $f$ towards $o$.
\end{lemma}

Let us see how \Cref{lem: most random set with small heavy set corrupts f} yields \Cref{lem: most functions from family can be corrupted with random and heavy set}:

\begin{proof}[Proof of \cref{lem: most functions from family can be corrupted with random and heavy set}]
    Let $G$ be the bipartite graph with left vertex set $U$ given by $\cF$ and right vertex set $V$ consisting of all size-$r$ subsets of $[\ell]$, where $r=\frac{100\ell\log(1/\delta)}{\gamma \log(h)}$.
    $G$ contains an edge between $f\in U$ and $B_R\in V$ if there exists $B_H = B_H(f, B_R)\subset [\ell]$ with $\abs{B_H}\le h$ such that $B_R\cup B_H$ can $(1-\gamma)$-bias $f$ towards $o_f$.
    By \cref{lem: most random set with small heavy set corrupts f}, the minimum left degree of this graph is at least $(1-\delta)\binom{\ell}{r}$.
    This implies that the average right degree of this graph is at least $(1-\delta)\abs{\cF}$, and hence, there exists $B_R^*\in V$ with degree at least $(1-\delta)\abs{\cF}$.
    Thus, there must exist some fixed $B_R^*$ such that for $(1-\delta)$ fraction of $f\in \cF$, there exists $B_H = B_H(f)$ with $|B_H| \leq h$ such that $B_R^*\cup B_H$ can $(1-\gamma)$-bias $f$ towards $o_f$.
\end{proof}

\subsection{Proving the Main Lemma}

In this subsection we will prove \cref{lem: most random set with small heavy set corrupts f}.

We will use the following concentration inequality regarding submartingales from \cite{RSZ02leader}:
\begin{lemma} [Lemma~9 of \cite{RSZ02leader}] \label{lem: azuma rsz}
Let $0 < \mu < 1, 0 < \eta < 1, \ell\in \N$ be arbitrary. Let $\bZ_0, \bZ_1, \dots, \bZ_\ell$ form a submartingale with $\bZ_0 = 0$, and suppose that for $i\in [\ell]$ we have $\bZ _i-\bZ_{i-1} \in [0,1]$ and $\E[\bZ_i - \bZ_{i-1}]\ge \mu$.
Then,
\[
    \Pr[\bZ_\ell < (1-\eta)\ell\mu] < e^{-\eta^2\mu \ell / 2}.
\]
\end{lemma}

We will also use the following result from \cite{RSZ02leader}, which follows as a slight extension of \cite{KKL88influence}:

\begin{lemma}\label{lem: kkl - heavy or average influence}
Let $0< \gamma < {\frac 1 2}, 0 < \theta < \frac{1}{8}, \ell\in \N$. Let $f: \zo^\ell\to \zo$ be such that $\gamma\le \Pr[f = 1]\le 1-\gamma$. If $\I_i(f) \le \theta$ for all $i\in [\ell]$, then 
\[
    \sum_{i=1}^\ell \I_i(f) \ge \frac{\gamma \log(1/\theta)}{20}.
\]
\end{lemma}

Let us see how \cref{lem: most random set with small heavy set corrupts f} follows using these results:

\begin{proof}[Proof of \cref{lem: most random set with small heavy set corrupts f}]
    Without loss of generality, assume $o = 1$.
    Let $r = \frac{100\ell\log(1/\delta)}{\gamma \log(h/2)}$.
    Consider the following semi-random procedure.  
    \begin{enumerate}
        \item Initialize $B_R\gets \emptyset, B_H\gets \emptyset$. Do the following for $r$ steps or until $\Pr[f|_{B_R\cup B_H} = 1] \ge 1-\gamma$:
        \begin{enumerate}
        \item (Heavy Case) If there exists $i\in [\ell]\setminus (B_R\cup B_H)$ with $I_i(f|_{B_R\cup B_H})\ge \frac{2}{h}$, then add $i$ to $B_H$.

        \item (Random Case) Otherwise, pick a random $i\in [\ell]\setminus (B_R\cup B_H)$ and add $i$ to $B_R$.
        \end{enumerate}
    \end{enumerate}
    We say that this procedure \emph{succeeds} if at the end of this process, $B_R\cup B_H$ can $(1-\gamma)$-bias $f$.
    We will show that this procedure succeeds with high probability:
    \begin{claim}\label{claim: process succeeds}
        With probability $\ge 1-\delta$ over the above process, $B_R\cup B_H$ can $(1-\gamma)$-bias $f$ towards 1.
    \end{claim}

    We will prove this claim later. Using this, we now show that for $1-\delta$ fraction of $B_R$, there exists $B_H$ with $\abs{B_H}\le h$ such that $B_R\cup B_H$ can $(1-\gamma)$-bias $f$ towards 1.
    
    Firstly, since every element that is added to $B_H$ increases the probability of outputting $1$ by $\frac{1}{h}$,
    it is always the case that $\abs{B_H}\le h$. We now prove the remaining statement.
    
    Consider a random set $\bR\subset [\ell]$ with $\abs{\bR} = r$ and a random permutation $\bpi$ of $[r]$ so that $(\bR, \bpi)$ determines an ordering of the elements from $\bR$.  
    For any fixed $(R, \pi)$, consider the modified random process from above where at each step in the Random case, instead of picking a truly random element, we pick the earliest element from $R$ that has not yet been added to $B_R\cup B_H$.
    Since the process always ends with $\abs{B_R} \le r$, this is a well defined operation and  we will always have that $B_R\subset R$.
    Hence if the process succeeds, we have that $R\cup B_H$ can $(1-\gamma)$-bias $f$ towards 1.
    
    By \cref{claim: process succeeds}, we know that for $(1-\delta)$ fraction of choices of $(R, \pi)$, the above process succeeds. By an averaging argument, this implies there exists a fixed permutation $\pi^*$ such that for $(1-\delta)$ fraction of choices of $R$, it holds that $R\cup B_H$ (where $B_H = B_H(R, \pi^*)$) $(1-\gamma)$-biases $f$ towards 1, as desired. 

    We finally prove the remaining claim:
    \begin{proof}[Proof of \cref{claim: process succeeds}]
    For $j\in [r]$, let the variable chosen at each step $j$ of the above semi-random process be $v_j$.
    For $j\in [r]$, let:
    \[
        \X_j = 
        \begin{cases}
            \I_{v_j}(f|_{v_1, \dots, v_{j-1}}) & \text{if~}\Pr[f|_{v_1, \dots, v_{j-1}} = 1] < 1-\gamma\\
            1 & \textrm{otherwise}
        \end{cases}
    \]
    At every step $j\in [r]$ where $\Pr[f|_{v_1, \dots, v_{j-1}} = 1] < 1-\gamma$ (note that since initially $\Pr[f = 1] \ge \gamma$, this also holds at all steps $j$ as well), by \cref{lem: kkl - heavy or average influence},
    either there exists a variable with influence $\frac{2}{h}$ or the sum of influences of all the variables is at least $\frac{\gamma \log(h/2)}{20}$. By our choice of $h$ we have that $\frac{2}{h}\ge \frac{\gamma \log(h/2)}{20 \ell}$, and so for all $j\in [r]$, $\E[\X_j | \X_1, \dots, \X_{j-1}] \ge \frac{\gamma \log(h/2)}{20 \ell}$.

    For $j\in [r]$, let $\bZ_j = \sum_{k\le j} \X_k$.
    We then see that $\bZ_1, \dots, \bZ_r$ form a submartingale with $\bZ_0=0$.
    We also have that
    \[
        \E[\bZ_r | \bZ_1, \dots, \bZ_{r-1}] = \frac{r\gamma \log(h/2)}{20\ell} \ge 4\log(1/\delta),
    \]
    where in the last inequality, we used the fact that $r = \frac{100\ell\log(1/\delta)}{\gamma \log(h/2)}$.
    Applying \cref{lem: azuma rsz} (with $\eta = \frac{1}{2})$, we infer that
    \[
        \Pr[\bZ_r < 2\log(1/\delta)] \le e^{-1/4 (4\log(1/\delta))} < \left(1/e\right)^{\log(1/\delta)}  < \delta.
    \]
    Since $\delta < \frac{1}{2}, 2\log(1/\delta) > 2$.
    Hence, with probability at least $1-\delta$, we have $\bZ_r > 2$.
    
    We claim that whenever this happens, we must be in the case that $\Pr[f|_{B_R\cup B_H} = 1] \ge 1-\gamma$.
    For $1\le j\le r$, let the random variables $\X_j, \bZ_j$ take on values $x_j, z_j$ respectively.
    Let $j^* \le r$ be such that at the end of that step, $\Pr[f|{v_1, \dots, v_{j^*}} = 1] \ge 1-\gamma$; if this doesn't happen by then, let $j^* = r+1$.
    Then, since we picked an influential variable and bias it towards $1$, for $1\le j\le j^*$ we have that
    \[
        \Pr[f|{v_1, \dots, v_{j-1}, v_j} = 1] \geq \Pr[f|{v_1, \dots, v_{j-1}} = 1] + x_j / 2.
    \]
    So, for all $1\le j\le j^*$, 
    \[
        \Pr[f|{v_1, \dots, v_{j-1}, v_j} = 1] \ge \gamma + z_j / 2
    \]
    Since the probability of any event is always less than $1$, and we know that $z_r \ge 2$, it must be the case that $j^* \le r$. So, we have that at the end of step $j^*$, $\Pr[f|{v_1, \dots, v_{j^*}} = 1] \ge 1-\gamma$ and the process succeeded.
    Hence, the process indeed succeeds with probability $1-\delta$ as desired.
    This concludes the proof of \cref{claim: process succeeds}.
    \end{proof}
This concludes the proof of \cref{lem: most random set with small heavy set corrupts f}.
\end{proof}

\dobib

\section{Biasing Coin Flipping Protocols} \label{sec:biasing-protocols}

We will prove the following lower bound regarding coin flipping protocols:

\begin{theorem}\label{thm: protocol lb}
There exist universal constants $C = 10^7, \ell_0 \in \N$ such that for all $k\in \N, \ell\in \N, 0 < \gamma < 1/4$, the following holds: For any $k$-round coin flipping protocol $\pi$ over $\ell \ge \ell_0$ players, sending $1$ bit per round, where $\Pr[\pi = 1]\ge \gamma$, there exists a set of bad players $B\subset [\ell]$ with $\abs{B}\le \frac{C \ell}{\gamma \log^{(k)}(\ell)}$ such that $\Pr[\pi|_B = 1] \ge 1-\gamma$.
\end{theorem}
From this, we obtain the following lower bound on the number of rounds required for coin flipping in the presence of a linear sized coalition of bad players:
\begin{corollary}
\label{cor: round lb}
There exist universal constants $\ell_0, C$ such that for all $\ell\in \N$ where $\ell \ge \ell_0$, the following holds: Let $\pi$ be a $k$-round coin flipping protocol over $\ell$ players, sending $1$ bit per round, where $k\le \log^{*}(\ell) - C$. Then, there exists a set of bad players $B\subset [\ell]$ with $\abs{B}\le 0.01\ell$ and $o\in \zo$ such that $\Pr[\pi|_{B} = o] \ge 0.999$.
\end{corollary}

From \cref{thm: protocol lb}, we also obtain the following lower bounds regarding leader election protocols:
\begin{corollary}\label{thm: leader election lb}
There exist universal constants $C, \ell_0 \in \N$ such that for all $k,\ell\in \N, 0 < \gamma < 1/4$, the following holds: For any $k$-round leader election protocol $\pi$ over $\ell \ge \ell_0$ players, there exists a set of bad players $B\subset [\ell]$ with $\abs{B}\le \frac{C \ell \log^{(k+1)}}{\gamma \log^{(k)}(\ell)}$ such that $\Pr[\pi|_B \in B] \ge 1-\gamma$.
\end{corollary}
We prove this corollary in \cref{subsec: leader election lb}.


To prove \cref{thm: protocol lb}, we will primarily utilize \cref{lem: most functions from family can be corrupted with random and heavy set}.
We will also use a simpler result 
that can be established by inductively applying the  KKL theorem \cite{KKL88influence}:
\begin{theorem}[KKL]\label{thm: kkl}
There exists a universal constant $C \le 10^7$ such that for all $\ell\in \N, \gamma > 0$ the following holds: Let $f: \zo^{\ell} \to \zo$ be such that $\Pr[f = 1] \ge \gamma$. Then, there exists a set of bad players $B$ with $\abs{B}\le \frac{C \ell}{\gamma \log (\ell)}$ such that $\Pr[f|_B = 1] \ge 1-\gamma$.
\end{theorem}

Using this, we finally prove our main theorem:

\begin{proof}[Proof of \cref{thm: protocol lb}]
We apply induction on $k\ge 1$.
For $k=1$, the result follows by \cref{thm: kkl}.
For $k\ge 2$, we present an `algorithm' which, given as input the protocol $\pi$,  produces a small coalition of bad players that $(1-\gamma)$-biases $\pi$ towards 1.

\paragraph{Setup.}
We will maintain two kinds of bad players: First, let $B_R, B_{H}\subset [\ell]$ be the sets of bad players that we will use to bias round $k$. Second, we let $B_I\subset [\ell]$ be the set of bad players that we will use to bias 
rounds $1$ to $k-1$ (these will be provided to us by induction). At the end, we will take the union of all these sets, and that will be the final set of bad players that we will use to bias $\pi$.
Let $h, c\in [\ell]$ be parameters that we set later (for intuition, $h$ will be set to be very small and $c$ will be set to be very large).

Given $\alpha \in (\zo^\ell)^{(k-1)}$, we write $\pi_\alpha(x)$ to denote the function from $\zo^\ell$ to $\zo$ that corresponds to the output of the protocol when the bits that the $\ell$ players output in rounds $1,\dots,k-1$ are given by $\alpha$ and the bits that the players output in round $k$ are given by $x.$

\paragraph{Informal description of algorithm.}
Since $\Pr[\pi = 1]\ge \gamma$, for a non-trivial fraction of $\alpha\in (\zo^\ell)^{k-1}$, the induced function $\pi_{\alpha}: \zo^{\ell}\to \zo$ has a non-trivial probability of outputting $1$. For the family of functions consisting of all such $\pi_\alpha$, we use \cref{lem: most functions from family can be corrupted with random and heavy set} to find a set $B_R$ such that for most such $\alpha$, there exists a set $H = g(\alpha)$ with $\abs{H}= h$ such that $B_R\cup g(\alpha)$ can bias $\pi_{\alpha}$ towards $1$.
We then use induction for the first time, biasing players from the first $k-1$ rounds, so that the overall fraction of $\alpha$ for which $B_R\cup g(\alpha)$ can bias $\pi_{\alpha}$ towards $1$ 
becomes at least $3/4$.

Note that the function $g: (\zo^\ell)^{k-1} \to 2^{[\ell]}$ can be viewed as $g(\alpha) = (g_1(\alpha), \dots, g_h(\alpha))$ where each $g_j$ maps $(\zo^\ell)^{k-1}$  to $[\ell]$ and the value $g_j(\alpha)$ is simply the $j$-th largest element of the set $g(\alpha)$.
We will inductively find bad players that can bias 
each of $g_1, \dots, g_h$ so that over most inputs, their image lies in a set of size $c$.

We maintain sets $C_1, \dots, C_h$ so that each of $g_1, \dots, g_h$ outputs an element from these respective sets with high probability.
Initially all these sets are set to equal $[\ell]$. We then iteratively find some $C_j$ such that $\abs{C_j} \ge c$ and cut $C_j$ in half. Doing this decreases  by a factor of $2$ the fraction of inputs $\alpha$ whose outputs fall into the sets $C_1, \dots, C_h$ ; so the fraction of inputs $\alpha$ for which $B_R\cup g(\alpha)$ can corrupt $\pi_{\alpha}$ is halved. 
To fix this, we use induction for a second time and find bad players that can corrupt the first $k-1$ rounds so that this fraction
again becomes at least $3/4$. 

At the end of this process, we are guaranteed that $\abs{C_1}\le c, \dots, \abs{C_h}\le c$. We then once again for a third time
use induction and find bad players to bias the first $k-1$ rounds so that the fraction of $\alpha$ for which $B_R\cup g(\alpha)$ can corrupt $\pi_{\alpha}$ is as large as desired. 
We collect the union of the set of bad players outputted by $g$, i.e.~the union of sets $C_1, \dots, C_h$, along with $B_R$ and the bad players from all our inductive calls and let that be our final set of bad players. 

We now briefly describe the parameters. Since we cut each of $C_1,\dots,C_h$ in half until its size is less than $c$, the second time we use induction we call the inductive hypothesis $h\log(\ell/c) + O(1)$ times and this will dominate the cost of our inductive calls.
Each call adds $\ell / \log^{(k-1)}(\ell)$ bad players. We also get $hc + \abs{B_r} = hc + O(\ell / \log(h))$ bad players that will bias the last round. Setting $h = \left(\log^{(k-1)}(\ell)\right)^{1/C_0}$ and $c = \ell / \left(\log^{(k-1)}(\ell)\right)^{C_0}$ for some large constant $C_0$ yields the desired bound.

\paragraph{Formal description of algorithm.}
Formally, our algorithm proceeds as follows: 
\begin{enumerate}
    \item 
    Let $B_R, B_H, B_I\gets \emptyset$.
    We begin by biasing the last round. 

    \begin{enumerate}
    \item
    For each $\alpha\in \left(\zo^{\ell}\right)^{k-1}$, let $\pi_{\alpha}: \zo^{\ell}\to \zo$ be the induced round $k$ protocol.

    \item
    Set $\cF = \{\pi_{\alpha}\}_{\E[\pi_\alpha] \ge \gamma / 2}$ and apply \cref{lem: most functions from family can be corrupted with random and heavy set} to $\cF$ with parameters $h, \gamma / 2$ and $\delta = 1/3$.
    Let $B_R$ be the set given by \cref{lem: most functions from family can be corrupted with random and heavy set}, and for $\alpha\in \left(\zo^{\ell}\right)^{k-1}$ let $g: \left(\zo^{\ell}\right)^{k-1} \to \left([\ell]\right)^{h} \cup \bot$ be such that (viewing $g(\alpha)$ as a set in the obvious way) $B_R\cup g(\alpha)$ can $(1-\gamma/2)$-bias $\pi_{\alpha}$ towards $1$ if possible, otherwise $g(\alpha) = \bot$.
    
    We assert that $\Pr[g\ne \bot] \ge \gamma / 6$ (this will be proven in \cref{claim: step 1b assertion}).

    \item 
    Let $\pi_{temp}^{(1)}$ be the $(k-1)$-round protocol such that $\pi_{temp}^{(1)}(\alpha) = 1$  iff $g(\alpha)\ne \bot$.
    We have that $\Pr[\pi_{temp}^{(1)} = 1] \ge \gamma / 6$.
    By induction (the first use of induction mentioned in the informal overview) 
    we can find a set of bad players, which we denote $B_{temp}^{(1)}$, so that $\Pr[\pi_{temp}^{(1)}|_{B_{temp}^{(1)}} = 1] \ge (1-\gamma/6) \ge 3/4$. Add all the bad players from $B_{temp}^{(1)}$ to $B_I$.
    \end{enumerate}

    \item 
    We now find bad players that can bias $g$.
    Initialize sets $C_1 = \dots = C_h = [\ell]$. 
    We will change these sets below while maintaining the following invariant: 
    \begin{quote}
        $(*)$~~~For every $\alpha$, if there exists $j\in [h]$ such that $g_j(\alpha)\not\in C_j$, then $g(\alpha) = \bot$. (Equivalently, if $g(\alpha) \neq \bot$, then $g_j(\alpha) \in C_j$ for all $j \in [h].$)
    \end{quote}
        
        We iteratively bias players from rounds $1$ to $k-1$ so that for all $j\in [h]$, $C_j$ will have $\abs{C_j}\le c$.
        At the end of each iteration, we maintain that $\Pr[g\ne \bot] \ge 3/4$ (this is equivalent to having $\Pr_{\alpha}[g_1(\alpha)\in C_1, \dots, g_h(\alpha)\in C_h] \ge 3/4$). This is done as follows:

    \begin{enumerate}
    
        \item 
        While there exists $j\in [h]$ with $\abs{C_j} > c$, do the following:
        \begin{enumerate}
            \item 
            Let $\X$ be a random subset of $C_j$ of size $\abs{C_j} / 2$.
            Since $\Pr_\alpha[g_j(\alpha)\in C_j]\ge 3/4$, we have that $\E_{\X}[\Pr[g_j(\alpha)\in \X]]\ge 3/8$.
            So, there exists $C'_j\subset [\ell]$ with $\abs{C'_j} = \abs{C_j}/2$ such that $\Pr_\alpha[g_j(\alpha)\in C'_j] \ge 3/8$.
    
            \item
            Update $C_j\gets C'_j$ and let $g_j(\alpha) = \bot$ if $g_j(\alpha)\not\in C_j$.
            We now have $\Pr_\alpha[g(\alpha)\ne \bot] \ge 3/8\ge 1/4$.
    
            \item 
            \label{item: Second use of induction}
            Let $\pi_{temp}^{(2)}$ be the $(k-1)$-round protocol such that $\pi_{temp}^{(2)}(\alpha) = 1$  iff $g(\alpha)\ne \bot$.
            We have that $\Pr[\pi_{temp}^{(2)} = 1] \ge 1/4$.
            By induction (the second use of induction mentioned in the informal overview), we can find a set of bad players, which we denote $B_{temp}^{(2)}$, so that $\Pr[\pi_{temp}^{(2)}|_{B_{temp}^{(2)}} = 1] \ge 3/4$. Add all the bad players from $B_{temp}^{(2)}$ to $B_I$.
        \end{enumerate}
    \end{enumerate}

    \item
    Let $\pi_{temp}^{(3)}$ be the $(k-1)$-round protocol such that $\pi_{temp}^{(3)}(\alpha) = 1$  iff $g(\alpha)\ne \bot$.
    We have that $\Pr[\pi_{temp}^{(3)} = 1] \ge 3/4\ge \gamma / 2$.
    By induction (the third use of induction mentioned in the informal overview), we can find a set of bad players, which we denote $B_{temp}^{(3)}$, so that $\Pr[\pi_{temp}^{(3)}|_{B_{temp}^{(3)}} = 1] \ge 1 - \gamma/2$. Add all the bad players from $B_{temp}^{(3)}$ to $B_I$.

    \item 
    Finally, let $B_H = \bigcup_{j\in [h]} C_j$.
\end{enumerate}

\paragraph{Correctness of the algorithm.}
We now prove the correctness of our procedure above.
We first prove the following claim that we informally asserted at end of step 1(b) above:
\begin{claim}
\label{claim: step 1b assertion}
At the end of step $1(b)$, $\Pr[g\ne \bot] \ge \gamma / 6$.
\end{claim}
\begin{proof}
Using the reverse Markov argument from \cref{claim:reverse-markov}, since $\Pr[\pi = 1]\ge \gamma$, we have that $\frac{\abs{\cF}}{2^{(k-1)\ell}} = \frac{\abs{\alpha\in (\zo^\ell)^{(k-1)}: \E[\pi_{\alpha}]\ge \gamma / 2}}{2^{(k-1)\ell}} \ge \gamma / 4$.
Since we set $\delta = 1/3$ while using \cref{lem: most functions from family can be corrupted with random and heavy set}, we indeed infer that $\Pr[g \ne \bot] \ge \gamma  /6$.
\end{proof}
Lastly, we show that the protocol is indeed $(1-\gamma)$ biased by our set of bad players. At the end of the last step, we have that $\Pr[g\ne \bot] \ge (1 - \gamma / 2)$. Moreover, by the choice of $B_R$ and $g$ in step $1$, we know that if for $\alpha\in \left(\zo^{\ell}\right)^{k-1}$ we have that $g(\alpha)\ne \bot$, then $B_R\cup g(\alpha)$ can bias $\pi_{\alpha}$ so that $\Pr[\pi_{\alpha}|_{B_R\cup B_H} = 1] \ge 1 - \gamma / 2$. Combining these, we conclude
\[
\Pr[\pi|_{B_R\cup B_H\cup B_I} = 1] \ge \Pr[g\ne \bot]\cdot (1-\gamma / 2) \ge (1 - \gamma / 2)^2 \ge 1 - \gamma
\]
as desired.

\paragraph{Setting parameters.}
We finally set parameters. 
We set $h = \left(\log^{(k-1)}(\ell)\right)^{1/10^4}, c = \frac{\ell}{\left(\log^{(k-1)}(\ell)\right)^{10^4}}$.
Recalling \Cref{lem: most functions from family can be corrupted with random and heavy set}, we see that 
\begin{equation} \label{eq:BRbound}
\abs{B_R} 
\le \frac{100\ell\log(3/2)}{\gamma \log(h)} 
\le \frac{10^6\ell}{\gamma \log^{(k)}(\ell)}.
\end{equation}
We also have that
\begin{equation} \label{eq:BHbound}
\abs{B_H}
\le c\cdot h
\le \frac{\ell}{\left(\log^{(k-1)}(\ell)\right)^{10^4}}\cdot \left(\log^{(k-1)}(\ell)\right)^{1/10^4}
\le \frac{\ell}{\left(\log^{(k-1)}(\ell)\right)^{10^4-1}}
\end{equation}

We see that the while loop of step~2(b) iterates for at most 
$h\cdot \log(\ell/c)\le 10^4\log^{(k)}(\ell)\cdot \left(\log^{(k-1)}(\ell)\right)^{1/10^4}$ 
steps. In each iteration, we inductively have  $\abs{B^{(2)}_{temp}}\le \frac{10^7\ell}{(1/4)\log^{(k-1)}(\ell)}$. We also call our inductive bound twice, once before the loop and once after, and for each of these calls the corresponding sets inductively satisfy $\abs{B^{(1)}_{temp}}, \abs{B^{(3)}_{temp}} \le \frac{10^7\ell}{(\gamma/6)\log^{(k-1)}(\ell)}$. This gives
\begin{equation}
\label{eq:BIbound}
\abs{B_I}
\le 
h\cdot \log(\ell/c)\cdot \frac{10^7\ell}{(1/4)\log^{(k-1)}(\ell)} + 2\cdot \frac{10^7\ell}{(\gamma/6)\log^{(k-1)}(\ell)}
\le 
\frac{10^{12}\log^{(k)}(\ell)}{\left(\log^{(k-1)}(\ell)\right)^{1 - 1/10^4}}
+
\frac{10^9 \ell}{\gamma \log^{(k-1)}(\ell)}.
\end{equation}

Hence, the total number of bad players we used is 
\begin{align*}
\abs{B_R} + \abs{B_H} + \abs{B_I} 
& \le \frac{10^6\ell}{\gamma \log^{(k)}(\ell)} + \frac{\ell}{\left(\log^{(k-1)}(\ell)\right)^{10^4-1}} + \frac{10^{12}\log^{(k)}(\ell)}{\left(\log^{(k-1)}(\ell)\right)^{1 - 1/10^4}} + \frac{10^9 \ell}{\gamma \log^{(k-1)}(\ell)}\\
& \le \frac{10^7\ell}{\gamma \log^{(k)}(\ell)}
\end{align*}
wherein for the last inequality, we set $\ell_0$ large enough and  use the fact that $\ell \ge \ell_0$.
\end{proof}

\subsection{Protocols with Longer Messages}

In this subsection, we show that coin flipping protocols in which each player is allowed to send more than one bit per round can also be biased by relatively few bad players. In particular, we show:
\begin{theorem}
\label{thm: protocol lb longer messages}
There exist universal constants $C = 10^7, \ell_0 \in \N$ such that for all $k\in \N, \ell\in \N, 0 < \gamma < 1/4, 0 < \delta < 1/10^6, \eps, r_1, \dots, r_k\in \N$ where $\eps \ge \frac{C\log(1/\gamma\delta)\log^{(k+2)}(\ell)}{\log^{(k+1)}(\ell)}$ and for $i\in [k]$, $1\le r_i\le (\log^{(i)}(\ell))^{1-\eps}$, the following holds: Let $\pi$ be a $k$-round coin flipping protocol over $\ell \ge \ell_0$ players where $\Pr[\pi = 1]\ge \gamma$ and in round $i$, each player is allowed to send $r_i$ bits. Then, there exists a set of bad players $B\subset [\ell]$ with $\abs{B}\le \delta \ell$ such that $\Pr[\pi|_B = 1] \ge 1-\gamma$.
\end{theorem}

Setting parameters, we obtain the following corollaries regarding corrupting protocols using $0.001\ell$ many bad players:

\begin{corollary}
Let $\Delta: \N \to \N$ be such that $\Delta(\ell)\ge \omega(1)$.
Then, there exists $\eps: \N\to [0, 1]$ where $\eps(\ell) \le o(1)$ and $\ell_0\in \N$ such that for all $k, \ell\in \N$ with $\ell\ge \ell_0$, and $k\le \log^*(\ell) - \Delta(\ell)$, the following holds:
Let $\pi$ be a $k$ round protocol over $\ell$ players where in round $i\in [k]$, the number of bits each player can send is at most $r_i = (\log^{(i)}(\ell))^{1-\eps(\ell)} = (\log^{(i)}(\ell))^{1-o(1)}$.
Then, there exists a set of bad players $B\subset [\ell]$ and an outcome $o\in \zo$ with $\abs{B}\le 0.001\ell$ such that $\Pr[\pi|_{B} = o] \ge 0.999$. 
\end{corollary}
In this first corollary, we let the number of rounds be $\log^*(\ell) - \Delta(\ell)$ where $\Delta(\ell) \ge \omega(1)$ and from that obtained a constraint on number of bits per round. We also note down the following corollary which follows from setting $\Delta$ to be a very large fixed constant:

\begin{corollary}
There exists universal constants $\ell_0, \Delta \in \N$ such that for all $k, \ell\in \N$ with $\ell\ge \ell_0$, and $k\le \log^*(\ell) - \Delta$, the following holds:
Let $\pi$ be a $k$ round protocol over $\ell$ players where in round $i\in [k]$, the number of bits each player can send is at most $r_i = (\log^{(i)}(\ell))^{0.999}$.
Then, there exists a set of bad players $B\subset [\ell]$ and an outcome $o\in \zo$ with $\abs{B}\le 0.001\ell$ such that $\Pr[\pi|_{B} = o] \ge 0.999$.
\end{corollary}


To prove     \cref{thm: protocol lb longer messages}, we will slightly modify the proof of \cref{thm: protocol lb} and set parameters differently. Here's how we proceed:

\begin{proof}[Proof of \cref{thm: protocol lb longer messages}]
We proceed by induction on $k$. For $k = 1$, we view the protocol $\pi: (\zo^{r_1})^{\ell}$ as a function over $\ell\cdot r_1$ bits and apply \cref{thm: kkl} to find a set $B\subset [\ell r]$ with $\abs{B}\le \frac{C\ell r_1}{\gamma \log(\ell r_1)}$ such that if all bits in $B$ are controlled by bad players, then they can $(1-\gamma)$-bias $\pi$ towards $1$. We let $B'\subset [\ell]$ denote the set of players that control all of the bits in $B$. We see that 
\[
\abs{B'}\le \abs{B}\le \frac{C\ell r_1}{\gamma \log(\ell r_1)}\le \frac{C\ell (\log (\ell))^{1-\eps}}{\gamma \log(\ell)} = \frac{C\ell}{\gamma (\log (\ell))^{\eps}}.
\]
We need to show $\abs{B'}\le \delta \ell$, or equivalently that $\frac{C}{\gamma (\log \ell)^{\eps}}\le \delta$. We rewrite this inequality as $\eps\ge \frac{\log(C/\gamma\delta)}{\log^{(2)}(\ell)}$ which holds by choice of $\eps$.

For $k\ge 2$, we proceed by using the same exact algorithm as in \cref{thm: protocol lb} - for $\alpha\in \zo^{r_1}\times \dots \times \zo^{r_{k-1}}$, we view $\pi_{\alpha}: (\zo^{r_k})^{\ell}$ as a function over $\ell r_k$ bits. We set some parameters $c, h$ and find sets $B_R, B_H, B_I$ where $B_R, B_H$ will find bad bits for the last round and $B_I$ will apply induction to find bad players that can corrupt the first $k-1$ rounds. When applying induction for $(k-1)$-round protocols, we call it with parameter $\delta_I$ that we set later. So, our inductive calls  to $k-1$ round protocol will be such that it corrupts at most $\delta_I \ell$ bad players. We will then choose bad players for the last round such that they will control the bits that are in $B_R\cup B_H$. We see that in that case, the number of bad players we control over last round is at most $\abs{B_R} + \abs{B_H}$.

Before setting parameters, we first recall from \Cref{eq:BRbound} and \Cref{eq:BHbound} that $\abs{B_R}\le \frac{100\ell r_k \log(3/2)}{\gamma \log (h)}$ and $\abs{B_H}\le c\cdot h$.
We also see, from \Cref{eq:BIbound} and the discussion preceding it, that $\abs{B_I}\le (h\cdot \log(\ell/c)+2)\cdot \delta_I \ell$. Also to enforce that we overall have $\delta \ell$ bad players, we will ensure $\abs{B_R}\le \frac{\delta\ell}{4}, \abs{B_H}\le \frac{\delta\ell}{4}$ and $\abs{B_I} \le \frac{\delta \ell}{4}$. 

We finally set parameters. We let $h$ be such that $\log h = \frac{400\log(3/2) (\log^{(k)}(\ell))^{1-\eps}}{\gamma \delta}$ and let $c = \frac{\delta \ell}{4h}$. We set $\delta_I = \delta / h^2$.
We now bound each of $B_R, B_H, B_I$ and also, we show that we can indeed set $\delta_I$ to the prescribed value for induction, checking that it satisfies the restriction on $\eps$.
First, we bound $\abs{B_R}$:
\[
    \abs{B_R}\le \frac{100\ell r_k \log(3/2)}{\gamma \log (h)} = \frac{\delta\ell r_k}{4 (\log^{(k)}(\ell))^{1-\eps}} \le \frac{\delta\ell}{4}.
\]
Second, we bound $\abs{B_H}$:
\[
    \abs{B_H}\le c\cdot h \le \frac{\delta\ell}{4}.
\]
Finally, we bound $\abs{B_I}$. To do that, we first observe that by choice of $h$, we have that $h > \frac{\log (h)}{10^6}$ and $h > \frac{\log(1/\delta)}{10^6}$. So, $2\log(4h/\delta) < h/4$. Using this, we see:
\begin{align*}
\abs{B_I}
& = (h\log(\ell/c) + 2) \delta_I \ell\\
& \le 2(h\log(\ell/c)) \frac{\delta}{h^2} \ell\\
& = \frac{2\log(4h/\delta)}{h}\cdot \delta \ell\\
& \le \frac{\delta \ell}{4},
\end{align*}
where in the last inequality we used the bound $2\log(4h/\delta) < h/4$.
Lastly, we show that we can indeed set $\delta_I$ to the desired value, i.e., that $\eps \ge \frac{C\log(1/\gamma \delta_I)\log^{(k+1)}(\ell)}{\log^{(k)}(\ell)}$. To show this inequality holds, we will make use of the fact that $\eps \ge \frac{C\log(1/\gamma\delta)\log^{(k+2)}(\ell)}{\log^{(k+1)}(\ell)}$.
We see that
\begin{align*}
\frac{C\log(1/\gamma \delta_I)\log^{(k+1)}(\ell)}{\log^{(k)}(\ell)}
& = \frac{C\log(h^2/\gamma\delta)\log^{(k+1)}(\ell)}{\log^{(k)}(\ell)}\\
& \le \frac{C\log(h^3)\log^{(k+1)}(\ell)}{\log^{(k)}(\ell)} & \textrm{(since $h\ge (1/\gamma \delta)$)}\\
& = \frac{3C\log(h)\log^{(k+1)}(\ell)}{\log^{(k)}(\ell)}\\
& = \frac{1200\log(3/2)C(\log^{(k)}(\ell))^{1-\eps}\log^{(k+1)}(\ell)}{\log^{(k)}(\ell)}\\
& \le \frac{1200C\log^{(k+1)}(\ell)}{\left(\log^{(k)}(\ell)\right)^{\eps}}.
\end{align*}
Hence, it suffices to show that $1200C\log^{(k+1)}(\ell) \le \eps (\log^{(k)}(\ell))^{\eps}$.
Equivalently, we want to show that $\log(1200C) + \log^{(k+2)}(\ell) \le \log(\eps) + \eps(\log^{(k+1)}(\ell))$.
This last inequality indeed holds since by assumption we have that $\eps \ge \frac{C\log^{(k+2)}(\ell)}{\log^{(k+1)}(\ell)}$.
\end{proof}

\subsection{Biasing leader election protocols}\label{subsec: leader election lb}

We here show how to bias $k$-round leader election protocols as specified in \cref{thm: leader election lb}. We do this by repeated applications of \cref{thm: protocol lb}.

\begin{proof}[Proof of \cref{thm: leader election lb}]
We fix a leader election protocol $\pi$.
We will iteratively decrease the output domain of $\pi$ by a factor of half, each time invoking \cref{thm: leader election lb} while doing so.
Let $H_0 = [\ell]$ and $B_0 = \varnothing$.
We proceed in $t = \log^{(k+1)}(\ell)$ steps. 
At the end of each step $i$ we will maintain the invariant that $\Pr[\pi|_{B_i}\in H_i] \ge 1 - \gamma$.
For $i = 1, \dots, t$, we proceed as follows:
\begin{enumerate}
\item 
The invariant property guarantees us that $\Pr[\pi|_{B_{i-1}}\in H_{i-1}] \ge 1 - \gamma$.
So by an averaging argument, there exists $H_i\subset H_{i-1}$ with $\abs{H_i} = \abs{H_{i-1}} / 2$ such that 
$\Pr[\pi|_{B_{i-1}}\in H_{i}] \ge (1 - \gamma) / 2 \ge \gamma$.

\item
We let $\pi^{(i)}$ be the $k$-round coin flipping protocol over $\ell - \abs{B_{i-1}}$ players that on the message transcript $\alpha$, outputs $1$ if $\pi|_{B_{i-1}}(\alpha) \in H_i$ and outputs $0$ otherwise (here $C_0$ is a universal constant).

\item
Since $\Pr[\pi|_{B_{i-1}}\in H_{i}] \ge \gamma$, we have that $\Pr[\pi^{(i)} = 1] \ge \gamma$.
We apply \cref{thm: protocol lb} to find bad players $B\subset [\ell]\setminus B_{i-1}$ with $\abs{B}\le \frac{C_0(\ell - \abs{B_{i-1}})}{\gamma \log^{(k)}(\ell - \abs{B_{i-1}})}$ such that  it can bias $\pi^{(i)}$ to output $1$ with probability $1 - \gamma$.

\item
We let $B_i = B \cup B_{i-1}$ where $B$ is from the previous step.

\item
We then indeed see that $\Pr[\pi|_{B_{i}}\in H_{i}] \ge 1 - \gamma$, maintaining the invariant property.
\end{enumerate}

We let the final set of bad players be $B_t\cup H_t$. 
By the invariant property at the end of step $t$, we do have that $\pi$ outputs from $H_t$ when corrupted by $B_t$ with probability $1 - \gamma$.
Since now both are part of the set of bad players, we conclude that a bad player is indeed elected the leader with probability $\ge 1 - \gamma$.

It remains to bound the number of bad players. We individually bound $H_t$ and $B_t$ as follows:
Since $\abs{H_i} = \abs{H_{i-1}} / 2$ for all steps $i$, we easily see that $\abs{H_t} = \abs{H_0} / 2^t = \ell / 2^t = \frac{\ell}{\log^{(k)}(\ell)}$.
To bound $\abs{B_t}$, we first observe that the size of the set of bad players $B$ added in each step is upper bounded by $\frac{C_0(\ell - \abs{B_{i-1}})}{\gamma \log^{(k)}(\ell - \abs{B_{i-1}})}\le \frac{C_0(\ell)}{\gamma \log^{(k)}(\ell)}$
Using this, we easily conclude that $\abs{B_t}\le t\cdot \frac{C_0(\ell)}{\gamma \log^{(k)}(\ell)} = \log^{(k+1)}(\ell)\cdot \frac{C_0(\ell)}{\gamma \log^{(k)}(\ell)}$.

Thus, the total number of bad players is at most $\log^{(k+1)}(\ell)\cdot \frac{2C_0(\ell)}{\gamma \log^{(k)}(\ell)}$. Setting our universal constant to be $2C_0$, we obtain the desired claim.
\end{proof}

\begin{remark}
We could also have obtained $1$ round leader election lower bounds directly from \cref{lem: improved coin flip condensing impossibility} by setting $\Delta = \log(\log(\ell))$. Our parameter dependence would have been slightly worse - we would have needed $\frac{\ell \log(\log(\ell))^2 \poly(\log(\log(\log(\ell))))}{\log(\ell)}$ many bad players had we done so. This would still be a doubly exponential improvement over all previous $1$-round leader election lower bounds. Moreover, our random selection lower bounds from \cref{lem: improved coin flip condensing impossibility} only rely on the Edge Isoperimetric inequality instead of the lower bound on influence due to KKL \cite{KKL88influence} as above.
\end{remark}

\dobib

\section{Constructing Improved Constant-Round Coin Flipping Protocols}\label{sec:explicit-protocols}

Our main results in this section are improved explicit constant-round coin flipping protocols,  where each player is allowed to send one bit per round. Formally, we prove:

\begin{theorem}
\label{thm: construct protocol}
For any $k, \ell\in \N, 0<\gamma<1/2$ with $k\ge 2$, there exists a $k$-round coin flipping protocol over $\ell$ players with each player sending one bit per round such that when the number of bad players is at most $\frac{\gamma \ell}{\log (\ell)(\log^{(k)} (\ell))^2}$, the output coin flip is $\eps$-close to uniform where $\eps = O\left(\gamma + \left(\log^{(k-1)}(\ell)\right)^{-0.2}\right)$.
\end{theorem}

To help construct our protocol, we need to introduce the notion of an \emph{assembly}: 

\begin{definition}[Assembly]
A \emph{\assembly[b, n, s] over $\ell$ players}, where each player is labeled either `good' or `bad,' is a collection of $n$ disjoint subsets $S_1, \dots, S_n\subset [\ell]$, each of size $s$, with the following property:
\begin{itemize}


    \item 
    There exists some fixed $B\subset [n]$ with $\abs{B} \le b$ such that if $i\not\in B$, then the set $S_i$ and all players within it are labeled as good. We say that such sets are `good' and that the remaining sets are `bad.' 
\end{itemize}
\end{definition}

We will use the following ways of transforming a given assembly into an assembly with different parameters (the proof will be given later):

\begin{lemma}[Transforming assemblies]\label{lem: assembly transformations}
Let $b, n, s$ be arbitrary and let $A$ be a \assembly[b, n, s]. Then  we can explicitly transform $A$, oblivious to which players are labelled good or bad, in any of the following ways (without changing the underlying labels of players being good or bad):
\begin{enumerate}
    \item (Grouping) 
    For any $t \ge 1$, in $0$ rounds, $A$ can be transformed into a \assembly[b, n/t, st].

    \item (Splitting)
    For any $t \ge 1$, in $0$ rounds, $A$ can be transformed into a \assembly[bt, nt, s/t].

    \item (Feige's Lightest Bin Protocol \cite{feige99lightestbin})
    For any $\beta \le 2^s, 0 < \delta < 1$, in $1$ round, $A$ can be transformed into a \assembly[\frac{b}{\beta} + \frac{\delta(n-b)}{\beta}, \frac{n}{\beta}, s], with success probability at least $1 - \beta \cdot\exp\left(\frac{-\delta^2(n - b)}{ 2\beta}\right)$.
    
\end{enumerate}
\end{lemma}

We will also use the following explicit resilient function. This can be viewed as a one-round operation which, given a suitable assembly, generates an almost-fair coin toss. Formally:
\begin{lemma}[Resilient function from  \cite{AL93resilient, IV25resilient}]
\label{lem: resilient function}
In one round, a \assembly[b, n, s] can be used to output a single bit $\in \zo$ that is $O\left( \frac{b (\log (n))^2}{n} + n^{-0.99}\right)$ close to uniform.
\end{lemma}

Let us show that using \Cref{lem: assembly transformations,lem: resilient function} we can obtain our desired $k$-round protocol. 

\begin{proof}[Proof of \cref{thm: construct protocol} using \Cref{lem: assembly transformations,lem: resilient function}]
We first informally describe our protocol and then we will formally describe it and analyze it.

\paragraph{Informal description}
First, we arbitrarily group players together and treat each group as a single entity (each group will correspond to a set of the assembly) which has access to multiple random bits. We will have many such entities, and most of them will contain solely  good players; call those `good' entities and the remaining entities `bad' entities. Then, we make these entities participate in one round of the lightest bin protocol with many bins. This vastly reduces the number of entities while roughly maintaining the absolute fraction of entities that are bad. We then again split each of the entities into many entities with fewer players in each one of them. This still maintains the absolute fraction of entities that are bad. We then repeat the lightest bin protocol and continue doing this until it is time for the last round. In the last round, we take one player from each entity and apply a resilient function to obtain our output coin flip. 

\paragraph{Formal protocol}
We formally proceed as follows. Since we know that at most $\frac{\gamma \ell}{\log(\ell)\log^{(k)}(\ell)}$ players are bad, our input set of players form a \assembly[\frac{\gamma \ell}{\log(\ell)\log^{(k)} (\ell)}, \ell, 1].
Our protocol will repeatedly perform transformations on this assembly as described below.  (We remark that the warmup case discussed in \Cref{sec:warmup-protocols} (the case of $k=2$) corresponds to performing only steps~1 and~3 below.)
\begin{enumerate}
    \item 
    In the first round we proceed as follows:
    \begin{enumerate}
        \item 
        We use transformation 1 from \cref{lem: assembly transformations} (grouping), setting $t = 3 \log(\ell)$, to transform our \assembly[\frac{\gamma \ell}{\log (\ell)\log^{(k)}(\ell)}, \ell, 1] into a \assembly[\frac{\gamma \ell}{\log (\ell)\log^{(k)} (\ell)}, \frac{\ell}{3\log (\ell)}, 3 \log(\ell)].

        \item 
        We use transformation 3 from \cref{lem: assembly transformations} (Feige's lightest bin protocol) with $\beta = \frac{\ell}{(\log (\ell))^3}, \delta = (\log (\ell))^{-0.25}$ to transform the $\assembly[\frac{\gamma \ell}{\log (\ell)\log^{(k)} (\ell)}, \frac{\ell}{3\log(\ell)}, 3 \log(\ell)]$ into a \assembly[\frac{\gamma(\log (\ell))^2}{(\log^{(k)} (\ell))^2} + (\log(\ell))^{1.76}, \frac{(\log(\ell))^2}{3}, 3\log(\ell)], succeeding with probability $\ge 1 - \ell\exp\left(-(\log(\ell))^{1.49}\right)$. 
    \end{enumerate}

    \item 
    For rounds $2$ to $k-1$, we proceed as follows:  In round $i$, we are given a \assembly[\frac{\gamma(\log^{(i-1)}(\ell))^2}{(\log^{(k)}(\ell))^2} + (\log^{(i-1)}(\ell))^{1.76}, \frac{(\log^{(i-1)}(\ell))^2}{3}, 3\log^{(i-1)}(\ell)]. 
    \begin{enumerate}
        \item 
        We use transformation 2 from \cref{lem: assembly transformations} (splitting), setting $t = \frac{\log^{(i-1)}(\ell)}{\log^{(i)}(\ell)}$, to transform our assembly above into a \assembly[\frac{\gamma (\log^{(i-1)} (\ell))^3}{\log^{(i)}(\ell)(\log^{(k)} (\ell))^2} + (\log^{(i-1)}(\ell))^{2.76},\frac{(\log^{(i-1)}(\ell))^3}{3\log^{(i)}(\ell)} , 3\log^{(i)}(\ell)].
        
        \item
        We use transformation 3 from \cref{lem: assembly transformations} (Feige's lightest bin protocol) with $\beta = \frac{(\log^{(i-1)}(\ell))^3}{(\log^{(i)} (\ell))^3}, \delta = (\log^{(i)} (\ell))^{-0.25}$ to transform the \assembly[\frac{\gamma (\log^{(i-1)} (\ell))^3}{(\log^{(i)}(\ell))(\log^{(k)}(\ell))^2} + (\log^{(i-1)}(\ell))^{2.76},\frac{(\log^{(i-1)}(\ell))^3}{3\log^{(i)}(\ell)} , 3\log^{(i)}(\ell)] into a \assembly[\frac{\gamma (\log^{(i)}(\ell))^2}{(\log^{(k)}(\ell))^2} + (\log^{(i)}(\ell))^{1.76},\frac{(\log^{(i)}(\ell))^2}{3},3\log^{(i)}(\ell)], succeeding with probability $\ge 1 - (\log^{(i-1)}(\ell))^3\exp\left(-(\log^{(i)}(\ell))^{1.49}\right)$.
    \end{enumerate}

    \item 
    In round $k$, we are given a \assembly[\frac{\gamma(\log^{(k-1)}(\ell))^2}{(\log^{(k)}(\ell))^2} + (\log^{(k-1)}(\ell))^{1.76}, \frac{(\log^{(k-1)}(\ell))^2}{3}, 3\log^{(k-1)}(\ell)].
    We use the resilient function from \cref{lem: resilient function} to obtain a single output bit $\in \zo$ that is $O\left(\gamma + (\log^{(k-1)}(\ell))^{-0.23}\right)$ close to uniform.
\end{enumerate}

By a union bound over all $k$ rounds (the failure probability is dominated by the $i=k-1$ iteration of step~2(b) above), we get that all our transformations succeed with probability
$1 - \exp(-(\log^{(k-1)}(\ell))^{1.48})$. 
In such an event, the error from our final round - round $k$ from applying the resilient function is $O\left(\gamma + (\log^{(k-1)} (\ell))^{-0.23}\right)$.
Since $(\log^{(k-1)}(\ell))^{-0.23} > \exp(-(\log^{(k-1)}(\ell))^{1.48})$, we get that the final error of our coin flip is at most $O\left(\gamma + (\log^{(k-1)}( \ell))^{-0.2}\right)$ as desired.
\end{proof}

It remains to prove that we can indeed obtain each of the transformations described earlier.

\begin{proof}[Proof of \cref{lem: assembly transformations}]

We present proofs of each of the claimed transformations.

\begin{enumerate}
    \item 
    To do this, we arbitrarily partition $[n]$ into parts of size $t$ and take the union of all sets that are in the same part. The number of sets indeed decreases to $n/t$ and the size of each set increases to $st$. Also, since $b$ sets were bad, there are at most $b$ parts which have at least one bad set. For the rest of the parts, all the players in each of the sets comprising the part are good, and hence after taking the union, the resulting set will also only consist of good players.

    \item
    To do this, we split each set $S_i$ arbitrarily into $t$ parts. We indeed obtain $tn$ sets of size $s/t$ each. Moreover, whenever we split a set consisting of only good players, all of the resulting sets only consist of good players. Hence, the number of such sets in the resultant assembly is $(n - b)t = nt - bt$ as desired.

    \item 
    To do this, we use Feige's lightest bin protocol from \cite{feige99lightestbin}. In particular, we set up $\beta$ bins $B_1, \dots, B_{\beta}$ which receive `votes' as follows. In one round, the players in each set $S_1, \dots, S_n$ each flip a random coin. For each set $S_i$, we interpret their $s$ coin flips together as generating a random number between $1$ and $\beta$ (since $\beta \le 2^s$) and that number is the `vote' for which bin the set should go to. After this voting process, we pick the lightest bin (the one that has received the fewest votes), say $B_{i^*}$, and let the sets in that bin (along with arbitrary other sets, chosen in some canonical way so that number of sets in the output assembly is $n / \beta$) form the output assembly .

    We analyze the process described above. 
    We treat all players in any bad set as bad, i.e., they are allowed to cast their votes after seeing the votes of all the good sets.
    We first note that the number of sets in the lightest bin $B_{i^*}$ is at most $n / \beta$ (since there are $n$ sets and $\beta$ bins). Hence, the above process always outputs an assembly consisting of $n / \beta$ many sets of size $s$ each. We now lower bound the number of good players in each of the bins.
    Fix any bin $i\in [\beta]$ and let $\X_i$ be the random variable representing the number of good sets that voted $i$.
    We think of the vote of each of the good sets as independently choosing whether to vote for bin $i$ or not, where the vote is cast for bin $i$ with probability $1 / \beta$. Hence $\E[\X_i] = (n - b) / \beta$, and by the Chernoff bound \cref{claim: lower tail chernoff}, we have that
    \[
        \Pr[\X_i \le (1-\delta)(n - b) / \beta] \le \exp\left(\frac{-\delta^2 (n-b)}{2\beta}\right).
    \]
    We take union bound over all $\beta$ bins to infer that with probability at least $1 - \beta\exp\left(\frac{-\delta^2 (n-b)}{2\beta}\right)$, every bin will have at least $\frac{(1-\delta)(n - b)}{\beta}$ good sets.
    As this is true for all bins, this is also true for the lightest bin $i^*$.
    Hence the number of bad sets in bin $i^*$ is at most 
    \[
        \frac{n}{\beta} - \frac{(1-\delta)(n-b)}{\beta} = \frac{b}{\beta} + \frac{\delta(n-b)}{\beta}
    \]
    as desired. \qedhere
\end{enumerate} 
\end{proof}

\dobib


\section{Multi-output One-Round Coin Flipping Protocols}
\label{sec: multioutput extract}

In this section, we will construct one-round coin flipping protocols that output many bits. In particular, we will prove

\begin{theorem}[Multi-output coin flipping protocol]
\label{thm: multioutput coin flipping protocol - general}
For any $0 < \eps < 1/4$, there exists a constant $C_0 = C_0(\eps)$ such that the following holds.
Let $b, \ell, m\in \N$ be such that $b \le \min\left(\frac{\ell}{C_0\cdot \log(\ell)^2}, \frac{\ell}{C_0 m}\right)$.
Then, there exists an explicit one-round coin flipping protocol $\pi: \zo^{\ell}\to \zo^m$ such that for any coalition of $b$ bad players corrupting $\pi$, the output distribution on $m$ bits has statistical distance at most $\eps$ from $\U_m$. \end{theorem}

As a direct corollary, we obtain the following optimal tradeoff when the number of output bits is more than $(\log(\ell))^2$.

\begin{corollary}
\label{cor: multioutput coin flipping protocol 1/polylog fraction bad}
For any $0 < \eps < 1/4$, there exists a constant $C_0 = C_0(\eps)$ such that the following holds.
For all $b, \ell, m \in \N$ such that $m \ge \log(\ell)^2$ and $b \le\frac{\ell}{C_0 m}$, there exists an explicit one-round coin flipping protocol $\pi: \zo^{\ell}\to \zo^m$ such that for any coalition of $b$ bad players corrupting $\pi$, the output distribution on $m$ bits has statistical distance at most $\eps$ from $\U_m$.
\end{corollary}

This also yields a one-round leader election protocol:
\begin{corollary}
\label{cor: leader election 1 round}
For any $0 < \eps < 1/4$, there exists a constant $C_0 = C_0(\eps)$ such that the following holds.
For all $\ell\in \N$, there exists a one-round leader election protocol $\pi: \zo^{\ell} \to [\ell]$ such that for any coalition of $b\le \frac{\ell}{C_0(\log(\ell))^2}$ bad players corrupting $\pi$, a good leader is elected with probability at least $1 - \eps$.
\end{corollary}

\begin{proof}
We apply \cref{thm: multioutput coin flipping protocol - general} with error parameter $\eps/2$ and $b = \frac{\ell}{C(\log(\ell))^2}$ where $C$ is a large enough universal constant to obtain $\pi: \zo^{\ell}\to \zo^m$ with $m = \log(\ell)$ such that in the presence of $b$ bad players, the output distribution has statistical distance at most $\eps/2$ from $\U_m$.
We interpret the $\log(\ell)$ bits obtained from $\pi$ as an integer in $[\ell]$ and elect that player as the leader.
The output distribution over the $[\ell]$ bits will still be $\eps/2$-close to uniform, showing that the probability that a bad player will be elected is at most $\frac{b}{\ell} + \eps/2 \le \eps$ as desired.
\end{proof}

We will devote the rest of this section to proving \cref{thm: multioutput coin flipping protocol - general}. 
To do that, we first need to introduce the slightly more general notion of \emph{resilient functions}. 
\begin{definition}[Resilient functions]
Let $b, \ell, m \in \N$ be such that $b\le \ell$.
We say $\Res: \zo^{\ell} \to \zo^m$ is \emph{$(b, \delta)$-resilient} if for set of $\ell-b$ good players $G\subset [\ell],$ we have
\[
\frac{\abs{F}}{2^{|G|}} \ge 1 - \delta,
\]
where $F:= \{y\in \zo^G :$ the value of $ \Res(y, \Adv(y))$ is fixed across all possible (randomized) adversarial functions $\Adv: \zo^G \to \zo^{[\ell]\setminus G}\}$.
\end{definition}

We will also use the following notion of the \emph{bias} of a multi-output function.  (Below we write $|\A-\B|$ to denote the statistical distance between distributions $\A$ and $\B.$)
\rnote{Added this parentheetical because I think we didn't define this in the paper? \mohit{Thanks! We could also move it to preliminiaries if necessary} {\bf Rocco:} Good point, probably this belongs in preliminaries?}

\begin{definition}[Bias of a multi-output function]
Let $\ell, m \in \N$ be arbitrary.
For $0 \le \eps \le 1$, we say a function $f: \zo^{\ell} \to \zo^m$ is \emph{$\eps$-biased} if $\abs{f(\U_{\ell}) - \U_m} \le \eps$.
\end{definition}
We will first construct a resilient function that outputs $O(\log(\ell))$ bits. We will do this by modifying the well known single-output resilient function of Ajtai and Linial \cite{AL93resilient}. Formally, we will show:
\begin{lemma}[Modified multi-output Ajtai-Linial]\label{lem: multioutput modified ajtai-linial}
For all $0 < \eps < 1/4$ and $C_0 \ge 1$, there exists a constant $C = C(\eps, C_0)$ such that the following holds.
For all $b, \ell\in \N$ with $b\le \ell$, there exists an explicit function $\Res: \zo^{\ell} \to \zo^m$ where $m = C_0\log(\ell)$ such that $\Res$ is $(b, \delta_b)$-resilient and $\eps$-biased where $\delta_b = b\cdot \frac{C(\log(\ell))^2}{\ell}$.
\end{lemma}

We will construct such a resilient function $\Res$ in \cref{subsec: multioutput ajtai linial}. 
Using this, we easily obtain our main theorem:
\begin{proof}[Proof of \cref{thm: multioutput coin flipping protocol - general}]
We let $C_0$ be large enough universal constant that we will set later.
Let $\Res: \zo^{\ell/2} \to \zo^{\log(\ell)}$ be the resilient function from  \cref{lem: multioutput modified ajtai-linial} that is $(b, \eps/4)$-resilient and $(\eps/4)$-biased.
We will let our universal constant $C_0$ be much much larger the universal constant $C$ required by $\Res$ to satisfy its resilience property.

Using this, we define our multi-output resilient function as follows:
On input $x\in \zo^{\ell}$, we partition $x$ into two halves $(x_1, x_2)\in (\zo^{\ell/2})^2$. We further divide $x_2$ into $\ell / 2m$ parts $y_1, \dots, y_{\ell/2m}\in \zo^m$.
We use $x_1$ to sample a random positive integer $j\in [\ell / 2m]$ and output $y_j$.

Let $G_y = \{i: y_i \textrm{ contains no bad players }$. 
As the number of bad players is at most $\frac{\ell}{C_0 m}$, size of $G$ is at least $\frac{\ell}{m}\left(\frac{1}{2} - \frac{1}{C_0}\right)$. 
So, if we truly randomly sampled $j\in [\ell/2m]$ and output $y_j$, then our output distribution will have distance at most $\frac{1}{2C_0}$ from the uniform distribution. We will let $C_0$ be large enough so that this is at most $\eps / 4$.
However, since we pick $j$ from $\Res(x_1)$, we infer that with probability at least $1 - \eps/4$, over the coin flips of the good players in $x_1$, the integer $j$ will be picked from a distribution that has statistical distance at most $\eps/2$ from the uniform distribution. Furthermore, this distribution will be independent from the outcomes of the coin flips of the players in $x_2$. Hence, we conclude that the overall output distribution of our algorithm will have statistical distance at most $\eps$ from $\U_m$ as desired.
\end{proof}

\subsection{Constructing multi-output Ajtai-Linial resilient function}\label{subsec: multioutput ajtai linial}

In this subsection we prove \cref{lem: multioutput modified ajtai-linial}, our multi-output resilient function outputting $O(\log(\ell))$ bits. 
We will use the following base construction that outputs $0.49\log(\ell)$ bits instead of $C_0 \log(\ell)$. 
\rnote{Changed this from $C \log (\ell)$ --- let's call this $C_0 \log (\ell)$ to align with the notation we are using in \cref{lem: multioutput modified ajtai-linial} \mohit{Thanks, looks great!}}
for any constant $C_0$ of our choice.
\begin{lemma}[Base multi-output Ajtai-Linial]\label{lem: base multioutput ajtai linial}
For all $0 < \eps < 1/4$, there exists a constant 
$C = C(\eps)$
\rnote{This was $C(\gamma)$ but then $\gamma$ never appears again in the statement - I assume it's meant to be $C(\eps)$? \mohit{Thanks for catching this - yes indeed should have been $C(\eps)$; I went back and forth several times here}} 
such that the following holds.
For all $\ell\in \N$, there exists an explicit function $\Res: \zo^{\ell} \to \zo^{m}$, where $m = 0.49\log(\ell)$, with the following properties:
for all $b\le \ell$, $\Res$ is $(b, \delta_b)$-resilient where $\delta_b = b\cdot \frac{C(\log (\ell))^2}{\ell}$; 
\rnote{This $0^m$ was ``$\bot$'' \mohit{My bad, there should be no such conditioning here. I made the required changes}} 
 $\Res$ is $\eps$-biased over $\zo^{m}$.
\end{lemma}

We will prove this in \cref{subsubsec: constructing base multioutput ajtai linial}. 
We will additionally need the following result that lets us increase the number of output bits given any multi-output resilient function (at the cost of increasing its bias).
\begin{lemma}[Getting more bits from any resilient function]\label{lem: more bits from resilient function}
Assume that for some $\ell, m\in \N$, $0 < \eps < 1$, there exists an explicit function $f: \zo^\ell \to \zo^m$ and $\alpha = \alpha(\ell, m)$ such that for all $b\le \ell$, $f$ is $(b, b\cdot \alpha)$-resilient and $f$ is $\eps$-biased.
Then, for any $t\in \N$, we can construct a function $\Res: \zo^{\ell\cdot t} \to \zo^{m\cdot t}$ that is $(b, b\cdot \alpha)$-resilient and is $\eps\cdot t$-biased.
\end{lemma}

We will prove this at the end of this subsection. Using these, we easily construct our desired resilient function:
\begin{proof}[Proof of \cref{lem: multioutput modified ajtai-linial}]
This directly follows by applying \cref{lem: more bits from resilient function} with $t = C_0 / 0.49$ to the resilient function from \cref{lem: base multioutput ajtai linial} with error $\eps / t$.
\end{proof}

Let us finally prove \cref{lem: more bits from resilient function}.

\begin{proof}[Proof of \cref{lem: more bits from resilient function}]
For $x\in \zo^{\ell\cdot t}$, we compute $\Res(x)$ as follows:
we split $x$ into $t$ equal parts $(x_1, \dots, x_t) \in (\zo^{\ell})^{t}$, and we output $\Res(x) := (f(x_1),\dots,f(x_t))$ where $f: \zo^{\ell}\to \zo^m$ is the resilient function provided by assumption of the lemma.

We now show the guarantees on resilience and bias.
Fix any set of $b$ bad players.
Let $b_1, \dots, b_t$ be the number of bad players in each of the $t$ parts, so  $\sum_{i=1}^t b_i = b$.
For each $i\in [t]$, by the guarantees on $f$, for a $1 - b_i\cdot \alpha$ fraction of  the outcomes of the good players in part $i$, the function $f$ becomes fixed. 
By a union bound, we infer that for $1 - \sum_{i=1}^t (b_i\cdot \alpha) = 1 - b\cdot \alpha
$ fraction of the outcomes of the good players, our final resilient function $\Res$ becomes fixed, and thus we have the desired resilience property.

Lastly, since $\abs{f(\U_{\ell}) - \U_m} \le \eps$, by applying the triangle inequality $t$ times, we have that $\abs{\Res(\U_{\ell\cdot t}) - \U_{m\cdot t}}\le \eps\cdot t$ as claimed.
\end{proof}

\subsubsection{Constructing base multi-output Ajtai-Linial resilient function}\label{subsubsec: constructing base multioutput ajtai linial}

We here prove \cref{lem: base multioutput ajtai linial}, a resilient function outputting $0.49\log(\ell)$ bits.
To obtain this, we will utilize an intermediate resilient function that is allowed to fail (i.e., output $\bot$) with probability bounded away from $1$:

\begin{lemma}[Modified Ajtai-Linial with failure mode]\label{lem: base multioutput ajtai linial with bot}
\rnote{Maybe here in \Cref{lem: base multioutput ajtai linial with bot} we can use a $'$ on the objects --- call the function $\Res'$, have the input length be $\ell'$, the output length $m'$, etc.  Otherwise I think there is potential for confusion because, for example, we have a function called $\Res$ in this lemma and also a function called $\Res$ in the \cref{lem: base multioutput ajtai linial} lemma that we are proving, but they are not the same function.

It looks like the proof of \cref{lem: base multioutput ajtai linial} given below goes some of the way towards doing this already, i.e.~it refers to the input length of the 
\Cref{lem: base multioutput ajtai linial with bot} function as $\ell$' --- maybe we can go further :) I'll take a stab at this below (but will probably make mistakes)
\mohit{That's fair yeah. I guess I was worried that when we actually write the proof of lemma 1.18, we would have to write ' versions everywhere. But yeah I agree this helps with the clarity of the proof here}
}
There exists a universal constant $C_0$ such that the following holds.
For all $\ell'\in\N$, there exists an explicit function $\Res': \zo^{\ell'} \to \zo^{m'}\cup \bot$ where $m' = 0.49\log(\ell')$ with the following properties.
$\Pr[\Res'\ne \bot] \ge \frac{1}{C_0}$; for all $b'\le \ell'$, $\Res'$ is $(b', \delta_{b'})$-resilient where $\delta_{b'} = b'\cdot \frac{C_0(\log (\ell'))^2}{\ell'}$; and conditioned on $\Res'$ not outputting $\bot$, $\Res'$ is $\eps$-biased over $\zo^{m'}$ where $\eps = C_0\cdot 2^{-{m'}}$.
\end{lemma}

We will prove this in \cref{subsubsec: base multioutput ajtai linial with bot}. Let us see how using this, \cref{lem: base multioutput ajtai linial} follows.

\begin{proof}[Proof of \cref{lem: base multioutput ajtai linial}]
Let $C_0$ be the universal constant from \cref{lem: base multioutput ajtai linial with bot}.
Let $t = C_0\log(4 / \eps)$
\rnote{Is this $\gamma$ meant to be $\eps$? Likewise for all the following $\gamma$'s in the proof...I'm asking because I don't see what role $\gamma$ plays in \Cref{lem: base multioutput ajtai linial} \mohit{I am so sorry, I made a last minute decision to change $\gamma$ in this entire proof to $\eps$ and did a poor job of percolating the change. Let me carefully make any further required changes here}} and let $\ell' = \ell / t$.
Let \rnote{Maybe we can just write $\Res'$ for $f$ throughout this if we use $\Res'$ for the function's name in \Cref{lem: base multioutput ajtai linial with bot} as suggested/modified above? I didn't change this yet in case all of my notation suggestions are bad :P \mohit{lol no, let's push through with your suggestion. I am making the changes here}} $\Res': \zo^{\ell'} \to \zo^{0.49\log(\ell')} \cup \bot$ be the function from \cref{lem: base multioutput ajtai linial with bot}. 
On input $x\in \zo^{\ell}$, we compute our function $\Res(x)$ as follows.
We first partition the input into $t$ parts so that $x = (x_1, \dots, x_t)\in \left(\zo^{\ell'}\right)^t$.
Next, for $1\le i\le t$, we compute $y_i = \Res'(x_i)$.
Finally, if $y_i = \bot$ for all $i\in [t]$, then we output $0^m$; otherwise, we pick the smallest $j\in [t]$ such that $y_j\ne \bot$ and output $y_j$.

We first show the desired resilience property.
Let $g: \zo^{\ell} \to \{0, 1, \bot\}^t$ \rnote{Is the range of $g$ $\zo^t$ or $\zo^t \cup \{\bot\}$ --- seems like the latter? \mohit{Thanks for catching this, my bad yes we definitely need $\bot$ in fact even inside the alphabet. I made the change}} be defined as follows: For all $i\in [t]$, $(g(x))_i = \Res'(x_i)$ where $x, x_i$, and $f$ are defined as above. Fix $b\le \ell$, and let  $G\subset [\ell]$ be such that $\abs{G} = \ell - b$.
Let $b_1, \dots, b_t \le \ell'$ 
\rnote{Should this \red{$\ell$} be $\ell'$, since each $b_i$ is the number of bad players in string $x_i$ and $x_i$ is an $\ell'$-bit string? \mohit{yes thanks, my bad. Fixed it}} be the number of bad players from $[\ell]\setminus G$ in strings $x_1, \dots, x_t$ respectively.
By the resilience property of $\Res'$, for each $i \in [t]$ we know that with probability $\ge 1 - b_i \cdot \frac{C_0 (\log(\ell'))^2}{\ell'}$ over the  output of the good players among the bits of $x_i$, the function $\Res'(x_i)$ will be fixed.
By a union bound over all $i \in [t]$, the function $g$ will be fixed for $\ge 1 - \sum_{i=1}^t b_i \cdot \frac{C_0 (\log(\ell'))^2}{\ell'} = 1 - b \cdot \frac{C_0 (\log(\ell'))^2}{\ell'}$ fraction of inputs.
We see that whenever $g$ will be fixed over a partial assignment of the good players, so will our function $\Res$ since $\Res(x)$ is completely determined from $g(x)$.
Hence, $\Res$ is $(b, \delta_b)$-resilient for $\delta_b = \frac{C_0 (\log(\ell'))^2}{\ell'} = t\cdot \frac{C_0 (\log(\ell/t))^2}{\ell} \le t\cdot \frac{C_0 (\log(\ell))^2}{\ell}$.
Since $t$ is a function of $\gamma$ and $C_0$ is a universal constant, we infer the desired resilience claim.

We now show the desired small bias property.
We think of our procedure as performing rejection sampling, where $t$ times we independently select a uniform random input for $\Res'$ and the first time we see an output not equal to $\bot$, we output it.
We see that the probability that  $y_1 = \dots =  y_t = \bot$ is at most $\left(1 - \frac{1}{C_0}\right)^t \le \exp(-t / C_0) \le \frac{\eps}{4}$, where the last inequality follows by our choice of $t$.
We are also guaranteed by $\Res'$ that conditioned on $y_i \ne \bot$, the output will be $C\cdot 2^{-m}$ close to the uniform distribution.
Since this still holds at each iteration of the rejection sampling procedure, we infer that conditioned on the procedure not resulting in every sample being equal to $\bot$, our final sample will be $\frac{C_0}{2^m}$ close to the uniform distribution.
Finally, if $y_1 =  \dots = y_t = \bot$ then the output of $\Res$ is $0^m$; this brings its bias to $\eps / 4 + C_0\cdot 2^{-m} \le \eps$, as claimed (the last inequality holds since $\eps$ is fixed while $m$ is growing.
\rnote{I'm definitely sympathetic to this sentiment, but I guess strictly speaking if we are doing this kind of thing then maybe \cref{lem: base multioutput ajtai linial} should be ``for all sufficiently large $\ell$ rather than ``for all $\ell \in \N$''? \mohit{Yeah that's fair. I was banking on the fact that the universal constant $C$ in our lemma statement is so large that the parameters become trivial for small $\ell$. Such a thing also holds for almost every lemma and claim in this section - that it holds for large enough $\ell$ (at least a certain constant); everywhere we implicitly have that restriction because we have a universal constant present that makes the statement trivial for small $\ell$}. But yeah happy to change all statements everywhere to say this explicitly if you think it will make things clearer}).
\end{proof}

\subsubsection{Constructing base multi-output Ajtai Linial with failure mode}\label{subsubsec: base multioutput ajtai linial with bot}

In this subsection we prove \cref{lem: base multioutput ajtai linial with bot}. To do this, we will closely follow the explicit version of the Ajtai-Linial function constructed in \cite{IV25resilient} with some slight modifications. To define the $\Res'$ function, let's recall the definition of the Tribes function.

\begin{definition}[Tribes function]
\label{def: tribes}
We say a function $f: \zo^n \to \zo$ is a \emph{Tribes function with width $w$} if $f$ can be written as a read-once DNF with $n/w$ terms of width $w$ each.
\end{definition}

Notice that even for fixed $n$ and $w$, there are many different tribes functions since we do not specify the order of the input variables in the read-once DNF. With this definition in hand we proceed to the proof.

\begin{proof}[Proof of \cref{lem: base multioutput ajtai linial with bot}] 
In this proof we will use $n$ for the number of variables instead of $\ell'$, $\Res$ for the name of the function instead of $\Res'$, and $m$ for the number of output variables instead of $m'$, in order to be consistent with the notation of \cite{IV25resilient}.
We let $M = 2^m$, and we view $\Res$ as outputting elements from $[M] \cup \{\bot\}$ rather than $\zo^m \cup \{\bot\}$. Notice that $M = n^{0.49}$.

We first review the construction of the $1$-output-bit function $f$ of \cite{IV25resilient}.
To compute $f$, they first construct $u = \poly(n) \ge n^2$ many Tribes functions $T_1, \dots, T_u: \zo^n \to \zo$ of width $w$ each.
(We remark that this tuple of Tribes functions is constructed in a careful pseudorandom way; we will return to this point later.)
Using this, they define $f(x) = \bigwedge_{i=1}^u T_i(x)$.

For our construction, we will consider the same set of Tribes functions as above. On input $x$, we compute $\Res(x) \in [M]$ as follows: For $j\in [M]$, we first compute 
\[
y_j = \bigwedge_{i = (j-1)\cdot(u / M) + 1}^{j\cdot(u / M)} T_i(x).
\] 
If there exists $z\in [M]$ such that $y_z = 0$ and for all $j\ne z$, $y_j = 1$, then we set $\Res(x)=z$; otherwise,  $\Res(x)=\bot$.

We first show that $\Res$ has the desired resilience property.
In \cite{IV25resilient} (Lemma 17 and its proof), they show that the map $h: \zo^n \to \zo^u$ defined by $h(x) = (T_1(x), \dots, T_u(x))$ is $(b, \delta_b)$-resilient where $b\le n$ is arbitrary and $\delta_b = b\cdot \frac{C\cdot (\log (n))^2}{n}$.
So, for any $b$ and $G\subset [n]$ with $\abs{G} = n - b$, for $1 - \delta_b$ fraction of the assignments of the bits in $G$, the function $h$ becomes fixed by that partial assignment.
We observe that our function $\Res$ can be uniquely computed given $h$, i.e., we can compute $\Res(x)$ if we are only given $h(x)$. So if $h(x)$ is fixed by a partial assignment of the good players, so is $\Res$. Hence, our function $\Res$ is indeed $(b,\delta_b)$-resilient as claimed (this is also the exact argument that \cite{IV25resilient} made for their function $f$ to prove its resilience).

We next show that our function $\Res$ is $\eps$-balanced and also that it outputs $\bot$ with not too high probability. 
Let $E$ be the event that $y_j = 1$ for all $j\in [M]$.
For $i\in [M]$, let $E_i$ be the event that $y_j = 1$ for all $j\in [M] \setminus \{i\}$.
Let $r = \Pr[E]$ and for $i\in [u]$, let $r_i = \Pr[E_i]$.
These events are relevant to us since $\Pr[\Res = i] + \Pr[E] = \Pr[E_i]$ for all $i\in [M]$, and so $\Pr[\Res = i] = r_i - r$.

To help bound these probabilities, we introduce some more quantities.
Let \[
p = \left(1 - \left(\frac{1}{2}\right)^w\right)^{n / w}, \quad 
q = (1 - p)^u,\quad \text{and} 
\quad q_i = (1 - p)^{u - u / M} \text{~for~}i \in [u].
\]
The parameters $q$ and $u$ in \cite{IV25resilient} are chosen carefully so that $\abs{q - \frac{1}{2}} \le \frac{C\cdot (\log(n))^2}{n}$ for some universal constant $C$.

We see that if all $u$ tribes had all independent inputs (meaning the overall function would be a read-once function depending on $u\cdot n$ many input bits), then $r$ would equal $q$ and for each $i\in [u]$, $r_i$ would equal $q_i$. We will show that in that scenario, $\Res$ would have very small bias. Formally, we  claim the following and prove it later:

\begin{claim}[Small bias under independent conditions]\label{claim: AL - small bias under independent conditions}
There exists a universal constant $C$ such that
\[
\abs{q_i - q}\le \frac{q\cdot \ln(1 / q)}{M} + C\cdot \left(\frac{1}{M^2}\right).
\]
\end{claim}

Since we only have $n$ inputs, we certainly don't have the luxury of such full independence. This is where the pseudorandom way in which \cite{IV25resilient} chose the $u$ tribes will help us. We will show that $q_i$ and $q$ are good approximations for $r_i$ and $r$ respectively:

\begin{claim}[Good approximation to the independent setup]\label{claim: AL - well approximation to independent conditions}
There exists a universal constant $C \ge 1$ such that for all $i\in [u]$, we have $\abs{q_i - r_i} \le \frac{(\log(n))^C}{n}$ and $\abs{q - r} \le \frac{(\log(n))^C}{n}$.
\end{claim}

Before we prove both these claims, lets see how we obtain the desired properties regarding bias and $\Pr[\Res \ne \bot]$ assuming them.
For $i\in [M]$, we upper and lower bound the probability that $\Res$ outputs $i$ as follows:
\begin{align*}
\Pr[\Res = i]
& = r_i - r\\
& = q_i\left(1 \pm \frac{(\log(n))^C}{n}\right) - q\left(1 \pm \frac{(\log(n))^C}{n}\right) & \tag{using \cref{claim: AL - well approximation to independent conditions}}\\
& = q_i - q \pm (q + q_i)\cdot\frac{(\log(n))^C}{n}\\
& = q_i - q \pm 2 \cdot\frac{(\log(n))^C}{n} \tag{using $0 \leq q,q_i \leq 1$}\\
& = \frac{q\ln(1/q)}{M} \pm C_1\cdot\frac{1}{M^2} \pm C_1\cdot\frac{(\log(n))^C}{n} & \tag{using \cref{claim: AL - small bias under independent conditions}}\\
& = \frac{q\ln(1/q)}{M} \pm C_2\cdot\frac{1}{M^2} & \tag{using that $M = n^{0.49}$},
\end{align*}
where $C_1$ and $C_2$ are universal constants.

So, we see that 
\[
\Pr[\Res \ne \bot] = \sum_{i\in [M]} \Pr[\Res = i] \ge q\ln(1/q) - C_2\cdot \frac{1}{M}.
\]
Since $q = \frac{1}{2} \pm o(1)$, we have that $\Pr[\Res \ne \bot] \ge 0.5\ln(2) - o(1)\ge 0.34$ as desired.
We also see that the statistical distance between $\Res$ conditioned on not outputting $\bot$ and $\U_m$ is
\[
\abs{(\Res(\U_n)|\Res(\U_n) \neq \bot) - \U_m} 
\le \frac{q\ln(1 / q) + \frac{C_3}{M}}{q\ln(1 / q) - \frac{C_3}{M})} - 1
= \frac{\frac{2C_3}{M}}{q\ln(1/q) - \frac{C_3}{M}}
\le \frac{C_4}{M},
\]
where $C_3$ and $C_4$ are some universal constants.
Hence, \cref{lem: base multioutput ajtai linial with bot} indeed follows by taking its final constant $C$ to be suitably chosen vis-a-vis 
all constants that appeared in our approximations and claims.

It remains to prove \cref{claim: AL - small bias under independent conditions} and
\cref{claim: AL - well approximation to independent conditions}.
We first show that if all inputs to all tribes were independent, then $\Res$ would have small bias:
\begin{proof}[Proof of \cref{claim: AL - small bias under independent conditions}]
We first obtain a more manageable expression for $q_i$:
\begin{align*}
q_i
& = (1 - p)^{u - u / M}\\
& = \left((1-p)^u\right)^{1 - 1 / M}\\
& = q^{1 - 1 / M} \\
& = q\cdot (q^{- 1 / M})\\
& = q\cdot e^{\ln(1/q) / M}.
\end{align*}

We use the fact that $e^x \ge 1 + x$ for all $x$ to obtain the lower bound:
\begin{align*}
q_i
& = q\cdot e^{\ln(1/q) / M}\\
& \ge q\cdot \left(1 + \ln(1 / q) / M\right)\\
& = q + \frac{q\cdot \ln(1 / q)}{M}.
\end{align*}

We similarly use the fact that $e^x \le 1 + x + x^2$ for all $x\in [0, 1]$ to obtain the desired upper bound:
\begin{align*}
q_i
& = q\cdot e^{\ln(1/q) / M}\\
& \le q\cdot \left(1 + \ln(q)/M + \left(\ln(q)/M\right)^2\right)\\
& = q + \frac{q\cdot \ln(1 / q)}{M} + \frac{q (\ln(q))^2}{M^2}\\
& \le q + \frac{q\cdot \ln(1 / q)}{M} + O\left(\frac{1}{M^2}\right),
\end{align*}
where in the last inequality we used the fact that $q \leq 1$.
\end{proof}

We now finally prove the last remaining claim, which states that $r_i$ and $r$ are well approximated by $q_i$ and $q$:
\begin{proof}[Proof of \cref{claim: AL - well approximation to independent conditions}]
This follows from inspection of the proof of Lemma 18 from \cite{IV25resilient}. 
We remark that the statement of Lemma 18 by itself only gives us that $\abs{q - r} \le \frac{1}{n^{1 - o(1)}}$.
However, tracing through the proof reveals that $\abs{q-r} \leq {\frac {O((\log(n))^6)}{n}}$, yielding our claim about $q$ well-approximating $r$.
To obtain that $q_i$ well-approximates $r_i$, we  consider the function that is the AND of the $u - u / m$ Tribes that do not include the ones that make up $y_i$ and we see that Lemma 18 applies to that function as well. It can be verified that even when restricting to these $u - u/m$ tribes, the collective pseudorandom properties of the tribes are not affected. In the paper's terminology, the ``design property'' is exactly preserved and the ``sampling property'' is scaled by a factor of at most $2$.
The sampling property is an expectation across the $u$ tribes of some non-negative function. Since now we take expectation over $u - u / m \ge u / 2$ tribes, this expectation can increase no more than by a factor of two which does not affect the proof of their Lemma.
\end{proof}

\end{proof}

\dobib


\section{Randomized Selection Impossibility Result}
\label{sec: random selection impossibility}

In this section, we will prove the following impossibility result regarding one-round randomized selection protocols.
\begin{theorem}\label{lem: improved coin flip condensing impossibility}
There exists a universal constant $C$ such that for all $\ell, M\in \N$, $0 < \Delta$, the following holds:
For any one round random selection protocol $f: \zo^{\ell} \to [M]$, there exists a set  of bad players $B\subset [\ell]$ with $\abs{B} \le C\cdot \frac{\ell}{\log(M)}\cdot \frac{\Delta^2 \log(\Delta / \eps)^2}{\eps}$ and $S\subset [M]$ with $\abs{S} \le M / 2^{\Delta}$ such that the players in $B$ can corrupt $f$ so that with probability $\ge 1 - \eps$, the corrupted protocol outputs an element from $S$.
\end{theorem} 

We will prove this result in \cref{subsec: improved coin flip condensing impossibility}. Let's first obtain a corollary regarding impossibility of random selection protocols outputting uniformly random bits.

\begin{corollary}\label{cor: technical coin flip extracting impossibility}
There exists a universal constant $C$ such that for $0 < \eps < 1/4$ and $\ell,m\in\N$ where $m > \log(2/\varepsilon)$, the following holds:
For any one round random selection protocol $f: \zo^{\ell} \to [M]$, there exists a set of bad players $B\subset [\ell]$ with $\abs{B} \le C\cdot \frac{\ell}{m}\cdot \frac{\log(1 / \eps)^4}{\eps}$ such that the players in $B$ can corrupt $f$ so that the output distribution on $\zo^m$ is $(1 - \eps)$-far, in statistical distance, from the uniform distribution.
\end{corollary}

\begin{proof}
We set $\Delta = \log(2 / \eps)$ and error parameter equal to $\eps / 2$ in \cref{lem: technical coin flip condensing impossibility} so that the theorem gives a set of bad players of the claimed size such that with probability $1 - (\eps/2)$, the protocol outputs from some fixed set of size $\frac{\eps}{2}\cdot 2^m$.
This implies statistical distance from the uniform distribution is at least $1 - \eps$ as desired.
\end{proof}

Towards proving \cref{lem: improved coin flip condensing impossibility}, we will first prove a weaker random selection impossibility result, with worse dependence on $\Delta$:
\begin{lemma}\label{lem: technical coin flip condensing impossibility}
There exists a universal constant $C$ such that for all $\ell, M\in \N$, $0 < \Delta$ and $0 < \eps, \gamma < 1/4$, the following holds:
For any one round random selection protocol $f: \zo^{\ell} \to [M]\cup \{?\}$ where $\Pr[f = ?] \le \gamma$, there exists a set  of bad players $B\subset [\ell]$ with $\abs{B} \le C\cdot \frac{\ell}{\log(M)}\cdot 2^{2\Delta}\cdot \frac{(\log(1 / \eps))^2}{\eps}$ and $S\subset [M]$ with $\abs{S} \le M / 2^{\Delta}$ such that the players in $B$ can corrupt $f$ so that with probability $\ge 1 - \eps - \gamma$, the corrupted protocol outputs an element from $S$.
\end{lemma}

We will prove this result in \cref{subsec: technical coin flip condensing impossibility}.
To prove \cref{lem: technical coin flip condensing impossibility}, we introduce the following notion of multi-output influence.
\begin{definition}[Multi-output influence]\label{def: multioutput influence}
Let $n, M\in \N$, $f:\zo^n\to [M]$ be arbitrary.
Then, for $i\in[n]$, we define the \emph{multi-output influence of $i$ on $f$} and the \emph{total multi-output influence} as 
\begin{align*}
    &\I_i(f)=\Pr_{x\sim\zo^n}[f(x)\neq f(x^{\oplus i})]\\
    &\I(f)=\sum_{i=1}^n\I_i(f).
\end{align*}
\end{definition}

For any multi-output function, we can lower bound the total multi-output influence.
\begin{theorem}[Multi-output Poincar\'e]\label{thm: general multioutput poincare}
Let $n, M\in \N$, $f:\zo^n\to [M]$ be arbitrary.
Then, $\I(f) \ge H(f(\U_n))$ where $H(\cdot)$ denotes Shannon entropy.
\end{theorem} 
We will prove this in \cref{subsec: general multioutput poincare}.

\subsection{Proving Random Selection Impossibility}\label{subsec: improved coin flip condensing impossibility}

We here prove our main result regarding random selection impossibility, using \cref{lem: technical coin flip condensing impossibility}.

\begin{proof}[Proof of \cref{lem: improved coin flip condensing impossibility}]
Let $m = \log(M)$.
Notice that for non-trivial setting of parameters, where $\abs{B} < \ell$, we must have $\Delta < m/100$. We will use this below.

We will repeatedly use \cref{lem: technical coin flip condensing impossibility} with parameter $\Delta = 1$ each time and use it to narrow down our final target set.
Let $S_0 = [M], B_0 = \emptyset$ and let $g_1: \zo^{\ell \setminus B_0} \to [M] \cup \{?\}$ be defined as $g_1(x) = f(x)$.
For $i = 1, 2, \dots$, we do the following:
\begin{enumerate}
    \item 
    If $\abs{S_i} \le 2^{m - \Delta}$, then we stop and end the loop. Otherwise, we continue to the next step.
    
    \item 
    We apply \cref{lem: technical coin flip condensing impossibility} with error parameter $\eps / \Delta$ and parameter $\Delta = 1$ to obtain new set of bad players $C_i \subset [\ell]\setminus B_{i-1}$, a function $\Adv_i: \zo^{[\ell]\setminus B_{i-1}}\to \zo^{C_i}$ specifying the behaviors of these bad players, and a set $S_{i+1} \subset S_i$ with $\abs{S_{i+1}} \le \abs{S_i} / 2$ specifying the set that the bad players output from with large probability.

    \item
    Let $B_i = B_{i-1} \cup C_i$ and let $g_{i+1}: \zo^{[\ell]\setminus B_i} \to S_{i+1}\cup \{?\}$ be defined as 
    \[
        g_{i+1}(x) = 
        \begin{cases}
        g_i(x, \Adv_i(x)) & g_i(x, \Adv_i(x))\in S_{i+1}\\
        ?                 & \textrm{otherwise}
        \end{cases},
    \]
   i.e., we fix the behavior of the bad players in $C_i$ and consider the induced function over the remaining variables in $[\ell]\setminus B_i$, outputting $?$ if the output is outside $S_{i+1}$.
\end{enumerate}

Above, we call \cref{lem: technical coin flip condensing impossibility} with output domain not necessarily equal to $[T]$ but we associate each $S_{i}$ with $\abs{S_i}$ in any canonical way and also interpret the resultant output set as being a subset of $S_i$ in the same way.

Using induction and guarantees from \cref{lem: technical coin flip condensing impossibility}, we see that $\abs{S_i} \le 2^{m-i}$ and so in at most $\Delta$ steps, the loop indeed terminates. Suppose the loop terminates on step $t \leq \Delta$.

Let our final small set be $S := S_{t} \subset [M]$ and let our final set of bad players be $B := B_{t-1}$.
By guarantees from \cref{lem: technical coin flip condensing impossibility}, and induction, we infer that we can use $B$ to ensure that $f$ outputs from $S$ with probability at least $1 - t\cdot \frac{\eps}{\Delta} \ge 1 - \eps$ as desired.

We lastly bound $\abs{B}$.
Whenever we apply \cref{lem: technical coin flip condensing impossibility}, we are guaranteed that the output domain size is at least $2^{m - \Delta} \ge 2^{m / 2}$.
Using this and the setting of parameters when we call \cref{lem: technical coin flip condensing impossibility}, we infer that for each step $i$, $\abs{C_i} \le C'\cdot \frac{\ell}{m}\cdot \frac{\Delta \log(\Delta / \eps)^2}{\eps}$ where $C'$ is a universal constant.
Since our final set $B$ is the union of all these $C_i$, we obtain that
\[
\abs{B} \le t\cdot C'\cdot \frac{\ell}{m}\cdot \frac{\Delta \log(\Delta / \eps)^2}{\eps} \le C'\cdot \frac{\ell}{m}\cdot \frac{\Delta^2 \log(\Delta / \eps)^2}{\eps}
\]
as desired.
\end{proof}

\subsection{Proving Weak Random Selection Impossibility}\label{subsec: technical coin flip condensing impossibility}

We prove our weak random selection impossibility result in this subsection. 

\begin{proof}[Proof of \cref{lem: technical coin flip condensing impossibility}]
Let $m = \log(M)$.
Note that in this proof, we will $1$-index everything; that is, when we count rounds and epochs below, our first round and epoch is indexed by a 1, not a 0. 
We will let our final universal constant $C$ be large enough and set it at the end.
For $\abs{B}$ to be less than the trivial bound of $\ell$, we must have that $2^{2\Delta} \le m$, implying $2^{\Delta} = o(m)$. Similarly, we must also have that $\eps \ge 1 / m$. We will use these without always explicitly stating this.
Also, for any partial assignment $x$, when we mention $f(x, \cdot)$ is unfixed over some outcomes, we only count the number of outcomes unfixed from $[M]$. In particular, if $f(x, \cdot)$ is only unfixed over $?$, then we treat this as $f$ being unfixed over $0$ outcomes.

We will first specify how to construct such $B$ and $S$ and later how to specify the behavior of the bad players so that with probability $1 - \eps - \gamma$ over $\zo^{[\ell]\setminus B}$, they can force the outcome to be from $S$. We will then show the desired bounds on size of $S$ and size of $B$ in \cref{subsubsec: bounding size of S} and \cref{subsubsec: bounding size of B} respectively.

\subsubsection{Setup and Definitions}
Our algorithm proceeds across 
\begin{align}
    u = \text{number of epochs} = 2^{\Delta + 3}\cdot \log(1/\eps),\label{eq:number epochs def}
\end{align}
where each epoch will consist of a variable number of rounds.
Let epoch $i$ last for $v_i$ rounds.
Our ``small set'' $S$ will be the union of two types of sets: the ``heavy set'' $S_H$, which we specify before all epochs, and a ``greedily'' built set that we pick after the last epoch $u$.
We will inductively build our coalition $B\subset [\ell]$ of bad players, adding one player per round in each epoch. 
We do not specify the behavior of any of the bad players until the end of epoch $u$.
At each epoch, we will ensure that the bad players can either force an outcome to be from $S_H$ or that the number of outcomes that the bad players can choose between must grow.

\paragraph{Definitions for every round and epoch}
For each round $r\in \N$ and epoch $i\in [u]$, we define the following quantities.
Let $B_i^r \subset [\ell]$ denote the coalition of bad players right before round $r$ in epoch $i$.\footnote{So we have $\abs{B_i^r}=r-1+\sum_{j=1}^{i-1}v_j$.}
Recall that $f:\zo^\ell\to[M]$, and let $U_i^r: \zo^{[\ell]\setminus B_i^r}\to \cP([M])$ be the function that maps $x\in \zo^{[\ell]\setminus B_i^r}$ to the set of all outcomes in $[M]$ that $f(x, \cdot)$ is unfixed over.\footnote{That is, for $x\in\zo^{[\ell]\setminus B_i^r}$, we define $U_i^r(x)=\left\{f(x,y): y\in\zo^{B_i^r}\right\} \setminus \{?\}$.}
We next define $g_i^r: \zo^{[\ell]\setminus B_i^r}\to \binom{[M]}{i} \cup \{*, \bot, \top\}$ as follows:

\[
g_i^r(x) = 
\begin{cases}
\top & \textrm{if $U_i^r(x) \cap S_H \ne \varnothing$ }\\
* & \textrm{if $U_i^r(x) \cap S_H = \varnothing$ and $(i+1)\le \abs{U_i^r(x)}$}\\
U_i^r(x)\in \binom{[M]}{i} =  & \textrm{if $U_i^r(x) \cap S_H = \varnothing$ and $\abs{U_i^r(x)} = i$}\\
\bot & \textrm{if $U_i^r(x) \cap S_H = \varnothing$ and $\abs{U_i^r(x)} < i$}\\
\end{cases}
\]
We finally define $h_i^r: \zo^{[\ell]\setminus B_i^r}\to \binom{[M]}{i} \cup \{@\}$ as follows:
\[
h_i^r(x) = 
\begin{cases}
@ & \textrm{if } g_i^r(x) \in \{\top, *, \bot\}\\
g_i^r(x) & \textrm{otherwise}\\
\end{cases}.
\]

As defined, $h_i^r$ only tells us whether $g_i^r$ outputs a special character. In general, we will try to accomplish one of the following: 1) bias $g_i^r$ away from outputting $\bot$ and towards outputting $\binom{[M]}{i}$, or  2) bias $g_i^r$ away from $\binom{[M]}{i}$ and towards $*$ and $\top$. 
To find the player that can most effectively do this, we will apply \cref{thm: general multioutput poincare} to $h_i^r$.

\subsubsection{Building the heavy set $S_H$ and coalition of bad players $B$}

We here describe our biasing procedure using the setup and definitions from above.

\paragraph{Specifying heavy set $S_H$}
We first specify $S_H$, our heavy set.
We let $S_H = \{y \in [M]: \Pr_{x\sim \U_n}[f(x) = h] \ge 2^{-m/2}\}$.

\paragraph{Building the coalition of bad players $B$ across rounds and epochs}
For the base case, we define $B_1^1 = \varnothing$.
With this, at every round $r\in \N$ and epoch $i\in [u]$, we proceed as follows:
\begin{step}
\item    
Based on $H$ and $B_i^r$, we define our functions $g_i^r$ and $h_i^r$ as declared above.

\item 
If $\Pr_{x\sim\zo^{[\ell]\setminus B_i^r}}\left[g_i^r(x) \in \binom{[M]}{i}\right] \le \frac{\eps}{4u}$, then we declare epoch $i$ as finished, setting $v_i = r-1$, and move on to the next epoch $i+1$.

\item\label{step:add bad player}
Otherwise, we apply our multioutput influence bound from \cref{thm: general multioutput poincare} to $h_i^r$ to find some $j\in [\ell]\setminus B_i^r$ such that 
\[
\I_j(h_i^r) 
\ge \frac{1}{\ell - \abs{B_i^r}}\cdot H(h_i^r(\U)).
\]

\item 
We let $B_i^{r+1} = B_i^r \cup \{j\}$ and continue to the next round $r+1$ in epoch $i$.
\end{step}
After epoch $u$ ends, we let $B\coloneqq B_u^{v_u}$.

\paragraph{Finishing up after the last epoch}\label{subsubsec: finishing after last epoch}

We now specify the behavior of the bad players and show how to pick our final set $S$.
Consider the bipartite graph $G$ with vertex sets $V_L = \zo^{[\ell]\setminus B}$ and $V_R = [M]$.
We add an edge between $x\in V_L$ and $y\in V_R$ if $f(x, \cdot)$ is unfixed over outcome $y$ (among other outcomes).
We will find a small almost dominating set $S\subset V_R$ such that the neighborhood of $S$ is a $1-\varepsilon$ fraction of $V_L$.
Once we do this, we specify the behavior of the bad players to always force the outcome to be from the set $S$ whenever $x\in\Nbr(S)$.
To find the almost dominating set $S$, we initially add all elements from $S_H$ to $S$. We then greedily add outcome $y\in V_R$ to $S$ such that it neighbors the maximum number of uncovered vertices from $V_L$, i.e., $y = \argmax_{z\in V_R}\{\Nbr(z)\setminus \Nbr(S)\}$. We do this until $1 - \eps$ fraction of the vertices from $V_L$ are covered by $S$.
This specifies the behavior of the bad players as well as the set $S$ as desired.

\subsubsection{Bounding the size of S}\label{subsubsec: bounding size of S}
In this subsubsection, we will show that $\abs{S} \le 2^{m-\Delta}$.
We need the following claim about the power of the bad players $B$ at the end of all the epochs:

\begin{claim}\label{claim: end guarantee of power of B}
For at least a $1 - \gamma - \eps / 2$ fraction of $x\in \zo^{[\ell] \setminus B}$, it holds that $f(x, \cdot)$ is either unfixed over some outcome from $S_H$ or is unfixed over $\ge u+1$ outcomes.
\end{claim}
We prove this claim towards the end of this subsubsection. Let's see how using this, we can bound the size of $S$.

We first claim that $\abs{S_H}\le 2^{m / 2}$. 
Indeed, since $\Pr_{x\sim \U_{\ell}}[f(x) \in S_H] \le 1$ and for all $y\in S_H$, $\Pr_{x\sim\U_{\ell}}[f(x) = y] \ge 2^{-m/2}$, we directly obtain that $\abs{S_H}\le 2^{m / 2}$.

Next, we consider the bipartite graph $G$ from \cref{subsubsec: finishing after last epoch}.
Suppose the greedy algorithm to find an almost dominating set runs for $t$ steps. 
Then, $\abs{S} = t + \abs{S_H} \le t + 2^{m / 2}\le t + 2^{m - \Delta - 1}$ where we use the fact that $\Delta = o(m)$.
Hence, it suffices to show that $t \le 2^{m - \Delta - 1}$ to obtain the desired bound on $\abs{S}$.

We now show the remaining claim that $t \le 2^{m - \Delta - 1}$. Note that after we add $S_H$ to $S$, the remaining uncovered vertices in $V_L$ have degree at least $u + 1$.
For $i = 0, \dots, t$, let the number of uncovered vertices in $V_L$ after $i$ steps be $\delta_i\cdot \abs{V_L}$.
Using \cref{claim: end guarantee of power of B}, we obtain that after step $i$, the number of outgoing edges from the  uncovered vertices in $V_L$ to $V_R$ is at least $(\delta_i - \gamma - \eps / 2)\cdot \abs{V_L}\cdot (u+1)$.
This implies that the chosen vertex after step $i$ from $V_R$ covers at least 
\[
(\delta_i - \gamma - (\eps / 2))\cdot\abs{V_L}\cdot (u)\cdot \frac{1}{\abs{V_R}} = (\delta_i - \gamma - (\eps / 2))\cdot\abs{V_L}\cdot (u)\cdot 2^{-m}
\]
So, $\delta_{i+1} \le \delta_i - (\delta_i - \gamma - (\eps / 2))\cdot u\cdot 2^{-m}$.
This implies $\delta_{i+1} - \gamma - (\eps / 2) \le (\delta_i - \gamma - (\eps / 2))\cdot (1 - (u\cdot 2^{-m}))$.
So, after $k$ steps, it holds that
\[
\delta_k - \gamma - (\eps / 2) \le (\delta_0 - \gamma - (\eps / 2))\cdot (1 - (u\cdot 2^{-m}))^k \le (\delta_0 - \gamma - (\eps / 2))\cdot \exp(-u\cdot 2^{- m}\cdot k).
\]
Setting $k = 2^{m - \Delta - 1}$, we see that
\begin{align*}
\delta_k - \gamma - (\eps/2)
& \le (\delta_0 - \gamma - (\eps / 2))\cdot \exp(-u\cdot 2^{-m}\cdot k)\\
& \le \exp(-u\cdot 2^{-m}\cdot k) = \exp(-u\cdot 2^{-\Delta-1})\\
& = \exp(-2^{\log(\log(1 / \eps)) + 1}) = \exp(-4\log(1 / \eps)) &\text{(by \cref{eq:number epochs def})}\\
& \le 2^{-2\log(1 / \eps)} = \frac{1}{\eps^2} \le \frac{\eps}{2}.
\end{align*}
Hence, after $k = 2^{m - \Delta - 1}$ steps of the greedy algorithm, we always have that $\delta_k \le \gamma + \eps$.
Since $t$ is the earliest step where $\delta_t \le \gamma + \eps$, we infer that $t\le k = 2^{m - \Delta - 1}$ as desired.

\paragraph{Claim about the bad players at the end of all epochs}\label{subsubsec: end guarantee of power of B}

We here prove \cref{claim: end guarantee of power of B}, that at the end of all epochs, for most fixings of the good bits, the players in $B$ can either force the outcome to be in $H$ or can choose amongst $(u+1)$ outcomes.

Towards this, we introduce the following quantity:
For $i\in [u]$ and $r\in \N$, 
let 
\begin{align*}
    \alpha_i^r(\bot) &= \Pr_{x\sim \zo^{[\ell]\setminus B_i^r}}[g_i^r = \bot].
\end{align*}
We will show the following claim regarding how this quantity evolves across each rounds and epoch:
\begin{claim}\label{claim: alpha i r bot evolve}
For all $i\in [u], r\in \N$, the following holds:
\begin{enumerate}
\item 
$\alpha_i^{r+1}(\bot) \le \alpha_i^r(\bot)$.
\item 
$\alpha_{i+1}^1(\bot) \le \alpha_i^{v_i}(\bot) + \frac{\eps}{4u}$.
\end{enumerate}
\end{claim}

We prove these claims at the end of this subsubsection. Let's see how using this our inductive step goes through.

\begin{proof}[Proof of \cref{claim: end guarantee of power of B}]
Recall that we are guaranteed that $\alpha_1^1(\bot) \le \gamma$.
Using \cref{claim: alpha i r bot evolve}, we induct and obtain that $\alpha_{i}^{v_i+1}(\bot)\le \alpha_{i}^1(\bot) \le \gamma + (i-1)\cdot \frac{\eps}{4u}$. 
Therefore, $\alpha_{u}^{v_u+1}(\bot) \le\gamma + \frac{\eps}{4}$.
This implies $\Pr[g_u^{v_u + 1}] \in \{\bot, *, \binom{[M]}{u}\} \ge 1 - \gamma - \eps/4$.
The end condition at epoch $u$ implies
$\Pr[g_u^{v_u + 1}] \in \binom{[M]}{u}] \le \frac{\eps}{4u}$.
Hence, 
\[
\Pr[g_u^{v_u + 1}] \in \{\bot, *\}] \ge 1 - \gamma - \frac{\eps}{4} - \frac{\eps}{4u} \ge 1 - \gamma - \frac{\eps}{2}.
\]
In other words, after all the rounds and epochs, for $1 - \gamma - \eps/2$ fraction of $x\sim \zo^{[\ell]\setminus B}$, the $f(x, \cdot)$ is unfixed over either some element from $H$ or over at least $u+1$ outcomes, as claimed.
\end{proof}

We now prove our remaining claims regarding how the value of $\alpha(\bot)$ evolves across each round and epoch.
\begin{proof}[Proof of \cref{claim: alpha i r bot evolve}]
We will show each of our claims, one after the other.
\begin{enumerate}
\item 
To show that $\alpha_i^{r+1}(\bot) \le \alpha_i^r(\bot)$, we first define the relevant quantities. 
Let $j$ be the player added at round $r$.
Define $A_i^r = \{x\in \zo^{\ell\setminus B_i^{r}}: g_i^r(x) = \bot\}$.
Similarly, we define $A_i^{r+1} = \{x\in \zo^{\ell\setminus B_i^{r+1}}: g_i^{r+1}(x) = \bot\}$.
For all assignments $x\in A_i^{r+1}$, consider the two assignments $x_0, x_1 \in \zo^{[\ell]\setminus B_i^r}$ obtained by assigning variables in $[\ell]\setminus B_i^{r+1}$ according to $x$ and assigning variable $j$ to $0$ and $1$ respectively. 
We claim that both $x_0, x_1\in A_i^r$.
Indeed, this is because the set of outcomes that $f(x, \cdot)$ is unfixed over is a superset of the outcomes that each of $f(x_0, \cdot)$ and $f(x_1, \cdot)$ are unfixed over.
Since $g_i^{r+1}(x) = \bot$, $f(x, \cdot)$ is unfixed over $< i$ outcomes, none of which are from $H$ and the same must be true for both $f(x_0, \cdot)$ and $f(x_1, \cdot)$.
This implies $2\cdot \abs{A_i^{r+1}}\le \abs{A_i^{r}}$.
Therefore,
\begin{align*}
\Pr[\alpha_i^{r+1} = \bot] 
& = \frac{A_i^{r+1}}{2^{\ell - \abs{B_i^{r+1}}}}
  = \frac{2\cdot A_i^{r+1}}{2^{\ell - \abs{B_i^{r}}}}\\
& \le \frac{A_i^{r}}{2^{\ell - \abs{B_i^{r}}}}
= \Pr[\alpha_i^{r+1} = \bot]
\end{align*}
as desired.

\item
Since the bad players at the beginning of round $v_i+1$ in epoch $i$ are the same as that at the beginning of round $1$ in epoch $i+1$, we have that
$\alpha_{i+1}^1(\bot)$ equals the fraction of $x\in \zo^{\ell\setminus B_i^{v_i+1}}$ such that $f(x, \cdot)$ is unfixed over $\le i$ outcomes none of which are from $H$.
By definition, this quantity equals $\Pr[g_i^{v_i+1} = \bot] + \Pr[g_i^{v_i+1} \in \binom{[M]}{i}] = \alpha_i^{v_i+1}(\bot) + \Pr[g_i^{v_i+1} \in \binom{[M]}{i}]$.
Since we ended epoch $i$ at round $v_i+1$, we have that  $\Pr\left[g_i^{v_i+1} \in \binom{[M]}{i}\right] \le \frac{\eps}{4u}$, showing the claim. \qedhere
\end{enumerate}
\end{proof}

\subsubsection{Bounding the size of B}\label{subsubsec: bounding size of B}

Since we add one player per round in each epoch, we see that $\abs{B} = \sum_{i=1}^u v_i$.
We claim that for all $i$, $\abs{v_i} \le \frac{64 u}{\eps}\cdot \frac{\ell}{m}$.
Proving this suffices to obtain our desired bound on $\abs{B}$ since then
\begin{align*}
\abs{B} 
& = \sum_{i=1}^u v_i\\
& \le u\cdot \frac{64 u}{\eps}\cdot \frac{\ell}{m}\\
& = 2^{2\Delta + 12}\cdot \frac{(\log(1 / \eps))^2}{\eps}\cdot \frac{\ell}{m}\\
& \le C\cdot 2^{2\Delta}\cdot \frac{(\log(1 / \eps))^2}{\eps}\cdot \frac{\ell}{m}
\end{align*}
where $C$ is a large enough universal constant.
This proves the claim as desired.

\paragraph{Bounding number of rounds in each epoch}
It remains to show that for all $i$, $v_i \le \frac{64 u}{\eps}\cdot \frac{\ell}{m}$.
Fix any such $i$.
We will use a potential function based argument.
For $r\in \N, r\le v_i+1$, we consider the quantity 
\[
\phi(r) := 1 - \Pr[g_i^r = \bot] + \Pr[g_i^r \in \{*, \top\}].
\]
We claim the following:
\begin{claim}\label{claim: phi always increases}
For all $r\in \N$, $\phi(r+1) \ge \phi(r) + \frac{\eps}{32u}\cdot \frac{m}{\ell}$.
\end{claim}
We prove this claim later.
Let's see how to use this to obtain our desired bound on $v_i$.
We first easily see that for all $r$, $\phi(r) \ge 0$ and in particular, this is also true for $r = 1$.
Then, using \cref{claim: phi always increases} we obtain that for all $r\le v_i+1$,
\begin{align}\label{eq: bound on B potential increases}
\phi(r) \ge (r-1)\cdot \frac{\eps}{32u}\cdot \frac{m}{\ell}.
\end{align}

Let $t = \frac{64 u}{\eps}\cdot \frac{\ell}{m}$ so that our claimed upper bound becomes $v_i \le t$.
Assume that $t\le v_i$.
By \cref{eq: bound on B potential increases}, we must have that $\phi(t+1) \ge t\cdot \frac{\eps}{32u}\cdot \frac{m}{\ell} = 2$.
By definition of $\phi$ and the fact that probabilities are non-negative and upper bounded by $1$, we obtain that $\Pr[g_i^{t+1} \in \{*, \top\}] = 1$.
This implies $\Pr[g_i^{t+1} \in \binom{[M]}{i}] = 0$, and we must meet the termination condition for the epoch by round $t+1$. This means $v_i + 1\le t + 1$ and hence, $v_i \le t$ as desired.

\paragraph{Claimed potential always increases}

We here prove \cref{claim: phi always increases}, that our desired potential function $\phi$ increases by every round.
To prove this, we will require the following lower bound on the influence of the variable added at every round an epoch.
\begin{claim}\label{claim: bad player each round and epoch influence}
For all $i\in [u], r\in \N$, if $j$ is the player added to $B_i^r$, then $\I_j(h_i^r) \ge \frac{\eps}{16u}\cdot \frac{m}{\ell}$.
\end{claim}
We prove this towards the end of this subsubsection.
With this, we are ready to prove our desired claim.

\begin{proof}[Proof of \cref{claim: phi always increases}]
We rewrite our claim in the following equivalent form:
\begin{align}\label{eq phi always increases: equiavalent claim}
\left(\Pr[g_i^r = \bot] - \Pr[g_i^{r+1} = \bot]\right) + \left(\Pr[g_i^{r+1} \in \{\top, *\}] - \Pr[g_i^r \in \{\top, *\}]\right) \ge \frac{\eps}{32u}\cdot \frac{m}{\ell}.
\end{align}
Fix such $r$.
Let $j$ be the player added in this round $r$.
We easily see that $\I_j(g_i^r) \ge \I_j(h_i^r)$.

We will abuse notation below as follows: for $x\in \zo^{[\ell]\setminus B_i^{r+1}}$, we will write $x\circ 0$ and $x\circ 1$ to mean assignments in $\zo^{[\ell]\setminus B_i^r}$ that assign all variables in $[\ell]\setminus B_i^{r+1}$ to the same value as $x$ and assign $j$ to $0$ and $1$ respectively.

We first define the set of assignments where variable $j$ was influential for $g_i^r$.
Let 
\[
T = \{x\in \zo^{[\ell]\setminus B_i^{r+1}}: g_i^r(x\circ 0) \ne g_i^r(x\circ 1)\}.
\]
We partition this set $T$ into the following sets.
For $s\in \{\top, *, \bot\}$, let
\[
T_s = \left\{x\in T: \{g_i^r(x\circ 0), g_i^r(x\circ 1)\} = \{s, z\} \textrm{ where $z\in \binom{[M]}{i}$}\right\}.
\]
Let 
\[
T_{\#} = \left\{x\in T: \{g_i^r(x\circ 0), g_i^r(x\circ 1)\} = \{z_1, z_2\} \textrm{ where $z_1\ne z_2\in \binom{[M]}{i}$}\right\}.
\]
Let $T_{\rest} = T\setminus (T_{\top}\cup T_{*}\cup T_{\bot} \cup T_{\#})$.
We will also define densities of these influence sets that are appropriately normalized.
Let $\eta = \frac{\abs{T}}{2^{\ell - \abs{B_i}^{r+1}}}$.
We similarly define $\eta_{\top}, \eta_{*}, \eta_{\bot}, \eta_{\#}, \eta_{\rest}$.

Using \cref{claim: bad player each round and epoch influence} and the definition of multi-output influence, we see that
\begin{align}\label{eq phi always increases: using influence}
\eta_{\top} + \eta_{*} + \eta_{\bot} + \eta_{\#} = \I_j(h_i^r) \ge \frac{\eps}{16 u}\cdot \frac{m}{\ell}.
\end{align}
We now make the following claim regarding how these influences change probabilities in $g_i^r$:
\begin{claim}\label{claim: bot decrease and top or star increases}
The following holds:
\begin{enumerate}
\item 
\[
    \Pr[g_i^r = \bot] - \Pr[g_i^{r+1} = \bot] \ge \eta_{\bot} / 2.
\]

\item
\[
    \Pr[g_i^{r+1} \in \{\top, *\}] - \Pr[g_i^r \in \{\top, *\}] \ge (\eta_{\top} + \eta_{*} + \eta_{\#}) / 2.
\]
\end{enumerate}
\end{claim}

We prove this claim later, let's first see how to finish off our current proof using them.
We compute
\begin{align*}
& \left(\Pr[g_i^r = \bot] - \Pr[g_i^{r+1} = \bot]\right) + \left(\Pr[g_i^{r+1} \in \{\top, *\}] - \Pr[g_i^r \in \{\top, *\}]\right)\\
& \ge \left(\eta_{\top} + \eta_{*} + \eta_{\#} + \eta_{\bot}\right) / 2 & \textrm{(by \cref{claim: bot decrease and top or star increases})}\\
& \ge \frac{\eps}{32 u}\cdot \frac{m}{\ell} & \textrm{(by \cref{eq phi always increases: using influence})}
\end{align*}
showing \cref{eq phi always increases: equiavalent claim} holds and hence proving \cref{claim: phi always increases}.

We now prove our remaining claim regarding how influence changes $g_i^r$.
\begin{proof}[Proof of \cref{claim: bot decrease and top or star increases}]
We prove the two claims one after the other.
\begin{enumerate}
\item 
Define $A_2 = \{x\in \zo^{[\ell]\setminus B_i^{r+1}}: g_i^r(x\circ 0) = g_i^r(x\circ 1) = \bot\}$.
Let $\eta_2 = \frac{\abs{A_2}}{2^{\ell - \abs{B_i^{r+1}}}}$.
We first observe that 
\[
    \Pr[g_i^r = \bot] \ge \eta_2 + \frac{\eta_{\bot}}{2}
\]
by definition of the sets in question.
We next observe
\[
    \Pr[g_i^{r+1} = \bot] \le \eta_2
\]
since if $f(x, \cdot)$ is unfixed over $< i$ outcomes none of which are from $S_H$, then it must be the case that both $f(x\circ 0, \cdot)$ and $f(x\circ 1, \cdot)$ are unfixed over $< i$ outcomes, none of which are from $S_H$.
Combining these two observations, our claim follows.

\item
We proceed similarly as above, defining more objects here.
Define $A_2 = \{x\in \zo^{[\ell]\setminus B_i^{r+1}}: g_i^r(x\circ 0)\in \{\top, *\} \land g_i^r(x\circ 1)\in \{\top, *\}\}$.
We also define $A_1 = \{x\in \zo^{[\ell]\setminus B_i^{r+1}}: g_i^r(x\circ 0)\in \{\top, *\} \land g_i^r(x\circ 1)\not\in \{\top, *\}\} \cup \{x\in \zo^{[\ell]\setminus B_i^{r+1}}: g_i^r(x\circ 1)\in \{\top, *\} \land g_i^r(x\circ 0)\not\in \{\top, *\}\}$.
Let $\eta_2 = \frac{\abs{A_2}}{2^{\ell - \abs{B_i^{r+1}}}}, \eta_1 = \frac{\abs{A_1}}{2^{\ell - \abs{B_i^{r+1}}}}$ be their densities.

We first observe that 
\[
    \Pr[g_i^r \in \{\top, *\}] = \eta_2 + \frac{\eta_1}{2}
\]
using definitions of the sets in question.
We next observe
\[
    \Pr[g_i^{r+1} \in \{\top, *\}] \ge \eta_2 + \eta_1 + \eta_{\#}.
\]
Combining these, we have that 
\[
     \Pr[g_i^{r+1} \in \{\top, *\}] - \Pr[g_i^{r} \in \{\top, *\}] \ge  \frac{\eta_1}{2} + \eta_{\#}.
\]
We lastly see that $\eta_1 \ge \eta_{\top} + \eta_{*}$ and obtain our claim. \qedhere
\end{enumerate}
\end{proof}
\end{proof}

\paragraph{Lower bound on the influence of each bad player}\label{subsubsec: bad player each round and epoch influence}
We here prove our claim that at each round and epoch, the player we add must have large influence. We will use multi-output Poincar\'e bound to prove this.
\begin{proof}[Proof of \cref{claim: bad player each round and epoch influence}]
Since player $j$ was added in round $r$ of epoch $i$, it must be the case that 
\begin{align}\label{eq: bad player each round and epoch influence - total prob}
\sum_{z\in \binom{[M]}{i}} \Pr[h_i^r = z] \ge \frac{\eps}{4u}.
\end{align}

We next claim that for all $z\in \binom{[M]}{i}$, it holds that $\Pr[h_i^r = z] \le 2^{-m/4}$.
Indeed, suppose this is not the case. 
We will contradict the fact that $z\not \in S_H$.
Then, we have that $\Pr_{x\sim \zo^{\ell}} [f(x) \in z] \ge 2^{-m/4}$.
This implies there exists some $y\in z$ such that $\Pr_{x\sim \zo^{\ell}} [f(x) = y] \ge 2^{-m/4}\cdot \frac{1}{i}$.
Since $i\le u$, we have that $i \le 2^{\Delta+3}\cdot \log(1 / \eps) \le \frac{1}{m^2}$ where we used the fact that $2^{\Delta} = o(m)$ and that $\eps \ge 1 / m$.
Therefore, this implies $\Pr_{x\sim \zo^{\ell}} [f(x) = y] \ge 2^{-m/4}\cdot \frac{1}{m^2} \ge 2^{-m/2}$.
This indeed contradicts the fact that $z\not \in S_H$.

Finally, we apply \cref{thm: general multioutput poincare} on $h_i^r$ to infer that 
\begin{align*}
\I(h_i^r) 
& \ge H(h_i^r)\\
& \ge \sum_{z\in \binom{[M]}{i}} \Pr[h_i^r = z]\log(1 / \Pr[h_i^r = z])\\
& \ge \frac{m}{4}\sum_{z\in \binom{[M]}{i}} \Pr[h_i^r = z] & \textrm{(since for all $z$, $\Pr[h_i^r = z] \le 2^{-m/4}$)}\\
& \ge \frac{m}{4}\cdot \frac{\eps}{4u} & \textrm{(using \cref{eq: bad player each round and epoch influence - total prob})}\\
& = \frac{\eps}{16 u}\cdot m.
\end{align*}
Since $j$ was the variable with the highest influence, we have that 
\[
    \I_j(h_i^r) \ge \frac{\I(h_i^r)}{\ell - \abs{B_i^r}} \ge \frac{\eps}{16 u}\cdot \frac{m}{\ell}
\]
as desired.
\end{proof}

\end{proof}

\subsection{Proving the multi-output influence bound}\label{subsec: general multioutput poincare}

Here, we prove our multioutput Poincar\'e inequality of \cref{thm: general multioutput poincare}.
We first recall the edge isoperimetric inequality
\cite{harper1966optimal, bernstein1967maximally, lindsey1964assignment, hart1976note}.
\begin{theorem}[Edge Isoperimetric Inequality]
\label{thm: edge isoperimetric inequality}
Let $n\in \N$ and $f: \zo^n\to \zo$ be arbitrary. Then,
\[
    \I(f) \ge 2\E[f]\log(1 / \E[f]).
\]
\end{theorem}
With this, we prove our desired multioutput influence bound as follows.

\begin{proof}[Proof of \cref{thm: general multioutput poincare}]
For all $y\in [M]$, we define a function $\chi_y: \zo^n \to \zo$ as $\chi_y(x) = 1 \iff f(x) = y$.
We apply the Edge Isoperimetric inequality from \cref{thm: edge isoperimetric inequality} to $\chi_y$ to obtain that $\I(\chi_y)\ge 2\Pr[\chi_y = 1]\log(1 / \Pr[\chi_y = 1])$.

We now claim that $\I(f) = \frac{1}{2}\sum_{y\in [M]}\I(\chi_y)$. We prove this claim later. Let's see how using this claim, we can finish our proof. We compute:
\begin{align*}
\I(f) 
& = \frac{1}{2}\sum_{y\in [M]}I(\chi_y)\\
& \ge \sum_{y\in [M]}\Pr[\chi_y = 1]\log(1 / \Pr[\chi_y = 1])\\
& = \sum_{y\in [M]}\Pr[f(\U_n) = y]\log(1 / \Pr[f(\U_n) = y])\\
& = H(f(\U_n)).
\end{align*}
We finally show that $\I(f) = \frac{1}{2}\sum_{y\in [M]}\I(\chi_y)$.
We prove something stronger that for any $i\in [n]$, 
$\I_i(f) = \frac{1}{2}\sum_{y\in [M]}\I_i(\chi_y)$. Fix any such i.
First, for $y\in [M]$, let $S_y = \{x\in \zo^n: \chi_y(x) \ne \chi_y(x^{\oplus i})\}$.
Let $T = \{x\in \zo^n: f(x)\ne f(x^{\oplus i})\}$.
We see that $\I_i(f) = \frac{\abs{T}}{2^n}$ and $\I_i(\chi_y) = \frac{\abs{S_y}}{2^n}$.
Hence, our claim is equivalent to showing that $\abs{T} = \frac{1}{2} \sum_{y\in [M]} \abs{S_y}$.
We observe that $x\in T$ if and only if there exist exactly two $y_1\ne y_2\in [M]$ such that $x\in S_{y_1}$ and $S_{y_2}$.
These $y_1, y_2$ must be such that $\{y_1, y_2\} = \{f(x), f(x^{\oplus i}\}$.
Hence, we count each such $x\in T$ exactly twice across all sets $S_y$, allowing us to conclude that $\abs{T} = \frac{1}{2} \sum_{y\in [M]} \abs{S_y}$ as claimed.
\end{proof}

\dobib

\section{Conclusion and Open Problems}\label{sec:open}

There still remains a gap between the best known upper bounds and lower bounds for coin flipping protocols. For instance, even when restricted to two-round protocols, our protocols from \cref{thm: construct protocol} can handle $\frac{\ell}{(\log \ell)\poly(\log^{(2)} \ell)}$ bad players while our lower bound from \cref{thm: protocol lb} requires $\frac{\ell}{\log^{(2)}\ell}$ bad players.
Towards proving stronger lower bounds to bridge this gap, we present the following question regarding simultaneously biasing functions. 
\begin{question}[Simultaneous biasing]
\label{conj: simultaneous biasing conjecture}
Prove or disprove the following:
Let $f = (f_1, \dots, f_m): \zo^{\ell}\to \zo^m$ where $m = \ell^{0.01}$ be an arbitrary map. Then, there exist $B\subset[\ell], P\subset [m], o\in \zo^{\abs{P}}$ with $\abs{B}\le O(\ell / \log \ell)$, $\abs{P} = 0.01 m$ such that for $g = f|_{B}$, and all $i\in [P]$, $\Pr[g_i = o_i] \ge 0.99$.
\end{question}

Note that a much weaker version of \cref{conj: simultaneous biasing conjecture} indeed holds by \cref{lem: most functions from family can be corrupted with random and heavy set}. The difference is that in \cref{lem: most functions from family can be corrupted with random and heavy set}, the (slightly different) sets of bad players can corrupt each $f_i$ differently while here we want each of the functions indexed by $P$ to be simultaneously corrupted not only by the same \emph{set} of bad players $B$, but also by the same \emph{behavior} of the bad players in that set. 

We note that when $m = O(1)$, by repeatedly applying the result of \cite{KKL88influence}, \cref{conj: simultaneous biasing conjecture} does hold. However, that strategy requires more than $n$ players once $m \ge \log \ell$, which is trivial. Answering the above question in the positive will lead to almost matching lower bounds for coin flipping; in particular, this would imply that $\frac{\ell \poly(\log^{(2)}\ell)}{\log \ell}$ bad players suffice to bias any two-round protocol. This bound would follow by using the same strategy as the proof of \cref{thm: protocol lb}, with the second induction step replaced by the simultaneous biasing conjecture. 
Our one-round random selection impossibility results from \cref{lem: improved coin flip condensing impossibility} can also be interpreted as a necessary step towards this conjecture.
For multi-round lower bounds nearly matching our constructions from \cref{thm: construct protocol}, we would need a stronger version of \cref{conj: simultaneous biasing conjecture} where we would want to simultaneously bias not just a function map but a $k$-round protocol map.

\subsection*{Acknowledgements}
 We thank the organizers of the Dagstuhl Seminar on Algebraic and Analytic Methods in Computational Complexity and Schloss Dagstuhl for providing a stimulating research environment, where early discussions between 
R.S. and E.C. 
 initiated this collaboration.

\dobib


\printbibliography

@inproceedings{KKL88influence,
  author       = {Jeff Kahn and
                  Gil Kalai and
                  Nathan Linial},
  title        = {The Influence of Variables on Boolean Functions (Extended Abstract)},
  booktitle    = {29th Annual Symposium on Foundations of Computer Science, White Plains,
                  New York, USA, 24-26 October 1988},
  pages        = {68--80},
  publisher    = {{IEEE} Computer Society},
  year         = {1988},
  url          = {https://doi.org/10.1109/SFCS.1988.21923},
  doi          = {10.1109/SFCS.1988.21923},
  bibsource    = {dblp computer science bibliography, https://dblp.org}
}

@article{Chattopadhyay2019TwoSource,
  author    = {Eshan Chattopadhyay and David Zuckerman},
  title     = {Explicit two-source extractors and resilient functions},
  journal   = {Annals of Mathematics},
  volume    = {189},
  number    = {3},
  pages     = {653--705},
  year      = {2019},
  doi       = {10.4007/annals.2019.189.3.1},
  publisher = {Mathematical Institute},
  url       = {https://projecteuclid.org/euclid.aom/1559895809}
}

@article{RSZ02leader,
  author       = {Alexander Russell and
                  Michael E. Saks and
                  David Zuckerman},
  title        = {Lower Bounds for Leader Election and Collective Coin-Flipping in the
                  Perfect Information Model},
  journal      = {{SIAM} J. Comput.},
  volume       = {31},
  number       = {6},
  pages        = {1645--1662},
  year         = {2002},
  url          = {https://doi.org/10.1137/S0097539700376007},
  doi          = {10.1137/S0097539700376007},
  bibsource    = {dblp computer science bibliography, https://dblp.org}
}

@inproceedings{feige99lightestbin,
  author       = {Uriel Feige},
  title        = {Noncryptographic Selection Protocols},
  booktitle    = {40th Annual Symposium on Foundations of Computer Science, {FOCS} '99,
                  17-18 October, 1999, New York, NY, {USA}},
  pages        = {142--153},
  publisher    = {{IEEE} Computer Society},
  year         = {1999},
  url          = {https://doi.org/10.1109/SFFCS.1999.814586},
  doi          = {10.1109/SFFCS.1999.814586},
  bibsource    = {dblp computer science bibliography, https://dblp.org}
}

@article{IV25resilient,
  author       = {Peter Ivanov and
                  Emanuele Viola},
  title        = {Resilient functions: Optimized, simplified, and generalized},
  journal      = {CoRR},
  volume       = {abs/2406.19467},
  year         = {2024},
  url          = {https://doi.org/10.48550/arXiv.2406.19467},
  doi          = {10.48550/ARXIV.2406.19467},
  eprinttype    = {arXiv},
  eprint       = {2406.19467},
  bibsource    = {dblp computer science bibliography, https://dblp.org}
}

@article{AL93resilient,
  author       = {Mikl{\'{o}}s Ajtai and
                  Nathan Linial},
  title        = {The influence of large coalitions},
  journal      = {Comb.},
  volume       = {13},
  number       = {2},
  pages        = {129--145},
  year         = {1993},
  url          = {https://doi.org/10.1007/BF01303199},
  doi          = {10.1007/BF01303199},
  bibsource    = {dblp computer science bibliography, https://dblp.org}
}

@article{RZ01leader,
  author       = {Alexander Russell and
                  David Zuckerman},
  title        = {Perfect Information Leader Election in log* n+O {(1)} Rounds},
  journal      = {J. Comput. Syst. Sci.},
  volume       = {63},
  number       = {4},
  pages        = {612--626},
  year         = {2001},
  url          = {https://doi.org/10.1006/jcss.2001.1776},
  doi          = {10.1006/JCSS.2001.1776},
  bibsource    = {dblp computer science bibliography, https://dblp.org}
}

@inproceedings{BL85coin,
  author       = {Michael Ben{-}Or and
                  Nathan Linial},
  title        = {Collective Coin Flipping, Robust Voting Schemes and Minima of Banzhaf
                  Values},
  booktitle    = {26th Annual Symposium on Foundations of Computer Science, Portland,
                  Oregon, USA, 21-23 October 1985},
  pages        = {408--416},
  publisher    = {{IEEE} Computer Society},
  year         = {1985},
  url          = {https://doi.org/10.1109/SFCS.1985.15},
  doi          = {10.1109/SFCS.1985.15},
  bibsource    = {dblp computer science bibliography, https://dblp.org}
}

@inproceedings{ORV94leader,
  author       = {Rafail Ostrovsky and
                  Sridhar Rajagopalan and
                  Umesh V. Vazirani},
  editor       = {Frank Thomson Leighton and
                  Michael T. Goodrich},
  title        = {Simple and efficient leader election in the full information model},
  booktitle    = {Proceedings of the Twenty-Sixth Annual {ACM} Symposium on Theory of
                  Computing, 23-25 May 1994, Montr{\'{e}}al, Qu{\'{e}}bec,
                  Canada},
  pages        = {234--242},
  publisher    = {{ACM}},
  year         = {1994},
  url          = {https://doi.org/10.1145/195058.195141},
  doi          = {10.1145/195058.195141},
  bibsource    = {dblp computer science bibliography, https://dblp.org}
}

@article{Zuckerman97sample,
  title={Randomness-optimal oblivious sampling},
  author={Zuckerman, David},
  journal={Random Structures \& Algorithms},
  volume={11},
  number={4},
  pages={345--367},
  year={1997},
  doi          = {10.1002/(SICI)1098-2418(199712)11:4\<345::AID-RSA4\>3.0.CO;2-Z},
  publisher={Wiley Online Library}
}

@article{AN93leader,
  author       = {Noga Alon and
                  Moni Naor},
  title        = {Coin-Flipping Games Immune Against Linear-Sized Coalitions},
  journal      = {{SIAM} J. Comput.},
  volume       = {22},
  number       = {2},
  pages        = {403--417},
  year         = {1993},
  url          = {https://doi.org/10.1137/0222030},
  doi          = {10.1137/0222030},
  bibsource    = {dblp computer science bibliography, https://dblp.org}
}

@article{BN00leader,
  author       = {Ravi B. Boppana and
                  Babu O. Narayanan},
  title        = {Perfect-Information Leader Election with Optimal Resilience},
  journal      = {{SIAM} J. Comput.},
  volume       = {29},
  number       = {4},
  pages        = {1304--1320},
  year         = {2000},
  url          = {https://doi.org/10.1137/S0097539796307182},
  doi          = {10.1137/S0097539796307182},
  bibsource    = {dblp computer science bibliography, https://dblp.org}
}

@article{Saks89leader,
  author       = {Michael E. Saks},
  title        = {A Robust Noncryptographic Protocol for Collective Coin Flipping},
  journal      = {{SIAM} J. Discret. Math.},
  volume       = {2},
  number       = {2},
  pages        = {240--244},
  year         = {1989},
  url          = {https://doi.org/10.1137/0402020},
  doi          = {10.1137/0402020},
  bibsource    = {dblp computer science bibliography, https://dblp.org}
}

@article{CL95leader,
  author       = {Jason Cooper and
                  Nathan Linial},
  title        = {Fast Perfect-Information Leader-Election Protocols with Linear Immunity},
  journal      = {Comb.},
  volume       = {15},
  number       = {3},
  pages        = {319--332},
  year         = {1995},
  url          = {https://doi.org/10.1007/BF01299739},
  doi          = {10.1007/BF01299739},
  bibsource    = {dblp computer science bibliography, https://dblp.org}
}

@inproceedings{Antonakopoulos06leader,
  author       = {Spyridon Antonakopoulos},
  editor       = {Jon M. Kleinberg},
  title        = {Fast leader-election protocols with bounded cheaters' edge},
  booktitle    = {Proceedings of the 38th Annual {ACM} Symposium on Theory of Computing,
                  Seattle, WA, USA, May 21-23, 2006},
  pages        = {187--196},
  publisher    = {{ACM}},
  year         = {2006},
  url          = {https://doi.org/10.1145/1132516.1132544},
  doi          = {10.1145/1132516.1132544},
  bibsource    = {dblp computer science bibliography, https://dblp.org}
}

@inproceedings{IMV23resilient,
  author       = {Peter Ivanov and
                  Raghu Meka and
                  Emanuele Viola},
  editor       = {Nikhil Bansal and
                  Viswanath Nagarajan},
  title        = {Efficient resilient functions},
  booktitle    = {Proceedings of the 2023 {ACM-SIAM} Symposium on Discrete Algorithms,
                  {SODA} 2023, Florence, Italy, January 22-25, 2023},
  pages        = {2867--2874},
  publisher    = {{SIAM}},
  year         = {2023},
  url          = {https://doi.org/10.1137/1.9781611977554.ch108},
  doi          = {10.1137/1.9781611977554.CH108},
  bibsource    = {dblp computer science bibliography, https://dblp.org}
}

@inproceedings{Meka17resilient,
  author       = {Raghu Meka},
  editor       = {Philip N. Klein},
  title        = {Explicit Resilient Functions Matching Ajtai-Linial},
  booktitle    = {Proceedings of the Twenty-Eighth Annual {ACM-SIAM} Symposium on Discrete
                  Algorithms, {SODA} 2017, Barcelona, Spain, Hotel Porta Fira, January
                  16-19},
  pages        = {1132--1148},
  publisher    = {{SIAM}},
  year         = {2017},
  url          = {https://doi.org/10.1137/1.9781611974782.73},
  doi          = {10.1137/1.9781611974782.73},
  bibsource    = {dblp computer science bibliography, https://dblp.org}
}

@misc{dodis2006fault,
  title={Fault-tolerant leader election and collective coin-flipping in the full information model},
  author={Dodis, Yevgeniy},
  year={2006},
  publisher={Survey}
}

@inproceedings{FHHHZ19,
  author       = {Yuval Filmus and
                  Lianna Hambardzumyan and
                  Hamed Hatami and
                  Pooya Hatami and
                  David Zuckerman},
  title        = {Biasing Boolean Functions and Collective Coin-Flipping Protocols over
                  Arbitrary Product Distributions},
  booktitle    = {46th International Colloquium on Automata, Languages, and Programming (ICALP)},
  series       = {LIPIcs},
  volume       = {132},
  doi          = {10.4230/LIPICS.ICALP.2019.58},
  year={2019},
  pages        = {58:1--58:13}
}

@article{harper1966optimal,
  title={Optimal numberings and isoperimetric problems on graphs},
  author={Harper, Lawrence H},
  journal={Journal of Combinatorial Theory},
  volume={1},
  number={3},
  pages={385--393},
  year={1966},
  publisher={Academic Press}
}

@article{bernstein1967maximally,
  title={Maximally connected arrays on the n-cube},
  author={Bernstein, Arthur J},
  journal={SIAM Journal on Applied Mathematics},
  volume={15},
  number={6},
  pages={1485--1489},
  year={1967},
  publisher={SIAM}
}

@article{lindsey1964assignment,
  title={Assignment of numbers to vertices},
  author={Lindsey, John H},
  journal={The American Mathematical Monthly},
  volume={71},
  number={5},
  pages={508--516},
  year={1964},
  publisher={Taylor \& Francis}
}

@article{hart1976note,
  title={A note on the edges of the n-cube},
  author={Hart, Sergiu},
  journal={Discrete Mathematics},
  volume={14},
  number={2},
  pages={157--163},
  doi          = {10.1016/0012-365X(76)90058-3},
  year={1976},
  publisher={Elsevier}
}

@article{GGL98selection,
  author       = {Oded Goldreich and
                  Shafi Goldwasser and
                  Nathan Linial},
  title        = {Fault-Tolerant Computation in the Full Information Model},
  journal      = {{SIAM} J. Comput.},
  volume       = {27},
  number       = {2},
  pages        = {506--544},
  year         = {1998},
  url          = {https://doi.org/10.1137/S0097539793246689},
  doi          = {10.1137/S0097539793246689},
  bibsource    = {dblp computer science bibliography, https://dblp.org}
}

@article{SV08selection,
  author       = {Saurabh Sanghvi and
                  Salil P. Vadhan},
  title        = {The Round Complexity of Two-Party Random Selection},
  journal      = {{SIAM} J. Comput.},
  volume       = {38},
  number       = {2},
  pages        = {523--550},
  year         = {2008},
  url          = {https://doi.org/10.1137/050641715},
  doi          = {10.1137/050641715},
  bibsource    = {dblp computer science bibliography, https://dblp.org}
}

@inproceedings{GVZ06selection,
  author       = {Ronen Gradwohl and
                  Salil P. Vadhan and
                  David Zuckerman},
  editor       = {Cynthia Dwork},
  title        = {Random Selection with an Adversarial Majority},
  booktitle    = {Advances in Cryptology - {CRYPTO} 2006, 26th Annual International
                  Cryptology Conference, Santa Barbara, California, USA, August 20-24,
                  2006, Proceedings},
  series       = {Lecture Notes in Computer Science},
  volume       = {4117},
  pages        = {409--426},
  publisher    = {Springer},
  year         = {2006},
  url          = {https://doi.org/10.1007/11818175\_25},
  doi          = {10.1007/11818175\_25},
  bibsource    = {dblp computer science bibliography, https://dblp.org}
}

@inproceedings{Li16TwoSource,
  author       = {Xin Li},
  editor       = {Irit Dinur},
  title        = {Improved Two-Source Extractors, and Affine Extractors for Polylogarithmic
                  Entropy},
  booktitle    = {{IEEE} 57th Annual Symposium on Foundations of Computer Science, {FOCS}
                  2016, Hyatt Regency, New Brunswick, New Jersey, USA, October 9-11,
                  2016},
  pages        = {168--177},
  publisher    = {{IEEE} Computer Society},
  year         = {2016},
  url          = {https://doi.org/10.1109/FOCS.2016.26},
  doi          = {10.1109/FOCS.2016.26},
  bibsource    = {dblp computer science bibliography, https://dblp.org}
}

@inproceedings{BCDT19,
  author       = {Avraham Ben{-}Aroya and
                  Gil Cohen and
                  Dean Doron and
                  Amnon Ta{-}Shma},
  editor       = {Dimitris Achlioptas and
                  L{\'{a}}szl{\'{o}} A. V{\'{e}}gh},
  title        = {Two-Source Condensers with Low Error and Small Entropy Gap via Entropy-Resilient
                  Functions},
  booktitle    = {Approximation, Randomization, and Combinatorial Optimization. Algorithms
                  and Techniques, {APPROX/RANDOM} 2019, September 20-22, 2019, Massachusetts
                  Institute of Technology, Cambridge, MA, {USA}},
  series       = {LIPIcs},
  volume       = {145},
  pages        = {43:1--43:20},
  publisher    = {Schloss Dagstuhl - Leibniz-Zentrum f{\"{u}}r Informatik},
  year         = {2019},
  url          = {https://doi.org/10.4230/LIPIcs.APPROX-RANDOM.2019.43},
  doi          = {10.4230/LIPICS.APPROX-RANDOM.2019.43},
  bibsource    = {dblp computer science bibliography, https://dblp.org}
}

@inproceedings{BDT17TwoSource,
  author       = {Avraham Ben{-}Aroya and
                  Dean Doron and
                  Amnon Ta{-}Shma},
  editor       = {Hamed Hatami and
                  Pierre McKenzie and
                  Valerie King},
  title        = {An efficient reduction from two-source to non-malleable extractors:
                  achieving near-logarithmic min-entropy},
  booktitle    = {Proceedings of the 49th Annual {ACM} {SIGACT} Symposium on Theory
                  of Computing, {STOC} 2017, Montreal, QC, Canada, June 19-23, 2017},
  pages        = {1185--1194},
  publisher    = {{ACM}},
  year         = {2017},
  url          = {https://doi.org/10.1145/3055399.3055423},
  doi          = {10.1145/3055399.3055423},
  bibsource    = {dblp computer science bibliography, https://dblp.org}
}

@inproceedings{CGL21affine,
  author       = {Eshan Chattopadhyay and
                  Jesse Goodman and
                  Jyun{-}Jie Liao},
  title        = {Affine Extractors for Almost Logarithmic Entropy},
  booktitle    = {62nd {IEEE} Annual Symposium on Foundations of Computer Science, {FOCS}
                  2021, Denver, CO, USA, February 7-10, 2022},
  pages        = {622--633},
  publisher    = {{IEEE}},
  year         = {2021},
  url          = {https://doi.org/10.1109/FOCS52979.2021.00067},
  doi          = {10.1109/FOCS52979.2021.00067},
  bibsource    = {dblp computer science bibliography, https://dblp.org}
}

@inproceedings{CL22sumset,
  author       = {Eshan Chattopadhyay and
                  Jyun{-}Jie Liao},
  editor       = {Stefano Leonardi and
                  Anupam Gupta},
  title        = {Extractors for sum of two sources},
  booktitle    = {{STOC} '22: 54th Annual {ACM} {SIGACT} Symposium on Theory of Computing,
                  Rome, Italy, June 20 - 24, 2022},
  pages        = {1584--1597},
  publisher    = {{ACM}},
  year         = {2022},
  url          = {https://doi.org/10.1145/3519935.3519963},
  doi          = {10.1145/3519935.3519963},
  bibsource    = {dblp computer science bibliography, https://dblp.org}
}

@inproceedings{Li23sumset,
  author       = {Xin Li},
  title        = {Two Source Extractors for Asymptotically Optimal Entropy, and (Many)
                  More},
  booktitle    = {64th {IEEE} Annual Symposium on Foundations of Computer Science, {FOCS}
                  2023, Santa Cruz, CA, USA, November 6-9, 2023},
  pages        = {1271--1281},
  publisher    = {{IEEE}},
  year         = {2023},
  url          = {https://doi.org/10.1109/FOCS57990.2023.00075},
  doi          = {10.1109/FOCS57990.2023.00075},
  bibsource    = {dblp computer science bibliography, https://dblp.org}
}

@article{KZ07NOSF,
  author       = {Jesse Kamp and
                  David Zuckerman},
  title        = {Deterministic Extractors for Bit-Fixing Sources and Exposure-Resilient
                  Cryptography},
  journal      = {{SIAM} J. Comput.},
  volume       = {36},
  number       = {5},
  pages        = {1231--1247},
  year         = {2007},
  url          = {https://doi.org/10.1137/S0097539705446846},
  doi          = {10.1137/S0097539705446846},
  bibsource    = {dblp computer science bibliography, https://dblp.org}
}

\appendix

\section{Comparison to Proof Strategy of \texorpdfstring{\cite{RSZ02leader}}{[RSZ02]}}
\label{sec: appendix compare to rsz}

Since our coin flipping impossibility results from \cref{thm: intro protocol lb} are inspired by the work of \cite{RSZ02leader} and our proofs build on their arguments, we here briefly sketch their strategy for 2 rounds and how we improve upon it. 

Let $\cF = \{\pi_{\alpha}\}$ where $\alpha\in \zo^{\ell}$ and let $\pi_{\alpha}$ be the induced second round protocol when in the first round the players output $\alpha$.
\cite{RSZ02leader} proceeds by finding a set of bad players in the second round to bias the functions in $\cF$.
To do this, as hinted at earlier, \cite{RSZ02leader} analyzes the same semi-random process as in \cref{lem:proof overview most random sets with small heavy set corrputs f}. At the end of the process, they also obtain a common random set $B_R$ and a specialized heavy set $B_H = B_H(f)$ such that for most functions $f\in \cF$, $B_R\cup B_H(f)$ can bias the function $f$. They then pick another random set $B_{R'}$ and bound the probability that $B_H(f)\subset B_{R'}$ for fixed $f$, and argue that in expectation, $\delta \approx \left(\frac{r}{\ell}\right)^{2^{\ell / r}}$ fraction of functions $f\in \cF$ will be such that $B_H(f)\subset B_{R'}$ where $\abs{B_{R'}} = r$.
They let $B_R\cup B_{R'}$ be the set of bad players that will bias the second round protocols.

Now, let $S\subset \zo^{\ell}$ be the set of $\alpha\in \zo^{\ell}$ such that $\pi_{\alpha}$ can be biased by $B_R\cup B_{R'}$, and let $g: \zo^\ell\to \zo$ be such that $g(\alpha) = 1 \iff \alpha\in S$.
We know from above that $\Pr[g = 1]\ge \delta$.
Now, the goal for the first round is to find bad players to bias $g$ towards $1$ so that with high probability, the resulting second round protocol can be biased using $B_R\cup B_{R'}$.
To do this, \cite{RSZ02leader} use the KKL theorem \cite{KKL88influence} which lets them do this by finding $O(\ell / (\delta \log \ell))$ bad players.
To minimize the total number of bad players, they end up using $O(\ell / \log^{(3)}\ell)$ bad players.
For multiple rounds, their result follows by an inductive argument, showing that $O\left(\ell / \log^{(2k-1)}\ell\right)$ bad players suffice to bias $k$ round protocols.

Our key insight is that instead of first committing to the set $B_{R'}$ that covers a small non-trivial fraction of $\cF$ and then biasing the first round to ensure that $B_{R}\cup B_{R'}$ can bias most of the resulting second round protocols, we can delay committing and instead \emph{commit incrementally}. We do this by initially setting $B_{R'} = [\ell]$, so that it can bias most of $\cF$; we then ``chisel away'' half of $B_{R'}$, so that we can only bias half of the functions from $\cF$. To fix this, we use the KKL theorem so that out of the set of $\pi_\alpha$ functions that are being considered, we again can bias most of them. We repeatedly do this until $B_{R'}$ becomes very small, while always maintaining that we can bias most of the resultant second round protocols. Doing this gives us an exponential improvement, with $O\left(\ell / \log^{(2)}\ell\right)$ bad players sufficing to bias two-round protocols.

To illustrate in a simple setting why delaying committing can give an exponential improvement, we consider the following scenario: Suppose that we are given a function $f: \zo^{\ell}\to \zo^t$ and our goal is to bias $f$ so that it outputs some single element $y$ from $\zo^t$ with high probability. One way to do this is to fix an element $y\in \zo^t$ that appears with probability at least $\frac{1}{2^t}$, then use the KKL theorem to obtain a set of bad players that can bias $f$ to ensure that $y$ is output with probability $0.99$. While this works, the upper bound on the size of the set of bad players resulting from this approach is $O\left(2^t\cdot \frac{\ell}{\log \ell}\right)$ (see \Cref{thm: kkl}).  In contrast, a ``delayed commitment'' approach is to maintain a set $S$ (initially setting $S = \zo^t$); cut $S$ in half; use KKL to find $O\left(\frac{\ell}{\log\ell}\right)$ bad players that will ensure that at least $0.99$ fraction of the inputs lie in $S$; and repeat. We do this cutting process $t$ times, so that at the end of it $\abs{S} = 1$, and we let its sole element be the final output string $y$.  With this approach, the number of bad players required used will be only $O\left(t\cdot \frac{\ell}{\log \ell}\right)$, which is an exponential improvement over $O\left(2^t\cdot \frac{\ell}{\log \ell}\right)$ in terms of the dependence on $t$.

\dobib

\end{document}